\documentclass[11pt]{article}
\usepackage{jheppub}
\usepackage{amsthm}
\usepackage{tikz}
\usepackage{circuitikz}
\usetikzlibrary{decorations.pathreplacing,decorations.markings}
\usepackage{mathrsfs}
\usepackage{mathtools}
\usepackage{dsfont}
\usepackage{blkarray}
\usepackage{todonotes}
\usepackage[T1]{fontenc}
\usepackage{subcaption}
\usepackage{braket}
\usepackage{booktabs}
\usepackage{multirow}
\usepackage{makecell}
\usepackage{cancel}
\usetikzlibrary{tikzmark}

\usepackage{amsmath,amssymb,mathrsfs,amsthm,tikz,shuffle,paralist,accents}
\usepackage{caption}
\usepackage[dvipsnames]{xcolor}

\usetikzlibrary{shapes.misc}
\usetikzlibrary{cd}
\usetikzlibrary{snakes}
\usetikzlibrary{calc}
\usetikzlibrary{fadings}
\usetikzlibrary{decorations.markings}
\usetikzlibrary{positioning}

\newtheorem{theorem}{Theorem}

\tikzset{
  midarrow/.style={
    postaction={decorate},
    decoration={
      markings,
      mark=at position 0.5 with {\arrow[scale=1.4]{stealth}}
    }
  }
}

\definecolor{charcoal}{HTML}{343837}

\title{\centering Steinmann Violation and Minimal Cuts}

\author{Holmfridur~S.~Hannesdottir,$^a$ Luke~Lippstreu,$^b$ Andrew~J.~McLeod,$^b$ and Maria~Polackova$^b$}
\affiliation{${}^a$ Institute for Advanced Study, Einstein Drive, Princeton, NJ 08540, USA}
\affiliation{${}^b$ Higgs Centre for Theoretical Physics, School of Physics and Astronomy,
The University of Edinburgh, Edinburgh EH9 3FD, Scotland, UK}

\abstract{The Steinmann relations are known to be violated with respect to some --- but not all --- two-particle momentum channels in massless Feynman integrals. We trace the source of this Steinmann violation to a special class of singularities, which arise from partially-overlapping minimal cuts. This allows us to propose an efficient graphical test for predicting which Steinmann relations will be violated by massless Feynman integrals of a given topology, which can be applied at any loop order. We provide evidence for this test by correctly predicting all instances of Steinmann violation in the complete set of known two-loop integrals that contribute to five-particle scattering with one or two external masses.}

\begin{document}

\maketitle

\section{Introduction}

The Steinmann relations, discovered over sixty years ago~\cite{Steinmann,Steinmann2}, place strong constraints on the analytic structure of scattering amplitudes. The original form of these relations states that the double discontinuities of scattering amplitudes must vanish when computed with respect to partially-overlapping momentum channels, namely
\begin{align} \label{eq:steinmann}
    \text{Disc}_{s_{\scalebox{.4}{$J$}}-m_{\scalebox{.4}{$J$}}^2}\big(  \text{Disc}_{s_{\scalebox{.4}{$I$}}-m_{\scalebox{.4}{$I$}}^2} \big( {\cal A} \big) \big) = 0  \qquad 
    \begin{cases}
    I \not\subset J \text{ and } J \not \subset I \\
    I \cap J \neq \emptyset
    \end{cases}
\end{align}
where $I$ and $J$ are sets of external particle indices, $s_I = (\sum_{i \in I} p_i)^2$ is the squared sum of external momenta labeled by $i \in I$, and $m_I$ is the sum of masses that cross the $s_I$ cut.\footnote{More generally, $m_I$ can involve differences of masses that cross the $s_I$ cut, making these pseudothreshold rather than threshold discontinuities.} Originally, the Steinmann relations were proven in the physical  region for amplitudes that describe the interaction of stable massive particles~\cite{Steinmann,Steinmann2,araki:1961,pham,Cahill:1973qp}. However, in recent years, a number of analogous relations have been discovered that apply to massless amplitudes, extend to longer sequences of discontinuities, and apply to discontinuities beyond multi-particle thresholds and pseudothresholds~\cite{Drummond:2017ssj,Caron-Huot:2019bsq,Bourjaily:2020wvq,Benincasa:2020aoj}. Beyond exposing striking universal properties of perturbative quantum field theory, these relations have proven instrumental for perturbative bootstrap approaches to computing quantities such as scattering amplitudes and form factors (see for instance~\cite{Caron-Huot:2020bkp,Dixon:2022xqh}).

In contrast to the original Steinmann relations, whose validity can be traced back to the requirement of causality, these newer versions of the Steinmann relations remain on less firm theoretical footing. This is particularly true for massless amplitudes, which are known to violate equation~\eqref{eq:steinmann} when one (or both) of the discontinuities is computed with respect to a two-particle momentum channel~\cite{Caron-Huot:2016owq}. Despite the many advances that have been made in our understanding of the analytic structure of massless amplitudes over the last five years (see for instance~\cite{Prlina:2018ukf,Mizera:2021icv,Lippstreu:2022bib,Hannesdottir:2022bmo,Lippstreu:2023oio,Fevola:2023kaw, Fevola:2023fzn, Caron-Huot:2023ikn,Hannesdottir:2024cnn,He:2024fij,Hannesdottir:2024hke}), no deep understanding of these instances of Steinmann violation has yet been developed. In fact, no systematic study has even been carried out to pinpoint exactly when Steinmann violation occurs, despite the fact that an abundance of theoretical data exists.

In this paper, we start to fill in this gap by characterizing when and why massless amplitudes violate the Steinmann relations. To do so, we make use of the Landau equations~\cite{nakanishi1959,Landau:1959fi,Bjorken:1959fd} --- which describe the kinematic locations at which Feynman integrals can become singular --- to systematically study when pairs of singularities with the right properties lead to Steinmann violation. Importantly, this includes the requirement that the corresponding double discontinuities are consistent with 
the hierarchical principle~\cite{boyling1968homological,Landshoff1966}, which states that discontinuities localize the integration contour of Feynman integrals to the on-shell locus of whichever propagators are involved in pinching the integration contour. As a result, whichever propagators are placed on-shell in order to compute the first discontinuity in~\eqref{eq:steinmann} must remain on shell when we compute the second discontinuity. 
(If we restrict our attention to the first discontinuity of Feynman integrals, the same result follows from Cutkosky's formula~\cite{Cutkosky:1960sp}.) We will review this phenomenon and its implications in the next section. 

At one loop, where an exhaustive analysis is possible, we can find all pairs of solutions to the Landau equations that satisfy this requirement. Restricting our attention to massless Feynman integrals, these pairs of solutions can all be associated with the following double cut, which involves at least two adjacent massless external legs: 
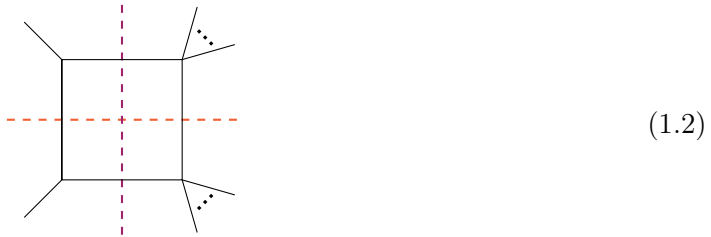
\begin{equation} 
\begin{gathered} \label{eq:one_loop_min_cut}
\begin{tikzpicture}[scale=1] 
        \coordinate (a) at (-.8,.8);
        \coordinate (b) at (.8,.8);
        \coordinate (c) at (.8,-.8);
        \coordinate (d) at (-.8,-.8);
        \coordinate (h) at (.8,0);

        \coordinate (a1) at (-1.3, 1.3);
        \coordinate (d1) at (-1.3, -1.3);
        \coordinate (b1) at (1, 1.5);
        \coordinate (b2) at (1.5, 1);
        \coordinate (c1) at (1, -1.5);
        \coordinate (c2) at (1.5, -1);

        \coordinate (cutL) at (-1.53, 0);
        \coordinate (cutR) at (1.53, 0);
        \coordinate (cutB) at (0,-1.53);
        \coordinate (cutT) at (0,1.53);
        
        \draw[] (a1) to (a);
        \draw[] (d1) to (d);
        \draw[] (b1) to (b);
        \draw[] (b2) to (b);
        \draw[] (c1) to (c);
        \draw[] (c2) to (c);

        \draw[dashed,thick,RedOrange] (cutL) to (cutR);
        \draw[dashed,thick,RedViolet] (cutB) to (cutT);

       \draw [dotted, line width=1.2pt] (1.2, 1) -- (1,1.2);
        \draw [dotted, line width=1.2pt] (1.2, -1) -- (1,-1.2);

        \draw[] (d) to node[midway, left] {}(a)
        to node[midway, above] {}(b)
        to node[midway, right] {}(c) 
        to node[midway, below] {}(d)
        to node[midway, right] {}(a);      
    \end{tikzpicture}
\end{gathered}
\end{equation}
\noindent (here, the ellipses indicate that the remaining external edges are allowed to be massive or massless). That is, given a pair of partially-overlapping channels $s_I$ and $s_J$, solutions to the Landau equations exist at $s_J=0$ that are consistent with the first discontinuity $\text{Disc}_{s_{\scalebox{.4}{$I$}}}(\mathcal{A})$ when at least one of $s_I$ or $s_J$ corresponds to a two-particle channel. As we will see, the offending solutions to the Landau equations all involve setting some of the virtual momenta to be either soft or collinear, giving us some indication of why Steinmann-violating solutions to the Landau equations don't generically arise for scattering processes involving massive internal particles, or for pairs of channels involving three or more momenta.

Having identified the two-mass hard box diagram (and its massless degenerations) as the unique cut diagram that leads to Steinmann violation at one loop, we next examine the broader class of cut diagrams in which all pairs of vertices are allowed to be connected by any number of internal lines:
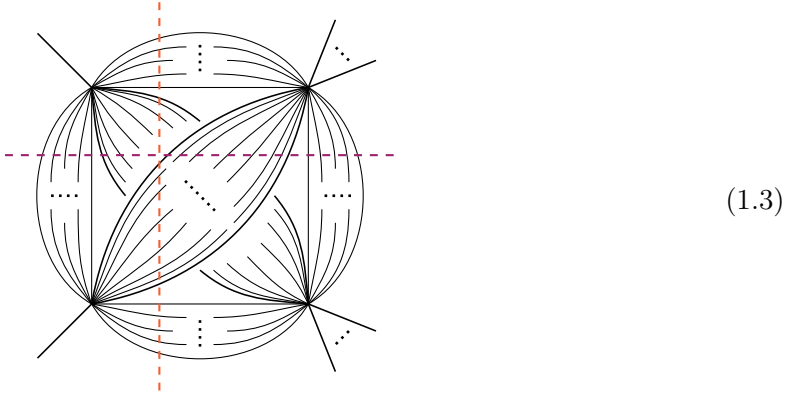
\begin{equation}
\begin{gathered} \label{eq:connected_minimal_cuts}
        \begin{tikzpicture}[scale=1.8]
        \coordinate (a) at (-.8,.8);
        \coordinate (b) at (.8,.8);
        \coordinate (c) at (.8,-.8);
        \coordinate (d) at (-.8,-.8);
        \coordinate (h) at (.8,0);

        \coordinate (a1) at (-1.2, 1.2);
        \coordinate (d1) at (-1.2, -1.2);
        \coordinate (b1) at (1.2, 1.2);
        \coordinate (c1) at (1.2, -1.2);

        \coordinate (i2) at (-0.3, 1.3);
        \coordinate (j2) at (0.3, 1.3);

        \draw[] (b) to (c);
        \draw[] (d) to (a);
        \coordinate (e1) at (1.2, 0.2);
        \coordinate (f1) at (1.2, -0.2);
        \coordinate (e2) at (1.3, 0.3);
        \coordinate (f2) at (1.3, -0.3);
        
        \draw[] (b) to [out=-80, in=90] (0.9,0.1);
        \draw[] (0.9,-0.1) to [out=-90, in=80] (c);
        \draw[] (b) to [out=-60, in=90] (1,0.2);
        \draw[] (1,-0.2) to [out=-90, in=60] (c);
        \draw[] (b) to [out=-40, in = 90] (1.1,0.1);
        \draw[] (1.1,-0.1) to [out=-90, in=40] (c);
        \draw[] (b) to [out=-30, in = 30] (c);
        
        \coordinate (g1) at (-1.2, 0.2);
        \coordinate (h1) at (-1.2, -0.2);
        \coordinate (g2) at (-1.3, 0.3);
        \coordinate (h2) at (-1.3, -0.3);

        \draw[] (a) to [out=-100, in = 90] (-0.9,0.1);
        \draw[] (-0.9,-0.1) to [out=-90, in = 110] (d);
        \draw[] (a) to [out=-120, in = 90] (-1,0.2);
        \draw[] (-1,-0.2) to [out=-90, in = 120] (d);
        \draw[] (a) to [out=-130, in = 90] (-1.1,0.1);
        \draw[] (-1.1,-0.1) to [out=-90, in=130] (d);
        \draw[] (a) to [out=-150, in=150] (d);

        \draw [dotted, line width=1pt] (0.92, 0) -- (1.12,0);
        \draw [dotted, line width=1pt] (-1.1, 0) -- (-0.9,0);
        
        \draw[] (a) to (b);
        \draw[] (a) to [out=10, in = 180] (-0.1, 0.9);
        \draw[] (0.1, 0.9) to [out=0, in = 170] (b);
        \draw[] (a) to [out=30, in= 180] (-0.2, 1);
        \draw[] (0.2, 1) to [out=0, in=150] (b);
        \draw[] (a) to [out=40, in=180] (-0.1, 1.1);
        \draw[] (0.1, 1.1) to [out=0, in=140] (b);
        \draw[] (a) to [out=60, in=120] (b);

        \draw [dotted, line width=1pt] (0, 0.92) -- (0,1.12);

        \draw[] (d) to (c);
        \draw[] (d) to [out=-10, in=180] (-0.1, -0.9);
        \draw[] (0.1, -0.9) to [out=0, in=-170] (c);
        \draw[] (d) to [out=-30, in=180] (-0.2, -1);
        \draw[] (0.2, -1) to [out=0, in=-150] (c);
        \draw[] (d) to [out=-40, in=180] (-0.1,-1.1);
        \draw[] (0.1,-1.1) to [out=0, in=-140] (c);
        \draw[] (d) to [out=-60, in=-120] (c);

        \draw[dotted, line width=1pt] (0,-0.92) -- (0,-1.12);

        \draw [dotted, line width=1pt] (-0.11, 0.11) -- (0.11,-0.11);
        \draw[line width=0.6pt] (d) to [out=80, in=-170] (b);
        \draw[line width=0.6pt] (d) to [out=10, in=-100] (b);
        \draw[] (d) to [out=50, in=-140] (-0.2, -0.1);
        \draw[] (0.1,0.2) to [out=45, in=-140] (b);
        \draw[] (d) to [out=40, in=-130] (-0.1, -0.2);
        \draw[] (0.2,0.1) to [out=45, in=-130] (b);
        \draw[] (d) to [out=60, in=-137] (-0.25, 0.08);
        \draw[] (-0.08, 0.25) to [out=45, in=-150] (b);
        \draw[] (d) to [out=30, in=-133] (0.08, -0.25);
        \draw[] (0.25, -0.08) to [out=45, in=-120] (b);
        \draw[] (d) to [out=70, in=-136] (-0.25,0.2);
        \draw[] (-0.2,0.25) to [out=44, in=-160] (b);
        \draw[] (d) to [out=20, in=-134] (0.2,-0.25);
        \draw[] (0.25,-0.2) to [out=46, in=-110] (b);

         \draw[line width=0.6pt] (a) to [out=-5, in=140] (0,0.55);
         \draw[line width=0.6pt] (a) to [out=-85, in=130] (-0.55,0);
         \draw[line width=0.6pt] (0.55,0) to [out=-50, in=95] (c);
         \draw[line width=0.6pt] (0,-0.55) to [out=-40, in=175] (c);
         \draw[] (a) to [out=-10, in=138] (-0.1, 0.55);
         \draw[] (a) to [out=-80, in=132] (-0.55, 0.1);
         \draw[] (0.1, -0.55) to [out=-42, in=170] (c);
         \draw[] (0.55, -0.1) to [out=-48, in=100] (c);
         \draw[] (a) to [out=-15, in=137] (-0.2, 0.5);
         \draw[] (a) to [out=-75, in=133] (-0.5,0.2);
         \draw[] (0.2,-0.5) to [out=-43, in=165] (c);
         \draw[] (0.5,-0.2) to [out=-47, in=105] (c);
         \draw[] (a) to [out=-30, in=136] (-0.35, 0.45);
         \draw[] (a) to [out=-60, in=134] (-0.45, 0.35);
         \draw[] (0.35, -0.45) to [out=-46, in=155] (c);
         \draw[] (0.45, -0.35) to [out=-44, in=115] (c);
         
        \coordinate (a1) at (-1.2, 1.2);
        \coordinate (d1) at (-1.2, -1.2);
        \coordinate (b1) at (1.2, 1.2);
        \coordinate (c1) at (1.2, -1.2);
        
        \draw[line width=0.6pt] (1,1.3) to (b);
        \draw[line width=0.6pt] (1.3,1) to (b);
        \draw[line width=0.6pt] (1,-1.3) to (c);
        \draw[line width=0.6pt] (1.3,-1) to (c);

        \draw[line width=0.6pt] (a1) to (a);
        \draw[line width=0.6pt] (d1) to (d);
        
        \draw[dotted, line width=1pt] (1.1,1) to (1,1.1);
        \draw[dotted, line width=1pt] (1.1,-1) to (1, -1.1);

        \draw[RedViolet,thick,dashed] (-1.44,0.3) to (1.44,0.3);
        \draw[RedOrange,thick,dashed] (-.3,-1.44) to (-.3,1.44);

\end{tikzpicture}
\end{gathered}
\end{equation}

\noindent (Note that, as before, at least two of the adjacent external particles are required to be massless.) We find that one can again find a pair of massless threshold solutions to the Landau equations on the support of these cuts, which we expect will lead to violations of the Steinmann relations. Notably, these diagrams constitute nothing other than the set of minimal cuts (as introduced in~\cite{Hannesdottir:2024cnn}) that can arise for a pair of partially-overlapping momentum channels, in which the cuts separate the diagram into four connected subgraphs. Correspondingly, this gives us an efficient graphical test to identify if these Steinmann-violating singularities can arise in a given Feynman integral topology --- one just needs to construct all the minimal cuts that exist for a given pair of partially-overlapping channels, and check whether any of them are isomorphic to~\eqref{eq:connected_minimal_cuts}.

While easy to implement, it is worth emphasizing that this graphical test only tells us whether or not this specific class of Steinmann-violating singularities can arise in a given Feynman integral topology.
We cannot currently rule out the possibility that additional Steinmann-violating singularities exist, that do not correspond to cuts of the form~\eqref{eq:connected_minimal_cuts}.  However, at two loops we conjecture that no such additional Steinmann-violating singularities occur. We provide evidence for this conjecture by analyzing all instances of Steinmann violation that appear in the complete set of integrals that contribute to five-particle scattering with one external mass~\cite{Abreu:2023rco}, as well as all known five-particle two-mass integrals~\cite{Abreu:2024yit} and certain massless six-particle integrals~\cite{Henn:2024ngj}. In particular, we find that  the Steinmann relations are violated at some order in the $\epsilon = \frac{d-4}{2}$ expansion in dimensional regularization if and only if one of the minimal cuts for the relevant partially-overlapping channels takes the form~\eqref{eq:connected_minimal_cuts}. Note that this graphical test makes different predictions for each integral topology, so these predictions can change when the internal edges of a diagram are contracted. Thus, even when Steinmann relations are violated by integrals in the top sector of a family, they are often restored in lower sectors, due to the fact that integrals in these lower sectors no longer support any Steinmann-violating cuts.

While we carry out our analysis in this paper in dimensional regularization, we make no attempt to predict the specific order in $\epsilon$ at which the Steinmann relations are first violated. While such predictions would indeed prove useful --- in particular, in examples where Steinmann relations are still obeyed by the finite part of an integral but are violated at $\mathcal{O}(\epsilon)$ --- we expect that making such predictions would rely on knowledge of the numerator that appears in a given Feynman integral, and would possibly require carrying out algebraic blowups (see for instance~\cite{pham,Berghoff:2022mqu,Fevola:2023kaw,Fevola:2023fzn}). This would significantly complicate the otherwise simple graphical test that we have proposed, which allows us to reliably predict when the Steinmann relations involving two-particle momentum channels will not be violated at any order in dimensional regularization. Even without more precise knowledge of the order in dimensional regularization at which the Steinmann relations are violated, these types of predictions are expected to prove useful for bootstrap methods, and for determining the boundary conditions of the systems of differential equations that describe families of Feynman integrals.

The remainder of the paper is organized as follows. In section~\ref{sec:landau_review}, we establish our conventions and review the tools from Landau analysis that we will make use of. We then go on in section~\ref{sec:steinmann_viiolating_solutions} to describe the properties that solutions to the Landau equations must have in order to lead to violations of the Steinmann relations, and identify all such solutions that can appear at one loop. This observation is then refashioned into a graphical test that allows us to easily identify when the Steinmann relations will be violated in one-loop integrals. In section~\ref{sec:all_loop_solutions}, we construct Steinmann-violating solutions to the Landau equations for the two-mass hard acnode and the two-mass hard crossed boxed, and ultimately show that similar solutions to the Landau equations can be identified to all loop orders. We then return our attention to two-loop integrals in section~\ref{sec:two_loops}, where we conjecture that we have identified all Steinmann-violating solutions to the Landau equations that can appear through this loop order. This allows us to extend our graphical test for Steinmann violation to all massless two-loop integrals. We present evidence for the validity of this test by checking its predictions against the full set of integrals that contribute to five-particle scattering with one external mass, as well as five-particle integrals that involve two external masses and massless six-particle integrals. Finally, we conclude by highlighting some open questions and interesting directions for future work.

\newpage

\section{Feynman Integrals and Landau Analysis}
\label{sec:landau_review}

The basic objects we study in this paper are Feynman integrals. We focus on these building blocks because --- although we would ultimately like to understand the analytic properties of more physically meaningful quantities such as scattering amplitudes --- it often proves possible to make more refined predictions about the discontinuities of individual integrals. These stronger constraints leverage information about how momentum flows within each integral, and thus distinguish between integrals that correspond to different diagram topologies. 

In more detail, given a graph $G$ and a choice of preferred vertex $v_\infty$ we begin by distinguishing a set of \emph{external edges}, which are those that are incident to $v_\infty$, from \emph{internal edges}, which are not. In practice, the vertex $v_\infty$ is generally omitted in favor of lines that do not end at any vertex. To this graph, we then associate the class of integrals
\begin{equation}
    \label{eq:Feynman_integral}
    {\cal I}_G(p, m) = \frac{(-1)^E}{(i \pi)^{L D/2}} \int {\rm d}^D k_1\cdots {\rm d}^D k_L\frac{\mathcal{N}}{A_1^{\nu_1} \cdots A_E^{\nu_E}}\, ,
\end{equation}
where $E$ and $L$ represent the number of internal edges and the number of loops formed by these internal edges, $D$ denotes the spacetime dimension, and each factor $A_i = q_i^2-m_i^2+i\delta$ is an inverse propagator that depends on the mass $m_i$ and the momentum $q_i$ flowing through the $i^{\text{th}}$ internal edge. The parameter $\delta \ll 1$ just keeps track of how the integration contour should be deformed if it intersects one of the singular hypersurfaces $A_i=0$. The value of the momentum $q_i$ is determined by the requirement that momentum is conserved at each vertex, and depends linearly on the loop momenta $k_j$ that are integrated over, as well as the momenta $p_k$ flowing through the external edges. We allow the propagators to take arbitrary powers $\nu_i$, and the numerator $\mathcal{N}$ to depend on the loop momenta; for more details, see for instance~\cite{Weinzierl:2022eaz}.

Feynman integrals are often divergent, due to singularities that arise when the momentum flowing through one or more of the propagators vanishes or becomes large. These infrared and ultraviolet divergences can be regulated by working in $D = d - 2 \epsilon$ dimensions, where $d$ is an integer and $\epsilon$ is treated as an expansion parameter. Feynman integrals can then be computed order-by-order in $\epsilon$, for instance using differential equation methods~\cite{Kotikov:1990kg,Kotikov:1991pm,Remiddi:1997ny,Gehrmann:1999as}. Unitarity dictates that the coefficients of this expansion must be multi-valued functions that develop discontinuities whenever certain physical situations arise, for instance when it becomes kinematically possible to produce new species of virtual particles. However, beyond the physical region (which describes scattering processes that can be realized in Lorentzian spacetime) and away from the principal branch of a given Feynman integral (as picked out by the Feynman $i \delta$ prescription) the physical interpretation of the discontinuities that appear in Feynman integrals is generally less clear.

Despite our incomplete understanding of the analytic structure of Feynman integrals, an abundance of theoretical data now attests to the existence of several universal principles. Chief among these are the Steinmann relations, which state that double discontinuities with respect to partially-overlapping momentum channels vanish. Although the Steinmann relations have only been proven for amplitudes involving massive external and virtual particles~\cite{Steinmann,Steinmann2,pham,araki:1961,Cahill:1973qp}, an analogous statement has been observed to hold in amplitudes and Feynman integrals that involve massless virtual particles:
\begin{align} \label{eq:steinmann_three_particles}
    \text{Disc}_{s_{\scalebox{.4}{$J$}}}\big(  \text{Disc}_{s_{\scalebox{.4}{$I$}}} \big( {\cal I}_G(p, m) \big) \big) = 0  \qquad 
    \begin{cases}
    I \not\subset J \ \wedge\ J \not \subset I \ \wedge\  I \cap J \neq \emptyset \\
    |I| \ge 3 \ \wedge\ |J| \ge 3
    \end{cases}
\end{align}
where $I$ and $J$ are sets of external particle indices, and the corresponding Mandelstam variables are defined by $s_I = (\sum_{i \in I} p_i)^2$. However, the situation is more complicated when at least one of the sets $I$ or $J$ consists of just a pair of (massless) particle indices:
\begin{align} \label{eq:steinmann_two_particles}
    \text{Disc}_{s_{\scalebox{.4}{$J$}}}\big(  \text{Disc}_{s_{\scalebox{.4}{$I$}}} \big( {\cal I}_G(p, m) \big) \big) \stackrel{?}{=} 0  \qquad 
    \begin{cases}
    I \not\subset J \ \wedge\ J \not \subset I \ \wedge\  I \cap J \neq \emptyset \\
    |I| = 2 \ \vee\ |J| = 2
    \end{cases}
\end{align}
No clear understanding yet exists for when the double discontinuities described by~\eqref{eq:steinmann_two_particles} are expected to be zero. Given the crucial role that the Steinmann relations for higher-particle momentum channels have played in bootstrap calculations --- which have now reached unprecedented orders in perturbation theory~\cite{Dixon:2016nkn,Drummond:2018caf,Caron-Huot:2018dsv,Dixon:2022rse,Dixon:2023kop,Basso:2024hlx} --- it seems clear that such a study should be undertaken.

It is worth briefly highlighting that the Steinmann relations are also known to be violated in Feynman integrals that involve massive propagators. One such example was presented in section 8.1 of~\cite{Hannesdottir:2022xki}. A more physically-relevant example is given by the diagram
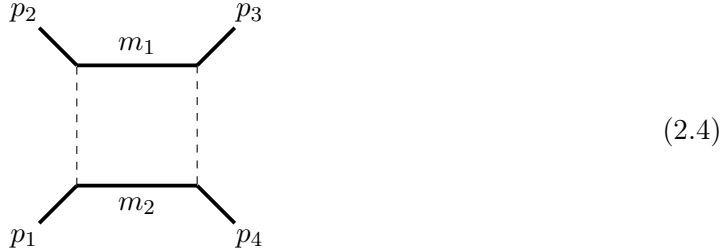
\begin{equation} 
\begin{gathered}
\label{eq:massive_steinmann_violation}
\begin{tikzpicture}[scale=1] 
        \coordinate (a) at (-.8,.8);
        \coordinate (b) at (.8,.8);
        \coordinate (c) at (.8,-.8);
        \coordinate (d) at (-.8,-.8);
        \coordinate (h) at (.8,0);

        \coordinate (a1) at (-1.3, 1.3);
        \coordinate (d1) at (-1.3, -1.3);
        \coordinate (b1) at (1.3, 1.3);
        \coordinate (c1) at (1.3, -1.3);

        \coordinate (cutL) at (-1.53, 0);
        \coordinate (cutR) at (1.53, 0);
        \coordinate (cutB) at (0,-1.53);
        \coordinate (cutT) at (0,1.53);
        
        \draw[line width=0.5mm] (a1) to (a);
        \draw[line width=0.5mm] (d1) to (d);
        \draw[line width=0.5mm] (b1) to (b);
        \draw[line width=0.5mm] (c1) to (c);

        \draw[dashed] (d) to node[midway, left] {}(a);
        \draw[line width=0.5mm] (a) to node[midway, above] {$m_1$}(b);
        \draw[dashed] (b) to node[midway, right] {}(c);
        \draw[line width=0.5mm] (c) to node[midway, below] {$m_2$}(d);

        \node[] at (-1.5,-1.5) {$p_1$};
        \node[] at (-1.5,1.5) {$p_2$};
        \node[] at (1.5,1.5) {$p_3$};
        \node[] at (1.5,-1.5) {$p_4$};
    \end{tikzpicture}
\end{gathered}
\end{equation}
where the dashed lines are massless, and the solid lines are massive. When the external particles are taken to be on-shell with $p_1^2 = p_4^2 = m_2^2$ and $p_2^2 = p_3^2 = m_1^2$, the Steinmann relations are violated by the thresholds $(p_1+p_2)^2=(m_1+m_2)^2$ and $(p_2+p_3)^2=0$, as can be seen for the expression for Box 14 in~\cite{Ellis:2007qk}. In this paper, we just focus on integrals that involve only massless virtual particles.

To refine our understanding of the phenomenon of massless Steinmann violation, we now characterize what properties the singularities of a Feynman integral must have in order to lead to a non-vanishing double discontinuity in~\eqref{eq:steinmann_two_particles}. In order to do this, we first review some of the existing algebraic tools that can be used to predict where singularities arise in Feynman integrals, and sketch various methods by which the sequences of discontinuities that appear in these integrals can be constrained from first principles.

\subsection{The Landau Equations and Landau Diagrams}

Feynman integrals can develop branch cuts when singularities in the integrand of~\eqref{eq:Feynman_integral}, which occur where propagators in the denominator vanish, intersect the integration contour in such a way that the contour cannot be deformed out of the way. Given a basis of cycles $C$ of the internal edges in $G$, the kinematic locations at which this phenomenon can occur are characterized by the \emph{Landau equations}:
\begin{align}
    \alpha_i A_i&=0 \qquad \forall i   \label{Leq1}  \\
        \sum_{i \in \text{ loop }l} \pm \alpha^{~}_i q_i^\mu &=0 \qquad \forall l
        \label{Leq2}
\end{align}
where at least
one of the $\alpha_i$ is required to be nonzero, and the sign in each term of equation~\eqref{Leq2} is determined by the relative orientation of the momenta $q_i$ and the cycle (or loop) $l \in C$. These equations constitute necessary (but not sufficient) conditions for the on-shell surfaces $A_i = 0$ to pinch the integration contour. Thus, a Feynman integral may or may not in fact become singular on the support of a given solution to the Landau equations, depending (for instance) on which Riemann sheet one considers the integral on. More generally, finding all of the singularities in Feynman integrals usually requires considering blown-up versions of the Landau equations (see for instance~\cite{Fevola:2023kaw, Fevola:2023fzn, Caron-Huot:2023ikn,Hannesdottir:2024hke}).

In addition to identifying where Feynman integrals can become singular, we are often interested in understanding what types of branch cuts are expected to arise. The numerator $\mathcal{N}$ plays a key role in determining this behavior, as $\mathcal{N}$ can soften a kinematic singularity if it vanishes on the corresponding kinematic loci. This can, in turn, modify the nature of singularities, for instance changing a simple pole to a square root, or a square root to a logarithm in the final expression (for a more in-depth description of this phenomenon, see~\cite{pham2011singularities,Hannesdottir:2021kpd,Hannesdottir:2024hke,Vergu:2025mag}). As a result, the numerator $\mathcal{N}$ can play a nontrivial role in determining whether a given Feynman integral violates the Steinmann relations. In this paper, we will only concern ourselves with where singularities can appear (in the combined space of integration and external kinematic variables), and not whether a branch cut in fact appears on these kinematic hypersurfaces. In practice, this means that our analysis will entirely ignore the presence of numerators in Feynman integrals; even so, we find that our predictions are respected in all examples considered, as long as we go to sufficiently high order in $\epsilon$.

Every solution to the Landau equations can also be associated with a so-called \emph{Landau diagram} by keeping track of which propagators are on-shell. Namely, given a Feynman diagram and a solution to the Landau equations, we can construct a new graph $\cancel{G}$ by contracting every internal edge for which $q_i^2 \neq m_i^2$. As an alternative perspective, we are free to think of Landau diagrams as the more primitive objects, to underscore the fact that a given singularity can arise in any integral whose corresponding Feynman diagram can be contracted to the relevant Landau diagram. In this way, Landau diagrams provide a graphical way of analyzing when specific singularities can be expected to arise in a Feynman integral. This latter perspective will be the one that we make use of throughout this paper.

\subsection{Hierarchical Constraints and Minimal Cuts}
\label{sec:minimal_cuts}

Once we start computing discontinuities of Feynman integrals, some of the singularities of the original integral will no longer appear. This can easily be seen by thinking about Cutkosky's formula~\cite{Cutkosky:1960sp}, which tells us that the (first) discontinuities of Feynman integrals can be computed as a sum of integrals in which some propagators have been put on shell. For instance, for a discontinuity with respect to the channel $s_I$, we have:
\begin{equation}
    \label{eq:cutkosky}
    \text{Disc}_{s_{\scalebox{.4}{$I$}}} {\cal I}_G(p, m) = \frac{(-1)^E}{(i \pi)^{L D/2}} \sum_{\cancel{G}_{s_{\scalebox{.4}{$I$}}}}  \Bigg[ \int {\rm d}^D k_1\cdots {\rm d}^D k_L   \frac{\mathcal{N}}{ \prod_{i \notin \, \cancel{G}_{s_{\scalebox{.4}{$I$}}}} A_i } \prod_{i \in \, \cancel{G}_{s_{\scalebox{.4}{$I$}}}} (-2\pi i) \theta(\pm q_i^0) \delta(A_i) \Bigg]
\end{equation}
where the  sum is over all the ways one can cut the diagram $G$ into two parts, such that the (square of the) momentum flowing across the cut is given by $s_I$. To each of these cuts, we associate an integral over the propagators that are not cut, times a product of delta functions that place the cut propagators on shell and theta functions that require the energy flowing through these cut propagators to be oriented in the same direction. Looking at~\eqref{eq:cutkosky}, it becomes clear that some of the singularities of the original integral $\mathcal{I}_G(p,m)$ will no longer arise in $\text{Disc}_{s_I} {\cal I}_G(p, m)$, as they will not be accessible on the support of the delta and theta functions. 

\begin{figure}[b]
\centering
\begin{tikzpicture}[scale=1]
        \coordinate (a) at (-.8,.8);
        \coordinate (b) at (.8,.8);
        \coordinate (c) at (.8,-.8);
        \coordinate (d) at (-.8,-.8);
        \coordinate (h) at (.8,0);

        \coordinate (a1) at (-1.3, 1.1);
        \coordinate (a2) at (-1.1, 1.3);
        \coordinate (d1) at (-1.1, -1.3);
        \coordinate (d2) at (-1.3, -1.1);
        \coordinate (b1) at (1.2, 1.2);
        \coordinate (c1) at (1.2, -1.2);
        
        \draw[] (a) to (b) to (c) to (d) to (a);

        \draw[] (a1) to (a);
        \draw[] (a2) to (a);
        \draw[] (d1) to (d);
        \draw[] (d2) to (d);
        \draw[] (b1) to (b);
        \draw[] (c1) to (c);

        \node[] at (1.4,-1.3) {$6$};
        \node[] at (1.4,1.3) {$5$};
        \node[] at (-1.5,1.2) {$3$};
        \node[] at (-1.2,1.5) {$4$};
        \node[] at (-1.5,-1.2) {$2$};
        \node[] at (-1.2,-1.5) {$1$};

        \draw[RedViolet,thick,dashed] (0.3,1.6) to [out=-100,in=0] (-1.6,.4);
        \draw[OliveGreen,thick,dashed] (-1.6,-.4) to [out=0,in=100] (0.3,-1.6);
        \draw[RedOrange,thick,dashed] (-1.6,-.8) to [out=20,in=-200] (1.6,-.8);
        \draw[MidnightBlue,thick,dashed] (.7,1.6) to [out=-110,in=110] (.7,-1.6);

        \node[] at (0.3,1.9) {\color{RedViolet} $s_{34}$};        
        \node[] at (2,-.9) {\color{RedOrange} $s_{345}$};        
        \node[] at (-2,.-.4) {\color{OliveGreen} $s_{12}$};        
        \node[] at (.8,-1.8) {\color{MidnightBlue} $s_{56}$};      

    \end{tikzpicture} \hspace{1cm} 
\begin{tikzpicture}[scale=1]
        \coordinate (a) at (-.8,.8);
        \coordinate (b) at (.8,.8);
        \coordinate (c) at (.8,-.8);
        \coordinate (d) at (-.8,-.8);
        \coordinate (e) at (2.4, .8);
        \coordinate (g) at (2.4, -.8);
        \coordinate (h) at (.8,0);

        \coordinate (a1) at (-1.2, 1.2);
        \coordinate (d1) at (-1.2, -1.2);
        \coordinate (e1) at (2.8, 1.2);
        \coordinate (g1) at (2.8, -1.2);
        \coordinate (h1) at (1.2,0);
        
        \draw[] (a) to (b) to (h) to (c) to (d) to (a);
        \draw[] (b) to (e) to (g) to (c);

        \draw[] (a1) to (a);
        \draw[] (d1) to (d);
        \draw[] (e1) to (e);
        \draw[] (g1) to (g);

        \node[] at (3.1,-1.2) {$4$};
        \node[] at (3.1,1.2) {$3$};
        \node[] at (-1.5,1.2) {$2$};
        \node[] at (-1.5,-1.2) {$1$};

        \draw[RoyalBlue,thick,dashed] (-.4,1.2) to (-.4,-1.2);
        \node[RoyalBlue] at (-0.4,1.5) {$s_{12}$};
        \draw[RedOrange,thick,dashed] (1.6,1.2) to (0,-1.2);
        \node[RedOrange] at (0,-1.5) {$s_{12}$};
        \draw[OliveGreen,thick,dashed] (0,1.2) to (1.6,-1.2);
        \node[OliveGreen] at (1.6,-1.5) {$s_{12}$};
        \draw[RedViolet,thick,dashed] (2,1.2) to (2,-1.2);
        \node[RedViolet] at (2,1.5) {$s_{12}$};
    \end{tikzpicture}
\caption{Examples of minimal cuts, for the two-mass hard box and the massless four-point double-box. In the box diagram, there exists a unique minimal cut for the threshold singularity in each momentum channel. Conversely, the $s_{12}$ channel of the double box diagram has four minimal cuts.}
\label{fig:minimalcuts}
\end{figure}
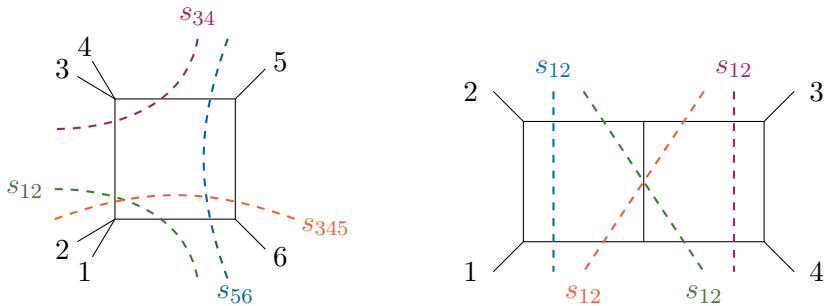

Although no universal formula like~\eqref{eq:cutkosky} yet exists for sequential discontinuities of Feynman integrals, the same basic idea continues to hold under the name of the \emph{hierarchical principle}~\cite{boyling1968homological,Landshoff1966}. Namely, after computing any discontinuity --- whether it be a first discontinuity or a seventh discontinuity --- the result can be expressed in terms of an integral in which the integration contour is localized to the on-shell locus of each propagator that was involved in pinching the integration contour. This places extremely powerful constraints on the sequences of discontinuities that can appear in Feynman integrals. However, deriving these constraints in practice can be difficult, as there generally exist many configurations of virtual momenta in which different on-shell surfaces pinch the integration contour, all of which give rise to a singularity at the same point in the space of external kinematics. Moreover, not all of these singular configurations will be accessible on any given Riemann sheet. Making use of the full power of the hierarchical principle thus requires keeping track of all the singularities that can be accessed on the current Riemann sheet, as well as which propagators are involved in pinching the contour in each case. Beyond the simplest examples, this proves to be a prohibitively complicated task.

As it turns out, however, the predictions made by the hierarchical principle can still be derived without explicitly working out all of these details. This was recently shown in~\cite{Hannesdottir:2024cnn}, where hierarchical predictions were derived by introducing \emph{minimal cuts}, which provide a conservative estimate of which propagators are placed on shell whenever one computes a discontinuity of a Feynman integral. In more detail:
{\tikzmark{minCut} 
\begin{quote}
    {\bf Minimal Cut:} Given a Feynman integral and a singular kinematic surface \\ $\lambda(\{s_I\}, \{m_i^2\})=0$, we refer to a set of cut propagators as a minimal cut if:
    \begin{enumerate}
    \item[\emph{(i)}] the cut propagators partition the external momenta into the combinations that appear in the Mandelstam variables $\{s_I\}$ 
    \item[\emph{(ii)}] each of the internal masses in $\{m_i^2\}$ appears in at least one of the cut propagators (unless this mass has already appeared as one of the Mandelstam variables)
    \item[\emph{(iii)}] one of the first two properties is no longer satisfied if any of the cut propagators are taken off shell
    \end{enumerate}
\end{quote}
\begin{tikzpicture}[overlay, remember picture,decoration={markings,mark=at position .99 with {\arrow[scale=1.4,>=stealth]{>}}}]
    \draw [solid,rounded corners=.2cm,color=Black] ([yshift=-.3cm,xshift=6cm]{pic cs:minCut}) rectangle ([yshift=-5.9cm,xshift=-8.3cm]{pic cs:minCut});
\end{tikzpicture}}
\vspace{-.5cm}

\noindent More intuitively, minimal cuts represent the smallest collections of propagators that must be put on shell in order to ensure that the corresponding Landau diagram $\cancel{G}^{\text{\,min}}_{\lambda}$ actually depends on the kinematic variables that are needed to resolve the singular hypersurface $\lambda(\{s_I\}, \{m_i^2\})=0$. It follows that the propagators that appear in $\cancel{G}^{\text{\,min}}_{\lambda}$ must all be put on shell by the discontinuity operator $\text{Disc}_{\lambda_i}$ (although more propagators may in fact be placed on shell).\footnote{This follows from general results in Picard--Lefschetz theory, and its relative version; see for instance~\cite{Berghoff:2022mqu}.} In the case of threshold singularities, which will be our focus in this paper, minimal cuts are nothing more than the cuts associated with the corresponding unitarity channel.\footnote{Note, however, that we do not place any requirements on the direction in which energy flows through the cut propagators, as one usually does for unitarity cuts.} Simple examples are shown in Figure~\ref{fig:minimalcuts}.

Minimal cuts can also be constructed for sequential discontinuities, since any propagators that are put on shell by one of the discontinuities in the sequence must still be on shell after all discontinuities have been computed. The simplest way to construct the minimal cut $\cancel{G}^{\text{\,min}}_{\lambda_1,\cdots,\lambda_n}$ corresponding to
\begin{equation}
    \text{Disc}_{\lambda_n} \cdots \text{Disc}_{\lambda_1} \mathcal{I}
\end{equation}
is just to take the union $\bigcup \cancel{G}^{\text{\,min}}_{\lambda_i}$ of the minimal cuts for each discontinuity $\text{Disc}_{\lambda_i} \mathcal{I}$. In cases where the individual discontinuities $\text{Disc}_{\lambda_i}$ have multiple minimal cuts, one should consider the union of all possible combinations of these minimal cuts. (Note that, in such cases, one will often end up constructing cuts for $\text{Disc}_{\lambda_n} \cdots \text{Disc}_{\lambda_1} \mathcal{I}$ that are no longer minimal, in the sense that condition \emph{(iii)} is no longer satisfied. These cuts can simply be discarded, since they can be contracted to other cuts that are generated by this procedure.) Correspondingly, we see that the hierarchical principle gives rise to a sequence of increasingly-stringent constraints on which singularities can still arise after additional discontinuities are computed (when these discontinuities place additional propagators on shell).

We can now draw out the important consequence this has for when Steinmann violation can occur. The hierarchical principle tells us that the second discontinuity in~\eqref{eq:steinmann_two_particles}, with respect to the $s_J$ channel, can only be nonzero if there exists a singularity at $s_J=0$ that arises on the support of the on-shell constraints represented by $\cancel{G}^{\text{\,min}}_{s_I}$. In fact, since any singularity at $s_J=0$ is itself expected to arise from a pinch involving the propagators in $\cancel{G}^{\text{\,min}}_{s_J}$, we can trivially upgrade this to the statement that the double discontinuity~\eqref{eq:steinmann_two_particles} will vanish unless there is a singularity at $s_J=0$ that arises on the support of all the on-shell conditions represented by $\cancel{G}^{\text{\,min}}_{s_I,s_J}$. Note that this does not mean all the lines in $\cancel{G}^{\text{\,min}}_{s_I,s_J}$ are required to participate in pinching the integration contour, just that the corresponding singularity must be kinematically accessible when all of these propagators are put on shell. As we will see in the next section, this constitutes a stringent constraint that will allow us to identify precisely when the Steinmann relations can be violated at one loop.


\newpage

\section{Steinmann Violation at One Loop}
\label{sec:steinmann_viiolating_solutions}

Already at one loop, Feynman integrals are known to violate the Steinmann relations. A classic example is given by the totally massless box, which (absent a loop-momentum-dependent numerator) evaluates in $d=4-2\epsilon$ dimensions to
\begin{equation}
\raisebox{-1.6cm}{\begin{tikzpicture}[scale=1]
        \coordinate (a) at (-.6,.6);
        \coordinate (b) at (.6,.6);
        \coordinate (c) at (.6,-.6);
        \coordinate (d) at (-.6,-.6);
        \node (a1) at (-1.3, 1.3) {$p_2$};
        \node (d1) at (-1.3, -1.3) {$p_1$};
        \node (b1) at (1.3, 1.3) {$p_3$};
        \node (c1) at (1.3, -1.3) {$p_4$};
        
        \draw[] (a) to (b) to (c) to (d) to (a);
        
        \draw[] (a1) to (a);
        \draw[] (d1) to (d);
        \draw[] (b1) to (b);
        \draw[] (c1) to (c);
    \end{tikzpicture} }
    \!\!\!\! \!\!\! \propto \
    \frac{1}{s t} \left(\frac{4}{\epsilon^2} - \frac{2}{\epsilon}\left(\log\frac{-s}{\mu^2} + \log\frac{-t}{\mu^2}\right) + 2\log\frac{-s}{\mu^2} \log\frac{-t}{\mu^2} - \pi^2 + \mathcal{O}(\epsilon) \right) \, , \label{eq:massless_box}
\end{equation}
where $s = (p_1+p_2)^2$ and $t = (p_2+p_3)^2$. It is easy to see that the double discontinuity of this integral with respect to the $s$ and $t$ channels will be nonzero, due to the appearance of the product $\log(-s) \log(-t)$ at $\mathcal{O}(\epsilon^0)$. Strictly speaking, this doesn't violate the Steinmann relations (as originally formulated), since $s$ and $t$ can simultaneously vanish only outside of the physical region~\cite{Stapp:1971hh,Bourjaily:2020wvq}. Regardless --- as our goal is to understand what sequences of discontinuities can appear in the symbol of Feynman integrals --- we will continue to refer to the existence of double discontinuities in partially-overlapping momentum channels as Steinmann violation, even if this double discontinuity cannot be reached in physical kinematics.

Our goal is now to understand this result from the point of view of Landau analysis. We know, from the last section, that such double discontinuities can only occur if the both singularities arise even after all four propagators are put on shell. Thus, we are interested in finding solutions to the equations
\begin{gather}
\ell^2 = (\ell+p_2)^2 = (\ell+p_2+p_3)^2 = (\ell - p_1)^2 = 0 \, , \label{eq:landau_massless_box_on_shell} \\
\alpha_1 \ell^\mu + \alpha_2 (\ell^\mu + p_2^\mu) + \alpha_3 (\ell^\mu + p_2^\mu + p_3^\mu) + \alpha_4 (\ell^\mu - p_1^\mu) = 0\, .
\end{gather}
One possible solution to these equations is given by setting \( l^{\mu} = -p_2^{\mu} \) and setting \( s = 0 \); doing so, we find that all four propagators go on shell, and the second Landau equation is satisfied with non-zero \( \alpha_2 \). (More Feynman parameters can be non-zero if one works with real external momenta, since in that case \( s = 0 \) implies \( p_1^{\mu} \propto p_2^{\mu} \), a property that does not hold when working with complexified kinematics.) Similarly, choosing \( l^{\mu} = 0 \) and setting \( t = 0 \), we again find that all four propagators go on shell, and the second Landau equation is satisfied with non-zero \( \alpha_1 \).

More generally, we can solve the on-shell conditions~\eqref{eq:landau_massless_box_on_shell} in general kinematics using spinor-helicity variables (see~\cite{Elvang:2013cua} for a review). Using the convention that \( p_i = -\ket{i}[i| \), the two complex solutions that put all four propagators on shell are  
\begin{gather}
    \ell^{\dot{\alpha}\alpha} = \frac{[23]}{[13]}\ket{2}^{\dot{\alpha}}[1|^{\alpha} \quad \text{and} \quad \ell^{\dot{\alpha}\alpha} = \frac{\braket{23}}{\braket{13}}\ket{1}^{\dot{\alpha}}[2|^{\alpha}\,.\label{eq:max cut sols}
\end{gather}
In the limit \( s = \braket{12}[12] = 0 \), at least one of the spinor brackets becomes collinear --- meaning \( \braket{12} = 0 \) and/or \( [12] = 0 \) --- which enables us to satisfy the second Landau equation on the support of the loop momenta~\eqref{eq:max cut sols} with non-vanishing Feynman parameters. For example, if we take $[12]=0$ by setting $|1]=\gamma |2]$ for some $\gamma \in \mathbb{C}$, then the first max-cut solution in (\ref{eq:max cut sols}) becomes $l=-p_2$, thus reproducing the earlier solution. (For real momenta, both brackets vanish simultaneously, and the two solutions (\ref{eq:max cut sols}) coalesce.) Similarly, in the limit \( t = \braket{23}[23] = 0 \), one of the max-cut solutions (\ref{eq:max cut sols}) becomes the zero vector, reproducing the earlier \( l^\mu = 0 \) solution.

Thus, as expected, the massless box indeed has singularities at $s=0$ and $t=0$, even with the requirement that all four propagators are on shell.\footnote{In this case, we could have also deduced the existence of these singularities by simply computing the max cut of the box, which is precisely $\frac{1}{st}$.} Despite the fact that these solutions seem nearly trivial in the space of integration variables, we have seen in~\eqref{eq:massless_box} that they hold important implications for the analytic structure of the final result. For comparison, one can check that this pair of solutions to the Landau equations disappears if we assign $p_1$ and $p_3$ nonzero masses, and thereby upgrade the massless box to the so-called two-mass easy box. It can also be checked that the two-mass easy box does not violate the Steinmann relations. To understand when Steinmann violation will occur more generally, then, we would like to understand when exactly such pairs of singularities exist at one loop. We now show that a general graphical answer to this question exists --- first, working in momentum space, and then also in Feynman parameter space.

\subsection{Momentum Space Analysis}

We begin our general investigation by considering what types of Landau diagrams can appear as minimal cuts at one loop. Since we are just considering threshold (or pseudothreshold) discontinuities at $s_I = 0$ for some set of particle indices $I$, it is easy to convince oneself that all possible minimal cuts $\cancel{G}^{\text{min}}_{s_I}$ (at one loop) involve putting precisely two propagators on shell. The fact that we restrict our attention to pairs of partially-overlapping momentum channels $s_I$ and $s_J$ further implies that the set of propagators that are cut in $\cancel{G}^{\text{min}}_{s_I}$ will not have any overlap with the set of propagators that are cut in $\cancel{G}^{\text{min}}_{s_J}$. Thus, all minimal cuts $\cancel{G}^{\text{min}}_{s_I, s_J}$ that we encounter will  involve four on-shell propagators, with at least one external particle attached to each remaining vertex; all remaining propagators in $G$, which are not cut, will be contracted.

With this in mind, the only cases we need to differentiate are the different configurations of external momenta coming into the corners of cut box diagrams. These cases can be checked exhaustively, and --- as we will now show --- singularities at $s=0$ and $t=0$ exist only when at least two adjacent corners of the diagram\, $\cancel{G}^{\text{min}}_{s_I, s_J}$ have a single (massless) external momentum attached to them. We now construct these solutions for the general configuration
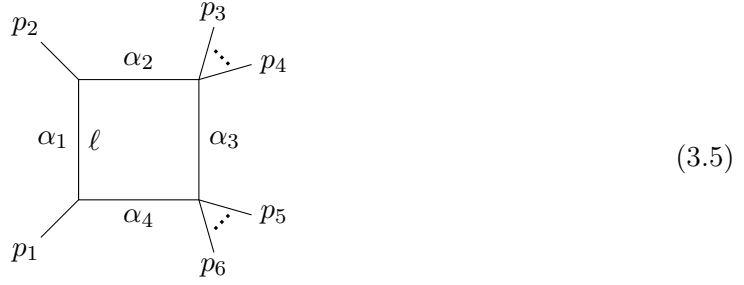
\begin{equation}\label{eq:one_loop_cut_conventions}
\raisebox{-1.6cm}{\begin{tikzpicture}[scale=1]
        \coordinate (a) at (-.8,.8);
        \coordinate (b) at (.8,.8);
        \coordinate (c) at (.8,-.8);
        \coordinate (d) at (-.8,-.8);
        \coordinate (h) at (.8,0);

        \coordinate (a1) at (-1.3, 1.3);
        \coordinate (d1) at (-1.3, -1.3);
        \coordinate (b1) at (1, 1.5);
        \coordinate (b2) at (1.5, 1);
        \coordinate (c1) at (1, -1.5);
        \coordinate (c2) at (1.5, -1);
        \coordinate (x) at (-.8,0);
        
        \draw[] (a1) to (a);
        \draw[] (d1) to (d);
        \draw[] (b1) to (b);
        \draw[] (b2) to (b);
        \draw[] (c1) to (c);
        \draw[] (c2) to (c);

        \draw [dotted, line width=1.2pt] (1.2, 1) -- (1,1.2);
        \draw [dotted, line width=1.2pt] (1.2, -1) -- (1,-1.2);

        \draw[] (d) to node[midway, left] {$\alpha_1$}(a)
        to node[midway, above] {$\alpha_2$}(b)
        to node[midway, right] {$\alpha_3$}(c) 
        to node[midway, below] {$\alpha_4$}(d)
        to node[midway, right] {$\ell$}(a); 

        \node[] at (-1.5,-1.5) {$p_1$};
        \node[] at (-1.5,1.5) {$p_2$};
        \node[] at (1,1.7) {$p_3$};
        \node[] at (1.8,1) {$p_4$};
        \node[] at (1.8,-1) {$p_5$};
        \node[] at (1,-1.7) {$p_6$};
                
    \end{tikzpicture}}
\end{equation}
in which an arbitrary number of external momenta are allowed to enter the two remaining corners. We continue to refer to the momentum flowing from the left to the right side of the diagram as $s_{12}= (p_1+p_2)^2$, and the momentum from the top to the bottom as $s_{234}=(p_2+p_3+p_4)^2$.

Let us first consider the $s_{12}$ channel. The Landau loop equation (\ref{Leq2}) reads
\begin{equation}
\alpha_1 \ell^\mu + \alpha_2 (\ell^\mu + p_2^\mu) + \alpha_3 (\ell^\mu + p_2^\mu + p_3^\mu + p_4^\mu) + \alpha_4 (\ell^\mu - p_1^\mu) = 0\, ,
\end{equation}
which we can solve by setting $\alpha_3 = 0$ and $\smash{\ell^{\mu}=\frac{\alpha_4p_1^{\mu}-\alpha_2p_2^{\mu}}{\alpha_1+\alpha_2+\alpha_4}}$. This solution satisfies three of the on-shell conditions thus far $\ell^2 = (\ell+p_2)^2 = (\ell-p_1)^2 = 0$ when $s_{12}=0$. The only nontrivial condition that remains is then the on-shell condition $(\ell+p_2+p_3+p_4)^2 = 0$. Plugging in the above value for $\ell^\mu$, we find that $\smash{\alpha_1 = -\frac{\alpha_2 s_{34}+\alpha_4 s_{56}}{s_{234}}}$. Altogether, then, we find the solution
\begin{equation}
\text{\bf $\boldsymbol{s_{12}=0}$ solution:} \quad\ell^{\mu}=\frac{\alpha_4p_1^{\mu}-\alpha_2p_2^{\mu}}{\alpha_1+\alpha_2+\alpha_4},\ \alpha_1=-\frac{\alpha_2 s_{34}+\alpha_4 s_{56}}{s_{234}},\, \alpha_3=0\,,\label{eq:momentumspace2mhardsol}
\end{equation}
where $\alpha_2$ and $\alpha_4$ remain unconstrained. Importantly, we see that no analogous solution exists if we attach additional edges to one of the left corners of the box in~\eqref{eq:one_loop_cut_conventions}, since setting $\alpha_3=0$ and solving for $\ell^\mu$ will no longer give us a momentum that satisfies $\ell^2=0$ when $s_{12}=0$. 


Now consider the $s_{234}$ channel. The relevant solution to the Landau equations can be constructed just as we did for the massless box, using the fact that setting $\ell^\mu$ to the zero vector solves the four on-shell constraints if and only if $s_{234} = 0$. As before, the loop equation (\ref{Leq2}) can then be satisfied if we choose $\alpha_2 = \alpha_3 = \alpha_4 = 0$. Thus, we have the solution:
\begin{equation}    
    \text{\bf $\boldsymbol{s_{234}=0}$ solution:} \quad \ell^\mu = 0,\ \alpha_2 = \alpha_3 = \alpha_4 =0,\label{eq:momentumspace2mhardsol2}
\end{equation}
where $\alpha_1$ is required to be non-vanishing. Note that we can also construct this solution using spinor helicity variables, by first setting the loop momentum to one of the values that sets all propagators on shell, which are given by
\begin{gather}
    \ell^{\dot{\alpha}\alpha}=-\frac{s_{234}}{\braket{31}[23]+\braket{41}[24]}\ket{1}^{\dot{\alpha}}[2|^{\alpha},\label{max cut hard box}
\end{gather}
and by the complex conjugate of this expression in which all angle and square brackets are interchanged. Clearly, at $s_{234}=0$ these two max cuts solutions will coalesce and $\ell^\mu$ will become the zero vector, thus allowing us to solve the Landau loop equation as before to reproduce~\eqref{eq:momentumspace2mhardsol2}. 

Now it just remains to show that there are no solutions to the Landau equations that set $s_{12} = 0$ or $s_{234} = 0$ when\, $\cancel{G}^{\text{min}}_{s_I, s_J}$ has no adjacent massless corners. This can be done using the Euler characteristic test, as described in~\cite{Fevola:2023fzn,Hannesdottir:2024cnn}. Namely, if there exists a solution to the Landau equations when either of these Mandelstam variables vanish, the topology of the space on which the Feynman integral is defined should degenerate. This degeneration can be probed for by checking whether the signed Euler characteristic (a topological invariant) decreases in either of these kinematic limits.\footnote{In more detail, Picard--Lefschetz theory tells us that, if there exists a singularity, there must be a vanishing cycle that computes the discontinuity of one's integral with respect to this singularity. Since this cycle vanishes in the strict singular limit, the number of independent cycles that the Feynman integrand can be integrated over should decrease; it follows that the Euler characteristic should also drop. For more details on the Euler characteristic test, see~\cite{Fevola:2023fzn,Hannesdottir:2024cnn}.} Computing the Euler characteristic for massless cut box diagrams with no adjacent massless corners, we find no change when either $s_{12}$ or $s_{234}$ are set to zero (as was also reported in~\cite{Hannesdottir:2024cnn}). This indicates that there is no longer anything singular happening in the integral on the $s_{12} = 0$ or $s_{234} = 0$ hypersurfaces, once we have set all four propagators on shell.

\begin{figure}[t]
    \centering
    \begin{subfigure}[c]{0.48\linewidth} 
        \begin{tikzpicture}[]
        \coordinate (a) at (-.8,.8);
        \coordinate (b) at (.8,.8);
        \coordinate (c) at (.8,-.8);
        \coordinate (d) at (-.8,-.8);
        \coordinate (h) at (.8,0);

        \coordinate (a1) at (-1.2, 1.2);
        \coordinate (d1) at (-1.2, -1.2);
        \coordinate (b1) at (1.1, 1.3);
        \coordinate (b2) at (1.3, 1.1);
        \coordinate (c1) at (1.1, -1.3);
        \coordinate (c2) at (1.3, -1.1);
        \coordinate (x) at (-.8,0);
        
        \draw[] (a) to node[midway, above] {$\alpha_2=0$}(b) 
                    to node[midway, right] {$\alpha_3=0$}(c) 
                    to node[midway, below] {$\alpha_4=0$}(d)
                    to node[midway, left] {$\ell_1^\mu = 0,\alpha_1 \neq 0$}(a);

        \draw[] (a1) to (a);
        \draw[] (d1) to (d);
        \draw[] (b1) to (b);
        \draw[] (b2) to (b);
        \draw[] (c1) to (c);
        \draw[] (c2) to (c);

        \node[] at (-1.4,-1.3) {$p_1$};
        \node[] at (-1.4,1.3) {$p_2$};
        \node[] at (1.2,1.5) {$p_3$};
        \node[] at (1.5,1.1) {$p_4$};
        \node[] at (1.5,-1.1) {$p_5$};
        \node[] at (1.3,-1.4) {$p_6$};
        \end{tikzpicture}
        \caption{(a)}
        \label{fig:2mhard1}
    \end{subfigure}
    \begin{subfigure}[c]{0.48\linewidth}
        \begin{tikzpicture}[]
        \coordinate (a) at (-.8,.8);
        \coordinate (b) at (.8,.8);
        \coordinate (c) at (.8,-.8);
        \coordinate (d) at (-.8,-.8);
        \coordinate (h) at (.8,0);

        \coordinate (a1) at (-1.2, 1.2);
        \coordinate (d1) at (-1.2, -1.2);
        \coordinate (b1) at (1.1, 1.3);
        \coordinate (b2) at (1.3, 1.1);
        \coordinate (c1) at (1.1, -1.3);
        \coordinate (c2) at (1.3, -1.1);
        \coordinate (x) at (-.8,0);
        
        \draw[] (a) to node[midway, above] {$\alpha_2\neq 0$}(b) 
                    to node[midway, right] {$\alpha_3=0$}(c) 
                    to node[midway, below] {$\alpha_4\neq 0$}(d)
                    to node[midway, left] {$\ell_1^\mu = \frac{\alpha_4p_1^{\mu}-\alpha_2p_2^{\mu}}{\alpha_1+\alpha_2+\alpha_4}, \alpha_1 \neq 0\,\,$}(a);
        \draw[line width =0] (d) to node[currarrow,rotate=90,scale=1.5] {}(a);

        \draw[] (a1) to (a);
        \draw[] (d1) to (d);
        \draw[] (b1) to (b);
        \draw[] (b2) to (b);
        \draw[] (c1) to (c);
        \draw[] (c2) to (c);

     \node[] at (-1.4,-1.3) {$p_1$};
        \node[] at (-1.4,1.3) {$p_2$};
        \node[] at (1.2,1.5) {$p_3$};
        \node[] at (1.5,1.1) {$p_4$};
        \node[] at (1.5,-1.1) {$p_5$};
        \node[] at (1.3,-1.4) {$p_6$};
        \end{tikzpicture}
        \caption{(b)}
        \label{fig:2mhard2}
    \end{subfigure}
    
    \caption{The two singular configurations that lead to Steinmann violation in the two-mass hard box. All momenta are taken to be incoming, and all propagators are on shell. (a) The solution in~\eqref{eq:momentumspace2mhardsol2} that gives rise to a singularity at $s_{234}=0$. Here, the momentum between the two massless corners is the zero vector and all Feynman parameters are zero except $\alpha_1$. (b) The solution in~\eqref{eq:momentumspace2mhardsol} that gives rise to a singularity at $s_{12}=0$. Only $\alpha_3$ is zero, and the momenta flowing through the edges associated with $\alpha_1$, $\alpha_2$, and $
    \alpha_4$ are all collinear for real external kinematics.}
    \label{fig:2mhardtotal}
\end{figure}
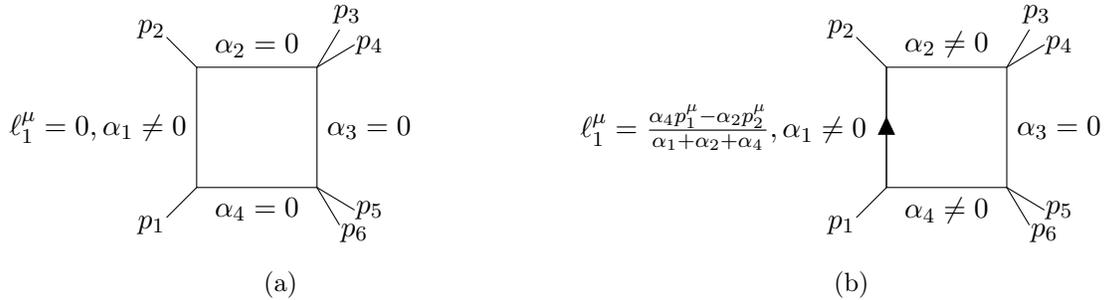

\subsection{Feynman Parameter Space Analysis}

Before going further, let us check that we can reproduce the singularities in~\eqref{eq:momentumspace2mhardsol} and~\eqref{eq:momentumspace2mhardsol2} directly in the Feynman parameter representation, which one gets by introducing Feynman parameters in~\eqref{eq:Feynman_integral} and integrating out the loop momentum. This procedure leaves us with an integral of the form
\begin{align}
   \sim \int_{\alpha_i \ge 0} d^E \alpha_i \, \delta \left(1 - \sum_i \alpha_i \right) \, \prod_i \alpha_i^{\nu_i -1}\, \frac{\mathcal{U}^{\nu - (L+1)D/2}}{\mathcal{F}^{\nu - LD/2}}
\end{align}
where $\nu = \sum_{i=1}^E \nu_i$, and $\mathcal{U}$ and $\mathcal{F}$ are the Symanzik polynomials  (see~\cite{Weinzierl:2022eaz} for more details). In general, singularities can arise in this representation when the integration contour gets pinched between the zeros of both the $\mathcal{U}$ and $\mathcal{F}$ polynomials; however, singularities associated with the $\mathcal{U} = 0$ hypersurface only occur when some of the components of the virtual momentum flowing through the diagram become infinite.\footnote{This can be seen explicitly in equation~(4.26) of~\cite{Hannesdottir:2022bmo}, which relates the momentum flowing through each propagator to the values taken by the Feynman parameters for a given solution to the Landau equations.} Since we know the singularities we are interested in occur for finite values of the loop momentum, we can restrict our attention to pinch singularities that just involve the $\mathcal{F}$ polynomial.\footnote{In addition to considering pinches involving $\mathcal{U}=0$, one also needs consider blowups to find all the locations at which singularities can arise --- see for example~\cite{Berghoff:2022mqu,Fevola:2023fzn,Fevola:2023kaw}. However, we will not have to carry out any blowups to find the solutions we are interested in. 
} The relevant analog of the Landau equations~\eqref{Leq1} and~\eqref{Leq2} thus becomes 
\begin{equation} \label{eq:feyn_param_landau}
    \alpha_i \frac{\partial}{\partial \alpha_i} \mathcal{F} = 0 \qquad \forall i \, .
\end{equation}
In principle, we should also require that $\mathcal{F} = 0$, but the vanishing of $\mathcal{F}$ is already guaranteed by the conditions in~\eqref{eq:feyn_param_landau}.

Each of the equations in~\eqref{eq:feyn_param_landau} can be thought of as enforcing one of two conditions. Either two (or more) of the roots of $\mathcal{F}$ with respect to $\alpha_i$ are coincident --- in which case these zeros have a chance of pinching the integration contour --- or $\alpha_i$ sits at the endpoint of integration (namely at zero). In particular, by comparing these equations with~\eqref{Leq1}, we expect that $\alpha_i=0$ whenever the propagator $A_i$ is not on shell (for a given solution to the Landau equations); conversely, the on shell requirement $A_i= 0$ now translates to the expectation that $\frac{\partial}{\partial \alpha_i} \mathcal{F} = 0$. This means we expect the singularities that we found above in momentum space will arise when
\begin{align} \label{eq:box_cut_landau}
    \frac{\partial}{\partial \alpha_i} \mathcal{F} &= 0 \qquad i \in \{a,b,c,d\}\, , \\
    \alpha_i &= 0 \qquad i \notin \{a,b,c,d\}\, ,
\end{align}
where $A_a$, $A_b$, $A_c$, $A_d$ are the four cut propagators in $\cancel{G}^{\text{min}}_{s_I, s_J}$.

Following the analysis in momentum space, we only expect the solutions of interest to exist for the two-mass hard configuration of external masses (and the massless limits of this configuration). Keeping with the labeling in~\eqref{eq:one_loop_cut_conventions}, the $\mathcal{F}$ polynomial for this graph is just
\begin{gather}
    \mathcal{F}= \alpha_1 \alpha_3\, s_{234} +  \alpha_2 \alpha_3\, s_{34}+  \alpha_2 \alpha_4 \, s_{12}+ \alpha_3 \alpha_4 \, s_{56} \, .
\end{gather}
Setting $\{a,b,c,d\} = \{1,2,3,4\}$ in~\eqref{eq:box_cut_landau}, one indeed finds two solutions that are codimension-one in the space of external kinematics. The first sets $s_{12}=0$, and requires that $\alpha_1=-\frac{\alpha_2 s_{34}+\alpha_4 s_{56}}{s_{234}}$ while $\alpha_3=0$, matching the solution we found in~\eqref{eq:momentumspace2mhardsol}. The second sets $s_{234}=0$, and simply requires that $\alpha_2 = \alpha_3 = \alpha_4 =0$ while $\alpha_1 \neq 0$, matching the solution in~\eqref{eq:momentumspace2mhardsol2}. This confirms that we can reproduce these singular configurations directly in the Feynman parameter representation. One can also check that these Feynman parameter solutions disappear if we make a third corner of the box massive.

\subsection{A Simple Graphical Test}

The upshot of the preceding analysis is that the Steinmann relations can only be violated in a pair of channels $s_I$, $s_J$ if there exists a minimal cut $\cancel{G}^{\text{min}}_{s_I, s_J}$ of the form~\eqref{eq:one_loop_cut_conventions}. This result allows us to formulate a simple graphical test for probing when Steinmann violation can occur: 

{\tikzmark{one_loop_test_start}\begin{quote}
{\bf Graphical Steinmann Violation Test (One Loop):} Given a massless one-loop Feynman graph $G$ and a pair of partially-overlapping channels $s_I$, $s_J$, construct all minimal cuts $\{\cancel{G}^{\text{min}}_{s_I, s_J}\}$. The Steinmann relations can be violated in this pair of channels only if at least one of these minimal cuts takes the form
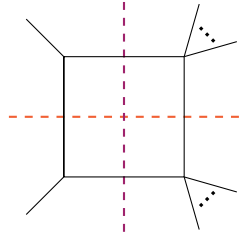
\begin{center}
\begin{tikzpicture}[scale=1]
        \coordinate (a) at (-.8,.8);
        \coordinate (b) at (.8,.8);
        \coordinate (c) at (.8,-.8);
        \coordinate (d) at (-.8,-.8);
        \coordinate (h) at (.8,0);

        \coordinate (a1) at (-1.3, 1.3);
        \coordinate (d1) at (-1.3, -1.3);
        \coordinate (b1) at (1, 1.5);
        \coordinate (b2) at (1.5, 1);
        \coordinate (c1) at (1, -1.5);
        \coordinate (c2) at (1.5, -1);

        \coordinate (cutL) at (-1.53, 0);
        \coordinate (cutR) at (1.53, 0);
        \coordinate (cutB) at (0,-1.53);
        \coordinate (cutT) at (0,1.53);
        
        \draw[] (a1) to (a);
        \draw[] (d1) to (d);
        \draw[] (b1) to (b);
        \draw[] (b2) to (b);
        \draw[] (c1) to (c);
        \draw[] (c2) to (c);

        \draw[dashed,thick,RedOrange] (cutL) to (cutR);
        \draw[dashed,thick,RedViolet] (cutB) to (cutT);

       \draw [dotted, line width=1.2pt] (1.2, 1) -- (1,1.2);
        \draw [dotted, line width=1.2pt] (1.2, -1) -- (1,-1.2);

        \draw[] (d) to node[midway, left] {}(a)
        to node[midway, above] {}(b)
        to node[midway, right] {}(c) 
        to node[midway, below] {}(d)
        to node[midway, right] {}(a);      
    \end{tikzpicture}
    \end{center}
    where at least one external particle attaches to each corner of the box.  
\end{quote}}
\begin{tikzpicture}[overlay, remember picture,decoration={markings,mark=at position .99 with {\arrow[scale=1.4,>=stealth]{>}}}]
    \draw [solid,rounded corners=.2cm,color=Black] ([yshift=-.3cm,xshift=-.1cm]{pic cs:one_loop_test_start}) rectangle ([yshift=-7.4cm,xshift=14.2cm]{pic cs:one_loop_test_start});
\end{tikzpicture}

\noindent At one loop, this test amounts to asking whether at least one of the channels is a two-particle channel (if the answer is yes, Steinmann can be violated). However, we highlight that the same partially-overlapping cut can also appear in higher-loop integrals, when only the propagators of one of the loops are cut (and all the other loops are contracted to one of the corners of the box). Thus, at higher loop orders, it will become necessary to phrase things in terms of minimal cuts, as done here. 

In principle, the mere existence of these solutions to the Landau equations does not guarantee that the Steinmann relations will be violated. For instance, it could be the case that one's integral only develops a simple pole due to these pinch singularities, and no branch cut. In particular, the nature of the singularity that arises at a given point in the space of integration variables can generally be modified by adding a numerator that vanishes on the same locus. We thus expect that any test purporting to establish the existence of Steinmann violation in a given integral would need to take this type of information into account. However, in practice, we observe that this graphical test can reliably be used to predict when the Steinmann relations will in fact be violated, as we have not found any examples in which a minimal cut of the above type exists but Steinmann is not violated (although sometimes this violation first occurs at higher orders in $\epsilon$). We will say more about this when we discuss the analogous test at two loops. 

\newpage
\section{Steinmann-Violating Minimal Cuts to All Loop Orders}
\label{sec:all_loop_solutions}

In the last section, we identified one-loop solutions to the Landau equations that allow the Steinmann relations to be violated. In this section, we identify similar configurations at every loop order, which can also lead to Steinmann violation. More specifically, we will prove the following result:
\begin{theorem}\label{con:2}
    Any Feynman diagram with a partially-overlapping minimal cut\, $\cancel{G}^{\text{min}}_{s_I, s_J}$ supports a pair of solutions to the Landau equations that allow for Steinmann violation in the $s_I$ and $s_J$ channels whenever\, $\cancel{G}^{\text{min}}_{s_I, s_J}$ involves exactly four vertices and has a pair of adjacent massless external legs.
\end{theorem}
\noindent We establish this result by induction. First, we construct the relevant pair of solutions to the Landau equations for the acnode and envelope diagrams (shown in Figures \ref{fig:acnode} and \ref{fig:envelope}). Next, we use the well-known fact that one can upgrade any solution to the Landau equations that involves a massless on-shell line to a higher-loop solution, in which the massless line has been replaced by a pair of massless lines. In this way, we identify the desired solutions to the Landau equations for any cut taking the form depicted in~\eqref{eq:connected_minimal_cuts}.

\subsection{Landau Analysis of the Acnode}\label{sect:landauacnode}

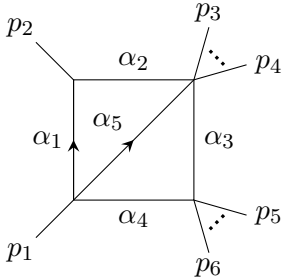
\begin{figure}[b]
\centering
 \begin{tikzpicture}[scale=1]
        \coordinate (a) at (-.8,.8);
        \coordinate (b) at (.8,.8);
        \coordinate (c) at (.8,-.8);
        \coordinate (d) at (-.8,-.8);
        \coordinate (h) at (.8,0);

        \coordinate (a1) at (-1.3, 1.3);
        \coordinate (d1) at (-1.3, -1.3);
        \coordinate (b1) at (1, 1.5);
        \coordinate (b2) at (1.5, 1);
        \coordinate (c1) at (1, -1.5);
        \coordinate (c2) at (1.5, -1);
        \coordinate (x) at (-.8,0);
        
        \draw[] (a1) to (a);
        \draw[] (d1) to (d);
        \draw[] (b1) to (b);
        \draw[] (b2) to (b);
        \draw[] (c1) to (c);
        \draw[] (c2) to (c);
        \draw[midarrow] (d) to (a);
        \draw[midarrow] (d) to node[midway, above left] {$\alpha_5$} (b);

        \draw [dotted, line width=1.2pt] (1.2, 1) -- (1,1.2);
        \draw [dotted, line width=1.2pt] (1.2, -1) -- (1,-1.2);

        \draw[] (d) to node[midway, left] {$\alpha_1$}(a)
        to node[midway, above] {$\alpha_2$}(b)
        to node[midway, right] {$\alpha_3$}(c) 
        to node[midway, below] {$\alpha_4$}(d)
        to node[midway, right] {}(a); 

        \node[] at (-1.5,-1.5) {$p_1$};
        \node[] at (-1.5,1.5) {$p_2$};
        \node[] at (1,1.7) {$p_3$};
        \node[] at (1.8,1) {$p_4$};
        \node[] at (1.8,-1) {$p_5$};
        \node[] at (1,-1.7) {$p_6$};
                
    \end{tikzpicture}
    \caption{The two-mass hard acnode graph. Arrows indicate the orientations of the loop momenta $\ell_1^\mu$ and $\ell_5^\mu$.} \label{fig:acnode}
\end{figure}

The first case we consider is the acnode graph, depicted in Figure \ref{fig:acnode}. Labeling the momentum flowing through the edge associated with $\alpha_i$ by $\ell_i^{\mu}$, we find that there exists a solution to the Landau equations at $s_{12}=0$ given by
\begin{gather}
\text{\bf $\boldsymbol{s_{12}=0}$ solution (acnode):} \quad \begin{cases}
    \ell_1^{\mu}=\frac{\alpha_4 \alpha_5 p_1^{\mu} - \alpha_2 (\alpha_4 + \alpha_5) p_2^{\mu}
}{D},\ \ell_5^{\mu}=\frac{\alpha_4 \left( \alpha_1 p_1^{\mu} + \alpha_2 (p_1^{\mu} + p_2^{\mu}) \right)
}{D}, \\[4pt]
    \alpha_1 = -\frac{s_{34} \alpha_2 \alpha_5 + s_{56} \alpha_4 (\alpha_2 + \alpha_5)}{s_{56} \alpha_4 + s_{234} \alpha_5},\ \alpha_3 = 0 ,
\end{cases} \label{eq:acnodemomentums12}  \\[0cm]
\text{where } D=(\alpha_1 + \alpha_2) \alpha_4 + (\alpha_1 + \alpha_2 + \alpha_4) \alpha_5 \, . \nonumber
\end{gather}
We highlight that this solution bears a striking resemblance to the $s_{12}=0$ solution~\eqref{eq:momentumspace2mhardsol} that we found at one loop, and that the momentum flowing through the new diagonal propagator is collinear with both $p_1^\mu$ and $p_2^\mu$.
An even simpler solution exists at $s_{234}=0$, namely
\begin{equation}
     \text{\bf $\boldsymbol{s_{234}=0}$ solution (acnode):} \quad \ell_1^\mu = \ell_5^\mu = 0 , \ \alpha_2 = \alpha_3 = \alpha_4 = 0 . \label{eq:acnodemomentums23} 
\end{equation}
This solution is trivially related to the one-loop solution in~\eqref{eq:momentumspace2mhardsol2}, since no momentum flows through the new diagonal propagator.\footnote{We highlight that the solutions in~\eqref{eq:acnodemomentums12} and~\eqref{eq:acnodemomentums23} can be derived using the $Y-\Delta$ move (see \cite{Prlina:2018ukf}), starting from the corresponding one-loop solutions in~\eqref{eq:momentumspace2mhardsol} and~\eqref{eq:momentumspace2mhardsol2}. However, one cannot use this move to construct solutions to the Landau equations for the envelope diagram, since the $Y-\Delta$ move preserves planarity.}

As a cross-check, we also identify these configurations as solutions to the Landau equations for the Feynman parameter representation. In our labeling, the second Symanzik polynomial is given by
\begin{gather}
    \mathcal{F}=s_{56} \alpha_1 \alpha_3 \alpha_4 + s_{56} \alpha_2 \alpha_3 \alpha_4 + s_{234} \alpha_1 \alpha_3 \alpha_5 + s_{34} \alpha_2 \alpha_3 \alpha_5 + s_{12} \alpha_2 \alpha_4 \alpha_5 + s_{56} \alpha_3 \alpha_4 \alpha_5.
\end{gather}
Since we are interested in configurations in which the integration contour over all five variables $\alpha_i$ is pinched, we require that
\begin{gather}
    \partial_{\alpha_i}\mathcal{F}=0,\qquad i=1,2,...,5 \, .
\end{gather}
Solving these equations, we reproduce the values of the Feynman parameters in~\eqref{eq:acnodemomentums12} when $s_{12}=0$, and the values of the Feynman parameters in~\eqref{eq:acnodemomentums23} when $s_{234}=0$.

\subsection{Landau Analysis of the Envelope}

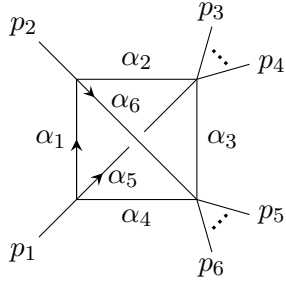
\begin{figure}[t]
\centering
 \begin{tikzpicture}[scale=1]
        \coordinate (a) at (-.8,.8);
        \coordinate (b) at (.8,.8);
        \coordinate (c) at (.8,-.8);
        \coordinate (d) at (-.8,-.8);
        \coordinate (h) at (.8,0);
        \coordinate (ccl) at (-0.1,-0.1);
        \coordinate (ccr) at (0.1,0.1);

        \coordinate (a1) at (-1.3, 1.3);
        \coordinate (d1) at (-1.3, -1.3);
        \coordinate (b1) at (1, 1.5);
        \coordinate (b2) at (1.5, 1);
        \coordinate (c1) at (1, -1.5);
        \coordinate (c2) at (1.5, -1);
        \coordinate (x) at (-.8,0);
        
        \draw[] (a1) to (a);
        \draw[] (d1) to (d);
        \draw[] (b1) to (b);
        \draw[] (b2) to (b);
        \draw[] (c1) to (c);
        \draw[] (c2) to (c);
        \draw[midarrow] (d) to (a);
        \draw[midarrow] (d) to node[pos=0.4, right] {$\alpha_5$} (ccl);
        \draw[] (ccr) to (b);
        
             \draw[
  postaction={decorate},
  decoration={
    markings,
    mark=at position 0.15 with {\arrow[scale=1.3]{stealth}}
  }
] (a) to node[pos=0.2, right] {$\alpha_6$} (c);

        \draw [dotted, line width=1.2pt] (1.2, 1) -- (1,1.2);
        \draw [dotted, line width=1.2pt] (1.2, -1) -- (1,-1.2);

        \draw[] (d) to node[midway, left] {$\alpha_1$}(a)
        to node[midway, above] {$\alpha_2$}(b)
        to node[midway, right] {$\alpha_3$}(c) 
        to node[midway, below] {$\alpha_4$}(d)
        to node[midway, right] {}(a); 

        \node[] at (-1.5,-1.5) {$p_1$};
        \node[] at (-1.5,1.5) {$p_2$};
        \node[] at (1,1.7) {$p_3$};
        \node[] at (1.8,1) {$p_4$};
        \node[] at (1.8,-1) {$p_5$};
        \node[] at (1,-1.7) {$p_6$};
                
    \end{tikzpicture}
    \caption{The two-mass hard envelope graph. Arrows indicate the orientations of the loop momenta $\ell_1^\mu$, $\ell_5^\mu$, and $\ell_6^\mu$.}  \label{fig:envelope}
\end{figure}

We next consider the envelope graph shown in Figure \ref{fig:envelope}. To solve the Landau equations, we first solve the loop equations~\eqref{Leq2} for the loop momenta, and then impose each on-shell constraint on the Feynman parameters and external kinematics. Doing so, we find a solution at $s_{12}=0$ when 
\begin{equation}\label{eq:s12 envelope}
\begin{gathered}
\text{\bf $\boldsymbol{s_{12}=0}$ solution (envelope):} \quad \begin{cases} \ell_{1}^{\mu} = \frac{-p_{2}^{\mu}\,\alpha_{2}(\alpha_{4}+\alpha_{5})\alpha_{6}
+ p_{1}^{\mu}\,\alpha_{4}\alpha_{5}(\alpha_{2}+\alpha_{6})}{D}, \\[4pt]
\ell_{5}^{\mu} = \frac{\alpha_{4}\Big(p_{1}^{\mu}\alpha_{1}\alpha_{2}
+ \big(p_{1}^{\mu}\alpha_{1}+(p_{1}^{\mu}+p_{2}^{\mu})\alpha_{2}\big)\alpha_{6}\Big)}{D}, \\[4pt]
\ell_{6}^{\mu} = \frac{\alpha_{2}\Big(p_{2}^{\mu}\alpha_{1}\alpha_{4}
+ \big(p_{2}^{\mu}\alpha_{1}+(p_{1}^{\mu}+p_{2}^{\mu})\alpha_{4}\big)\alpha_{5}\Big)}{D}, \\[4pt]
\alpha_1 = \frac{-\alpha_2  \alpha_5  (\alpha_4 + \alpha_6)  s_{34} - \alpha_4 (\alpha_2 + \alpha_5) \alpha_6 s_{56}}{\alpha_2 \alpha_5 s_{34} + \alpha_2 \alpha_4 (-s_{234} + s_{34} + s_{56}) + \alpha_6 (\alpha_5 s_{234} + \alpha_4 s_{56})} , \\[4pt]
\alpha_3 = 0  ,
\end{cases} \hspace{-.3cm} \end{gathered}
\end{equation}
where $D=\alpha_{4}\alpha_{5}\alpha_{6}
+ \alpha_{1}(\alpha_{4}+\alpha_{5})(\alpha_{2}+\alpha_{6})
+ \alpha_{2}\left(\alpha_{4}\alpha_{5}+(\alpha_{4}+\alpha_{5})\alpha_{6}\right)$ is the first Symanzik polynomial $\mathcal{U}$ evaluated at $\alpha_3=0$.

We also find two solutions to the Landau equations that set $s_{234}=0$. The first solution takes the form
\begin{gather}
\text{\bf $\boldsymbol{s_{234}=0}$ solution 1 (envelope):} \quad   \begin{cases}
    \ell_1^{\mu}=-\ell_5^{\mu}=\ell_6^{\mu}=\frac{\alpha_3}{\alpha_1-\alpha_3}(p_1+p_4)^{\mu},  \\
    \alpha_2=\alpha_4=0, \\
    \alpha_6=\alpha_5=-\alpha_1.
\end{cases}
    \label{eq:s23 envelope 1}
\end{gather}
The second solution closely resembles the solutions we found for the box and the acnode:
\begin{gather}
\text{\bf $\boldsymbol{s_{234}=0}$ solution 2 (envelope):} \quad \begin{cases}
    \ell_1^{\mu}=\ell_5^{\mu}=\ell_6^{\mu}=0, \\
    \alpha_2 = \alpha_3 = \alpha_4 =0.
\end{cases} \label{eq: s23 envelope 2}
\end{gather}
Although these two solutions intersect in the space of Feynman parameters and loop momenta, the fact that they are distinct can be seen from the fact that $\alpha_{1},\alpha_5$ and $\alpha_6$ can take arbitrary values for the latter solution but not the former. 

Finally, in the Feynman parameter representation we have
\begin{equation}
\begin{aligned}
\mathcal{F} ={}& (s_{56}-s_{12}-s_{234})\,\alpha_{1}\alpha_{2}\alpha_{3}\alpha_{4}\\
&+ \Big((s_{234}\,\alpha_{1}\alpha_{3}+s_{12}\,\alpha_{2}\alpha_{4})\,\alpha_{5}
+ s_{56}\,\alpha_{3}\alpha_{4}(\alpha_{1}+\alpha_{2}+\alpha_{5})\Big)\alpha_{6}\\
&+ s_{34}\,\alpha_{2}\alpha_{3}\Big(\alpha_{1}(\alpha_{4}+\alpha_{5})
+ \alpha_{5}(\alpha_{4}+\alpha_{6})\Big) \, .
\end{aligned}
\end{equation}
Solving the equations given by
\begin{gather}
    \partial_{\alpha_i}\mathcal{F}=0,\qquad i=1,2,...,6 \, ,\label{eq:max cut crossed box}
\end{gather}
we reproduce the configurations of Feynman parameters reported in~\eqref{eq:s12 envelope},~\eqref{eq:s23 envelope 1}, and~\eqref{eq: s23 envelope 2}.

\subsection{Landau Singularities at Higher Loops}
Having found solutions to the Landau equations that set $s_{12}=0$ or $s_{234}=0$ in the acnode and envelope diagrams when all propagators are on shell, we now upgrade these solutions to on-shell configurations involving additional loops. To do this, we simply use the following graphical rule:
\vspace{-.6cm}

\begin{equation}
\begin{gathered}
\begin{tikzpicture}[thick, line cap=round,scale=.8]
  \coordinate (L) at (-1.5,0);
  \coordinate (R) at ( 1.5,0);

  \draw[line width=1.0] (L) -- (R);
  \node at (0.04,.3) {$\ell^\mu$};
  \node at (0,-.26) {$\alpha$};
  \node at (-.25,-.89) {\color{white} $k_2^\mu$};

  \fill (L) circle (2pt);
  \fill (R) circle (2pt);

  \draw (L) -- ++(130:1);
  \draw (L) -- ++(160:.5);
  \draw (L) -- ++(-160:.5);
  \draw (L) -- ++(-130:1);

  \draw (R) -- ++(50:1);
  \draw (R) -- ++(20:.5);
  \draw (R) -- ++(-20:.5);
  \draw (R) -- ++(-50:1);

  \fill (-2.2,.3) circle (1pt);
  \fill (-2.3,0) circle (1pt);
  \fill (-2.2,-.3) circle (1pt);
  
  \fill (2.2,.3) circle (1pt);
  \fill (2.3,0) circle (1pt);
  \fill (2.2,-.3) circle (1pt);

\end{tikzpicture}
\qquad \qquad \qquad
\begin{tikzpicture}[thick, line cap=round,scale=.8]
  \coordinate (L) at (-1.5,0);
  \coordinate (R) at ( 1.5,0);

  \draw[line width=1.0] (L) to [bend right=35] (R);
  \draw[line width=1.0] (L) to [bend left=35] (R);
  \node at (.27,.84) {$k_1^\mu$};
  \node at (-.25,-.89) {$k_2^\mu$};
  \node at (.25,.2) {$\gamma_1$};
  \node at (-.25,-.22) {$\gamma_2$};

  \fill (L) circle (2pt);
  \fill (R) circle (2pt);

  \draw (L) -- ++(130:1);
  \draw (L) -- ++(160:.5);
  \draw (L) -- ++(-160:.5);
  \draw (L) -- ++(-130:1);

  \draw (R) -- ++(50:1);
  \draw (R) -- ++(20:.5);
  \draw (R) -- ++(-20:.5);
  \draw (R) -- ++(-50:1);

  \fill (-2.2,.3) circle (1pt);
  \fill (-2.3,0) circle (1pt);
  \fill (-2.2,-.3) circle (1pt);
  
  \fill (2.2,.3) circle (1pt);
  \fill (2.3,0) circle (1pt);
  \fill (2.2,-.3) circle (1pt);

\end{tikzpicture} 
\\[-1.6cm]
\text{\Large \vspace{.2cm} $\Rightarrow$}
\end{gathered}
\end{equation}
\vspace{.4cm}

\noindent where the new momenta are given by $k_1^\mu  = \beta \ell^\mu$ and $k_2^\mu = (1-\beta) \ell^\mu$, and the new Feynman parameters take the values $\gamma_1=\frac{\alpha_i}{\beta}$ and $\gamma_2=\frac{\alpha_i}{1-\beta}$. Stated in words, choosing any on-shell propagator with momentum $\ell^{\mu}$ and Feynman parameter $\alpha$, we can replace the propagator with a bubble that has the momentum and Feynman parameter assignments given above. Since the new momenta are proportional to the original momentum $\ell^\mu$, both new propagators are on shell. Moreover, by construction, the sum of these new momenta adds up to $\ell^\mu$, so the remaining momenta in the diagram remain unperturbed. It can be easily checked that this new (higher-loop) configuration of momenta and Feynman parameters also solves the Landau equations.\footnote{For another  presentation of this rule, see Appendix A of \cite{Dennen_2017}.}

Using this rule, and starting from the two-mass hard configurations of the box, acnode, and envelope diagrams, we can iteratively construct solutions to the Landau equations in which an arbitrary number of massless on-shell propagators connect any pair of vertices:
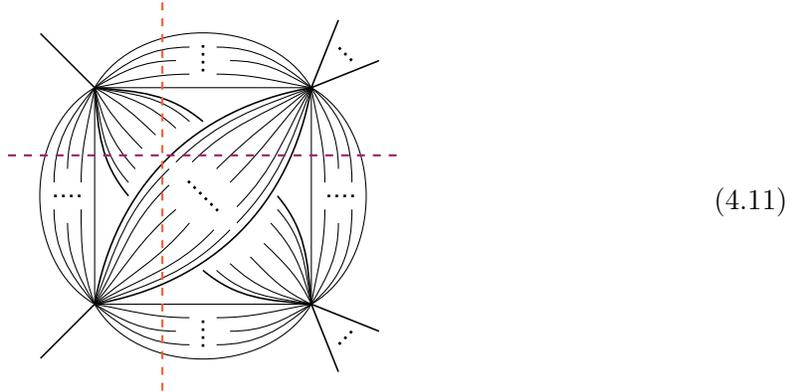
\begin{equation} \label{eq:violating_minimal_cuts}
\begin{gathered}
    \begin{tikzpicture}[scale=1.8]
        \coordinate (a) at (-.8,.8);
        \coordinate (b) at (.8,.8);
        \coordinate (c) at (.8,-.8);
        \coordinate (d) at (-.8,-.8);
        \coordinate (h) at (.8,0);

        \coordinate (a1) at (-1.2, 1.2);
        \coordinate (d1) at (-1.2, -1.2);
        \coordinate (b1) at (1.2, 1.2);
        \coordinate (c1) at (1.2, -1.2);

        \coordinate (i2) at (-0.3, 1.3);
        \coordinate (j2) at (0.3, 1.3);

        \draw[] (b) to (c);
        \draw[] (d) to (a);
        \coordinate (e1) at (1.2, 0.2);
        \coordinate (f1) at (1.2, -0.2);
        \coordinate (e2) at (1.3, 0.3);
        \coordinate (f2) at (1.3, -0.3);
        
        \draw[] (b) to [out=-80, in=90] (0.9,0.1);
        \draw[] (0.9,-0.1) to [out=-90, in=80] (c);
        \draw[] (b) to [out=-60, in=90] (1,0.2);
        \draw[] (1,-0.2) to [out=-90, in=60] (c);
        \draw[] (b) to [out=-40, in = 90] (1.1,0.1);
        \draw[] (1.1,-0.1) to [out=-90, in=40] (c);
        \draw[] (b) to [out=-30, in = 30] (c);
        
        \coordinate (g1) at (-1.2, 0.2);
        \coordinate (h1) at (-1.2, -0.2);
        \coordinate (g2) at (-1.3, 0.3);
        \coordinate (h2) at (-1.3, -0.3);

        \draw[] (a) to [out=-100, in = 90] (-0.9,0.1);
        \draw[] (-0.9,-0.1) to [out=-90, in = 110] (d);
        \draw[] (a) to [out=-120, in = 90] (-1,0.2);
        \draw[] (-1,-0.2) to [out=-90, in = 120] (d);
        \draw[] (a) to [out=-130, in = 90] (-1.1,0.1);
        \draw[] (-1.1,-0.1) to [out=-90, in=130] (d);
        \draw[] (a) to [out=-150, in=150] (d);

        \draw [dotted, line width=1pt] (0.92, 0) -- (1.12,0);
        \draw [dotted, line width=1pt] (-1.1, 0) -- (-0.9,0);
        
        \draw[] (a) to (b);
        \draw[] (a) to [out=10, in = 180] (-0.1, 0.9);
        \draw[] (0.1, 0.9) to [out=0, in = 170] (b);
        \draw[] (a) to [out=30, in= 180] (-0.2, 1);
        \draw[] (0.2, 1) to [out=0, in=150] (b);
        \draw[] (a) to [out=40, in=180] (-0.1, 1.1);
        \draw[] (0.1, 1.1) to [out=0, in=140] (b);
        \draw[] (a) to [out=60, in=120] (b);

        \draw [dotted, line width=1pt] (0, 0.92) -- (0,1.12);

        \draw[] (d) to (c);
        \draw[] (d) to [out=-10, in=180] (-0.1, -0.9);
        \draw[] (0.1, -0.9) to [out=0, in=-170] (c);
        \draw[] (d) to [out=-30, in=180] (-0.2, -1);
        \draw[] (0.2, -1) to [out=0, in=-150] (c);
        \draw[] (d) to [out=-40, in=180] (-0.1,-1.1);
        \draw[] (0.1,-1.1) to [out=0, in=-140] (c);
        \draw[] (d) to [out=-60, in=-120] (c);

        \draw[dotted, line width=1pt] (0,-0.92) -- (0,-1.12);

        \draw [dotted, line width=1pt] (-0.11, 0.11) -- (0.11,-0.11);
        \draw[line width=0.6pt] (d) to [out=80, in=-170] (b);
        \draw[line width=0.6pt] (d) to [out=10, in=-100] (b);
        \draw[] (d) to [out=50, in=-140] (-0.2, -0.1);
        \draw[] (0.1,0.2) to [out=45, in=-140] (b);
        \draw[] (d) to [out=40, in=-130] (-0.1, -0.2);
        \draw[] (0.2,0.1) to [out=45, in=-130] (b);
        \draw[] (d) to [out=60, in=-137] (-0.25, 0.08);
        \draw[] (-0.08, 0.25) to [out=45, in=-150] (b);
        \draw[] (d) to [out=30, in=-133] (0.08, -0.25);
        \draw[] (0.25, -0.08) to [out=45, in=-120] (b);
        \draw[] (d) to [out=70, in=-136] (-0.25,0.2);
        \draw[] (-0.2,0.25) to [out=44, in=-160] (b);
        \draw[] (d) to [out=20, in=-134] (0.2,-0.25);
        \draw[] (0.25,-0.2) to [out=46, in=-110] (b);

         \draw[line width=0.6pt] (a) to [out=-5, in=140] (0,0.55);
         \draw[line width=0.6pt] (a) to [out=-85, in=130] (-0.55,0);
         \draw[line width=0.6pt] (0.55,0) to [out=-50, in=95] (c);
         \draw[line width=0.6pt] (0,-0.55) to [out=-40, in=175] (c);
         \draw[] (a) to [out=-10, in=138] (-0.1, 0.55);
         \draw[] (a) to [out=-80, in=132] (-0.55, 0.1);
         \draw[] (0.1, -0.55) to [out=-42, in=170] (c);
         \draw[] (0.55, -0.1) to [out=-48, in=100] (c);
         \draw[] (a) to [out=-15, in=137] (-0.2, 0.5);
         \draw[] (a) to [out=-75, in=133] (-0.5,0.2);
         \draw[] (0.2,-0.5) to [out=-43, in=165] (c);
         \draw[] (0.5,-0.2) to [out=-47, in=105] (c);
         \draw[] (a) to [out=-30, in=136] (-0.35, 0.45);
         \draw[] (a) to [out=-60, in=134] (-0.45, 0.35);
         \draw[] (0.35, -0.45) to [out=-46, in=155] (c);
         \draw[] (0.45, -0.35) to [out=-44, in=115] (c);
         
        \coordinate (a1) at (-1.2, 1.2);
        \coordinate (d1) at (-1.2, -1.2);
        \coordinate (b1) at (1.2, 1.2);
        \coordinate (c1) at (1.2, -1.2);
        
        \draw[line width=0.6pt] (1,1.3) to (b);
        \draw[line width=0.6pt] (1.3,1) to (b);
        \draw[line width=0.6pt] (1,-1.3) to (c);
        \draw[line width=0.6pt] (1.3,-1) to (c);

        \draw[line width=0.6pt] (a1) to (a);
        \draw[line width=0.6pt] (d1) to (d);
        
        \draw[dotted, line width=1pt] (1.1,1) to (1,1.1);
        \draw[dotted, line width=1pt] (1.1,-1) to (1, -1.1);

        \draw[RedViolet,thick,dashed] (-1.44,0.3) to (1.44,0.3);
        \draw[RedOrange,thick,dashed] (-.3,-1.44) to (-.3,1.44);

\end{tikzpicture}
\end{gathered}
\end{equation}
This concludes our proof of Theorem \ref{con:2}.

\newpage

\section{Steinmann Violation at Two Loops}
\label{sec:two_loops}

We now refocus our attention on violations of the Steinmann relations that occur at two loops. Although it is not (currently) feasible to find all two-loop solutions to the Landau equations, we can still explore when Steinmann is violated in Feynman integrals that have already been computed. From the analysis of the last section, we know that such violations can arise in conjunction with the on-shell configurations shown in~\eqref{eq:violating_minimal_cuts}. We now conjecture that these are the \emph{only} on-shell configurations that give rise to Steinmann violation at two loops.

{\tikzmark{two_loop_test_start}\begin{quote}
{\bf Two-Loop Steinmann Violation Conjecture:} Given a massless Feynman integral $\mathcal{I}_G$ and a pair of partially-overlapping momentum channels $s_I$, $s_J$, construct the complete set of minimal cuts $\smash{\big\{ \cancel{G}^{\text{min}}_{s_I, s_J} \big\}}$. If all of these minimal cuts correspond to one- and two-loop diagrams, the Steinmann relations for this pair of channels will be violated only if one of the minimal cuts $\cancel{G}^{\text{min}}_{s_I, s_J}$ takes the form of (any of) the following cuts:
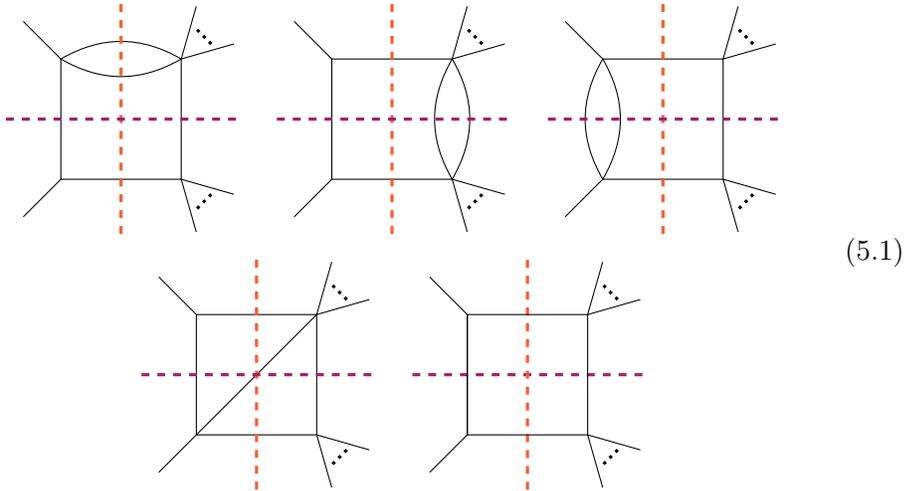
\begin{equation}
\begin{gathered}
\begin{tikzpicture}[scale=1] 
        \coordinate (a) at (-.8,.8);
        \coordinate (b) at (.8,.8);
        \coordinate (c) at (.8,-.8);
        \coordinate (d) at (-.8,-.8);
        \coordinate (a1) at (-1.3, 1.3);
        \coordinate (d1) at (-1.3, -1.3);
        \coordinate (b1) at (1, 1.5);
        \coordinate (b2) at (1.5, 1);
        \coordinate (c1) at (1, -1.5);
        \coordinate (c2) at (1.5, -1);
        
        \draw[] (a) to [out=-30,in=-150] (b);
        \draw[] (a) to [out=30,in=150](b);
        \draw[] (b) to (c) to (d) to (a);
        
        \draw[] (b1) to (b) to (b2);
        \draw[] (c1) to (c) to (c2);
        \draw[] (a1) to (a);
        \draw[] (d1) to (d);
        \draw [dotted, line width=1.2pt] (1.2, 1) -- (1,1.2);
        \draw [dotted, line width=1.2pt] (1.2, -1) -- (1,-1.2);

        \draw[RedOrange,thick,dashed,line width=1.2pt] (0,1.53) to (0,-1.53);
        \draw[RedViolet,thick,dashed,line width=1.2pt] (-1.53,0) to (1.53,0);
          
    \end{tikzpicture}\hspace{0.5cm}
    \begin{tikzpicture}[scale=1]
        \coordinate (a) at (-.8,.8);
        \coordinate (b) at (.8,.8);
        \coordinate (c) at (.8,-.8);
        \coordinate (d) at (-.8,-.8);
        \coordinate (a1) at (-1.3, 1.3);
        \coordinate (d1) at (-1.3, -1.3);
        \coordinate (b1) at (1, 1.5);
        \coordinate (b2) at (1.5, 1);
        \coordinate (c1) at (1, -1.5);
        \coordinate (c2) at (1.5, -1);
        
        \draw[] (c) to (d) to (a) to (b);
        \draw[] (b) to [out=-120,in=120] (c);
        \draw[] (b) to [out=-60,in=60] (c);
        
        \draw[] (b1) to (b) to (b2);
        \draw[] (c1) to (c) to (c2);
        \draw[] (a1) to (a);
        \draw[] (d1) to (d);
        \draw [dotted, line width=1.2pt] (1.2, 1) -- (1,1.2);
        \draw [dotted, line width=1.2pt] (1.2, -1) -- (1,-1.2);

        \draw[RedOrange,thick,dashed,line width=1.2pt] (0,1.53) to (0,-1.53);
        \draw[RedViolet,thick,dashed,line width=1.2pt] (-1.53,0) to (1.53,0);
          
    \end{tikzpicture}\hspace{0.5cm}
\begin{tikzpicture}[scale=1]
        \coordinate (a) at (-.8,.8);
        \coordinate (b) at (.8,.8);
        \coordinate (c) at (.8,-.8);
        \coordinate (d) at (-.8,-.8);
        \coordinate (a1) at (-1.3, 1.3);
        \coordinate (d1) at (-1.3, -1.3);
        \coordinate (b1) at (1, 1.5);
        \coordinate (b2) at (1.5, 1);
        \coordinate (c1) at (1, -1.5);
        \coordinate (c2) at (1.5, -1);
        
        \draw[] (a) to (b) to (c) to (d);
        \draw[] (d) to [out=120, in = -120] (a);
        \draw[] (d) to [out=60, in = -60] (a);
        
        \draw[] (b1) to (b) to (b2);
        \draw[] (c1) to (c) to (c2);
        \draw[] (a1) to (a);
        \draw[] (d1) to (d);
        \draw [dotted, line width=1.2pt] (1.2, 1) -- (1,1.2);
        \draw [dotted, line width=1.2pt] (1.2, -1) -- (1,-1.2);

        \draw[RedOrange,thick,dashed,line width=1.2pt] (0,1.53) to (0,-1.53);
        \draw[RedViolet,thick,dashed,line width=1.2pt] (-1.53,0) to (1.53,0);
          
    \end{tikzpicture} \\
    \begin{tikzpicture}[scale=1]
        \coordinate (a) at (-.8,.8);
        \coordinate (b) at (.8,.8);
        \coordinate (c) at (.8,-.8);
        \coordinate (d) at (-.8,-.8);
        \coordinate (a1) at (-1.3, 1.3);
        \coordinate (d1) at (-1.3, -1.3);
        \coordinate (b1) at (1, 1.5);
        \coordinate (b2) at (1.5, 1);
        \coordinate (c1) at (1, -1.5);
        \coordinate (c2) at (1.5, -1);
        
        \draw[] (a) to (b) to (c) to (d) to (a);
        \draw[] (d) to (b);
        
        \draw[] (b1) to (b) to (b2);
        \draw[] (c1) to (c) to (c2);
        \draw[] (a1) to (a);
        \draw[] (d1) to (d);
        \draw [dotted, line width=1.2pt] (1.2, 1) -- (1,1.2);
        \draw [dotted, line width=1.2pt] (1.2, -1) -- (1,-1.2);

        \draw[RedOrange,thick,dashed,line width=1.2pt] (0,1.53) to (0,-1.53);
        \draw[RedViolet,thick,dashed,line width=1.2pt] (-1.53,0) to (1.53,0);
          
    \end{tikzpicture}
\hspace{0.5cm}
\begin{tikzpicture}[scale=1]
        \coordinate (a) at (-.8,.8);
        \coordinate (b) at (.8,.8);
        \coordinate (c) at (.8,-.8);
        \coordinate (d) at (-.8,-.8);
        \coordinate (h) at (.8,0);

        \coordinate (a1) at (-1.3, 1.3);
        \coordinate (d1) at (-1.3, -1.3);
        \coordinate (b1) at (1, 1.5);
        \coordinate (b2) at (1.5, 1);
        \coordinate (c1) at (1, -1.5);
        \coordinate (c2) at (1.5, -1);

        \coordinate (cutL) at (-1.53, 0);
        \coordinate (cutR) at (1.53, 0);
        \coordinate (cutB) at (0,-1.53);
        \coordinate (cutT) at (0,1.53);
        
        \draw[] (a1) to (a);
        \draw[] (d1) to (d);
        \draw[] (b1) to (b);
        \draw[] (b2) to (b);
        \draw[] (c1) to (c);
        \draw[] (c2) to (c);

        \draw[dashed,thick,RedViolet,line width=1.2pt] (cutL) to (cutR);
        \draw[dashed,thick,RedOrange,line width=1.2pt] (cutB) to (cutT);

       \draw [dotted, line width=1.2pt] (1.2, 1) -- (1,1.2);
        \draw [dotted, line width=1.2pt] (1.2, -1) -- (1,-1.2);

        \draw[] (d) to node[midway, left] {}(a)
        to node[midway, above] {}(b)
        to node[midway, right] {}(c) 
        to node[midway, below] {}(d)
        to node[midway, right] {}(a);      
\end{tikzpicture}
\end{gathered} \label{eq:2loopViolations}
\end{equation}
where at least one external particle is attached to each corner of the box. 
\end{quote}}
\begin{tikzpicture}[overlay, remember picture,decoration={markings,mark=at position .99 with {\arrow[scale=1.4,>=stealth]{>}}}]
    \draw [solid,rounded corners=.2cm,color=Black] ([yshift=-.3cm,xshift=-.1cm]{pic cs:two_loop_test_start}) rectangle ([yshift=-11.6cm,xshift=14.2cm]{pic cs:two_loop_test_start});
\end{tikzpicture}

\noindent Note that two-loop integrals that involve exactly four external momenta will always give rise to one of the cuts in~\eqref{eq:2loopViolations} for any pair of partially-overlapping channels, as long as no more than one of the incoming momenta is massive (or, equivalently, is comprised of a sum of massless momenta); a similar statement holds for two-loop integrals involving exactly five massless external momenta.\footnote{We have assumed here that the corresponding diagram is one-particle-irreducible.} Therefore, this conjecture first becomes interesting when considering diagrams that involve at least six massless external momenta (or fewer external momenta, if one trades a pair of massless momenta for a massive one). 

This still leaves us with a wealth of data against which we can test our two-loop conjecture. For instance, there exist known results for five-point one-mass pentaboxes~\cite{Abreu:2020jxa}, five-point one-mass hexaboxes~\cite{Abreu:2021smk}, five-point one-mass double-pentagons~\cite{Abreu:2023rco}, five-point two-mass pentaboxes~\cite{Abreu:2024yit}, and planar six-point diagrams~\cite{Henn:2024ngj}. Altogether, these families of integrals furnish us with 1,626 master integrals whose Steinmann properties we can study.\footnote{The actual number of independent master integrals will be smaller than this, since there will be some overlap between the master integrals in lower sectors.} To test our conjecture in each of these integrals, we simply construct the complete set of minimal cuts that exist for each pair of partially-overlapping momentum channels, and see if any of these minimal cuts can be found in~\eqref{eq:2loopViolations}. If not, we predict that the Steinmann relations in the corresponding pair of channels should be respected.

\begin{figure}
    \centering
    \tikzmark{minimal_cuts}
        \begin{tikzpicture}
        \coordinate (a) at (-.8,-.2);
        \coordinate (b) at (.8,-.2);
        \coordinate (c) at (.8,-1.8);
        \coordinate (d) at (-.8,-1.8);
        \coordinate (e) at (2.2, .2);
        \coordinate (f) at (3, -1);
        \coordinate (g) at (2.2, -2.2);

        \coordinate (d1) at (-1.1, -2.3);
        \coordinate (d2) at (-1.3, -2.1);
        \coordinate (e1) at (2.6, .6);
        \coordinate (f1) at (3.4, -1);
        \coordinate (g1) at (2.6, -2.6);
        
        \draw[] (a) to (b) to (c) to (d)to (a);
        \draw[] (b) to (e) to (f) to (g) to (c);
        
        \draw[] (-1.2,.2) to (a);
        \draw[] (d1) to (d);
        \draw[] (d2) to (d);
        \draw[] (e1) to (e);
        \draw[] (f1) to (f);
        \draw[] (g1) to (g);

        \node[] at (-1.5,-2.2) {$6$};
        \node[] at (-1.2,-2.5) {$1$};
        \node[] at (2.8,-2.8) {$2$};
        \node[] at (3.6,-1) {$3$};
        \node[] at (2.8,.8) {$4$};
        \node[] at (-1.4, .4) {$5$};

        \draw[RedOrange,thick,dashed] (-0.2,.4) to (-0.2,-2.4);
        \draw[RedOrange,thick,dashed] (2,.4) to (2,-2.4);
        \draw[RedOrange,thick,dashed] (0.1,.4) to (1.7,-2.4);
        \draw[RedOrange,thick,dashed] (1.7,.4) to (0.1,-2.4);
        \draw[RoyalBlue,thick,dashed, line width=1pt] (-1.2,-1.1) to [out=0,in=160] (3,-1.7);
        \draw[Green,thick,dashed, line width=1pt] (-1.2,-0.7) to [out=30,in=180] (0.8,.2)
                                               to [out=0, in=130] (3,-1.5);
        \draw[Purple,thick,dashed, line width=1pt] (-1.2,-1.5) to [out=0,in=-180] (0.6,-2.2)
                                               to [out=0, in=-180] (3,-1.9);

        \node[RedOrange] at (-0.2,0.7) {$a$};
        \node[RedOrange] at (0,-2.6) {$c$};
        \node[RedOrange] at (1.7,-2.6) {$d$};
        \node[RedOrange] at (2,0.75) {$b$};

        \begin{scope}[shift={(-0.6,0)}]
        \draw[] (5,1.4) to (5.8, 1.4) to [out=-30, in=30] (5.8, 0.6) to [out=150, in=-150] (5.8, 1.4);
        \draw[] (5.8, 0.6) to (5, 0.6) to (5, 1.4);
        \draw[] (4.8, 1.6) to (5, 1.4);
        \draw[] (4.9, 0.3) to (5, 0.6);
        \draw[] (4.7, 0.5) to (5, 0.6);
        \draw[] (5.9, 1.7) to (5.8, 1.4) to (6.1, 1.5);
        \draw[] (6, 0.4) to (5.8, 0.6);

        \node[] at (4.55,0.5) {\scriptsize $6$};
        \node[] at (4.8,0.2) {\scriptsize $1$};
        \node[] at (6.1,0.3) {\scriptsize $2$};
        \node[] at (6.2,1.6) {\scriptsize $3$};
        \node[] at (6,1.9) {\scriptsize $4$};
        \node[] at (4.7,1.7) {\scriptsize $5$};
        
        \draw[RedOrange,thick,dashed] (5.4,1.6) to (5.4,0.4);
        \draw[RoyalBlue,thick,dashed] (4.8,1) to (6.2,1);

        \node[RedOrange] at (5.4,2.4) {$a$};
        \end{scope}

        \begin{scope}[shift={(-0.4,0)}]
        \draw[] (7,1.4) to (7.8, 1.4) to (7.8, 0.6) ;
        \draw[] (7.8, 0.6) to (7, 0.6) to [out= 150, in= -150] (7, 1.4) to [out=-30, in=30](7, 0.6);
        \draw[] (6.8, 1.6) to (7, 1.4);
        \draw[] (6.7, 0.5) to (7, 0.6);
        \draw[] (6.9, 0.3) to (7, 0.6);
        \draw[] (7.9, 1.7) to (7.8, 1.4) to (8.1, 1.5);
        \draw[] (8, 0.4) to (7.8, 0.6);

        \node[] at (6.55,0.5) {\scriptsize $6$};
        \node[] at (6.8,0.2) {\scriptsize $1$};
        \node[] at (8.1,0.3) {\scriptsize $2$};
        \node[] at (8.2,1.6) {\scriptsize $3$};
        \node[] at (8,1.9) {\scriptsize $4$};
        \node[] at (6.7,1.7) {\scriptsize $5$};

        \draw[RedOrange,thick,dashed] (7.4,1.6) to (7.4,0.4);
        \draw[RoyalBlue,thick,dashed] (6.6,1) to (8,1);

        \node[RedOrange] at (7.4,2.4) {$b$};
        \end{scope}

        \begin{scope}[shift={(0.2,0)}]
        \draw[] (11,0.6) to (11.8, 0.6) to (11.8, 1.4);
        \draw[] (11.8, 1.4) to (11, 1.4) to (11, 0.6);
        \draw[] (11.8, 1.4) to (11, 0.6);
        \draw[] (10.7, 0.5) to (11, 0.6);
        \draw[] (10.9, 0.3) to (11, 0.6);
        \draw[] (10.8,1.6) to (11, 1.4);
        \draw[] (12, 0.4) to (11.8, 0.6);
        \draw[] (11.9, 1.7) to (11.8, 1.4) to (12.1,1.5);

        \node[] at (10.55,0.5) {\scriptsize $6$};
        \node[] at (10.8,0.2) {\scriptsize $1$};
        \node[] at (12.1,0.3) {\scriptsize $2$};
        \node[] at (12.2,1.6) {\scriptsize $3$};
        \node[] at (11.9,1.9) {\scriptsize $4$};
        \node[] at (10.7,1.7) {\scriptsize $5$};

        \draw[RedOrange,thick,dashed] (11.4,0.4) to (11.4,1.6);
        \draw[RoyalBlue,thick,dashed] (10.8,1) to (12,1);

        \node[RedOrange] at (11.4,2.4) {$d$};
        \end{scope}

        \begin{scope}[shift={(-0.2,0)}]
        \draw[] (9,0.6) to (9.8, 0.6) to (9.8, 1.4);
        \draw[] (9.8, 1.4) to (9, 1.4) to (9, 0.6);
        \draw[] (9.8, 0.6) to (9, 1.4);
        \draw[] (8.7, 0.5) to (9, 0.6);
        \draw[] (8.9, 0.3) to (9, 0.6);
        \draw[] (8.8, 1.6) to (9, 1.4);
        \draw[] (10, 0.4) to (9.8, 0.6);
        \draw[] (9.9, 1.7) to (9.8, 1.4) to (10.1,1.5);

        \node[] at (8.55,0.5) {\scriptsize $6$};
        \node[] at (8.8,0.2) {\scriptsize $1$};
        \node[] at (10.1,0.3) {\scriptsize $2$};
        \node[] at (10.2,1.6) {\scriptsize $3$};
        \node[] at (9.9,1.9) {\scriptsize $4$};
        \node[] at (8.7,1.7) {\scriptsize $5$};

        \draw[RedOrange,thick,dashed] (9.4,0.4) to (9.4,1.6);
        \draw[RoyalBlue,thick,dashed] (8.8,1) to (10,1);

        \node[RedOrange] at (9.4,2.4) {$c$};
        \end{scope}

        \begin{scope}[shift={(-0.6,0)}]
        \draw[] (5,-1.4) to (5.4,-1) to (5,-0.6) to (5,-1.4);
        \draw[] (5.4,-1) to [out=60, in=120](5.8,-1) to [out=-120, in=-60](5.4,-1);
        \draw[] (5.4,-1) to (5.4, -1.4);
        \draw[] (4.9, -1.7) to (5, -1.4);
        \draw[] (4.7, -1.5) to (5, -1.4);
        \draw[] (4.8,- 0.4) to (5, -0.6);
        \draw[] (6,-0.8) to (5.8,-1) to (6,-1.2);

        \node[] at (4.55,-1.5) {\scriptsize $6$};
        \node[] at (4.8,-1.8) {\scriptsize $1$};
        \node[] at (5.4,-1.6) {\scriptsize $2$};
        \node[] at (6.1,-1.3) {\scriptsize $3$};
        \node[] at (6.1,-0.7) {\scriptsize $4$};
        \node[] at (4.7,-0.3) {\scriptsize $5$};
        
        \draw[RedOrange,thick,dashed] (5.2,-0.5) to (5.2,-1.5);
        \draw[Green,thick,dashed] (4.8,-1) to [out=30, in=180](5.4,-0.7) to [out=0, in=100](5.7,-1.4);
        \end{scope}

        \begin{scope}[shift={(-0.4,0)}]
        \draw[] (7.8,-0.6) to (7.4,-1) to (7.8, -1.4) to (7.8,-0.6);
        \draw[] (7.4,-1) to [out=-120, in=-60] (7,-1) to [out=60, in=120] (7.4,-1);
        \draw[] (6.7,-1) to (7,-1);
        \draw[] (7.4,-1) to (7.5,-1.4);
        \draw[] (7.4,-1) to (7.3,-1.4);
        \draw[] (8,-1.6) to (7.8,-1.4);
        \draw[] (7.9, -0.3) to (7.8,-0.6) to (8.1,-0.5);

        \node[] at (7.5,-1.6) {\scriptsize $6$};
        \node[] at (7.3,-1.6) {\scriptsize $1$};
        \node[] at (8.1,-1.7) {\scriptsize $2$};
        \node[] at (8.2,-0.5) {\scriptsize $3$};
        \node[] at (7.9,-0.1) {\scriptsize $4$};
        \node[] at (6.6,-1) {\scriptsize $5$};

        \draw[RedOrange,thick,dashed] (7.7,-0.5) to (7.7,-1.5);
        \draw[Green,thick,dashed] (7.1,-1.4) to [out=80, in=-180](7.4,-0.7) to [out=0, in=150](8.1,-1.1);
        \end{scope}

        \begin{scope}[shift={(-0.1,0)}]
        \draw[] (8.8,-1.4) to (9.6, -1.4) to (9.6, -0.6);
        \draw[] (9.6, -0.6) to (8.8, -0.6) to (8.8, -1.4);
        \draw[] (8.5, -1.5) to (8.8, -1.4);
        \draw[] (8.7, -1.7) to (8.8, -1.4);
        \draw[] (8.6,- 0.4) to (8.8, -0.6);
        \draw[] (9.8, -1.6) to (9.6, -1.4);
        \draw[] (9.6,-1.4) to (10,-1) to (9.6,-0.6);
        \draw[] (10.2,-0.8) to (10,-1) to (10.2,-1.2);

        \node[] at (8.35,-1.5) {\scriptsize $6$};
        \node[] at (8.6,-1.8) {\scriptsize $1$};
        \node[] at (9.9,-1.7) {\scriptsize $2$};
        \node[] at (10.3,-1.3) {\scriptsize $3$};
        \node[] at (10.3,-0.7) {\scriptsize $4$};
        \node[] at (8.5,-0.3) {\scriptsize $5$};

        \draw[RedOrange,thick,dashed] (9.1,-1.7) to (10.1,-.7);
        \draw[Green,thick,dashed] (8.5,-1.1) to [out=30, in=-160](9.6,-0.4) to [out=0, in=130](10,-1.3);
        \end{scope}

        \begin{scope}[shift={(0.1,0)}]
        \draw[] (11.4,-1.4) to (11,-1) to (11.4, -0.6) to (11.4,-1.4);
        \draw[] (11.4,-1.4) to (12.2,-1.4) to (12.2,-0.6) to (11.4,-0.6);
        \draw[] (10.7,-1) to (11,-1);
        \draw[] (11.1,-1.5) to (11.4,-1.4);
        \draw[] (11.3,-1.7) to (11.4,-1.4);
        \draw[] (12.2,-1.4) to (12.4,-1.6);
        \draw[] (12.3,-0.3) to (12.2,-0.6) to (12.5,-0.5);

        \node[] at (10.95,-1.5) {\scriptsize $6$};
        \node[] at (11.2,-1.8) {\scriptsize $1$};
        \node[] at (12.5,-1.7) {\scriptsize $2$};
        \node[] at (12.6,-0.5) {\scriptsize $3$};
        \node[] at (12.3,-0.1) {\scriptsize $4$};
        \node[] at (10.6,-1) {\scriptsize $5$};

        \draw[RedOrange,thick,dashed] (10.9,-0.7) to (11.9,-1.7);
        \draw[Green,thick,dashed] (11,-1.3) to [out=50, in=-180](11.6,-0.4) to [out=-30, in=150](12.5,-1.1);
        \end{scope}

        \begin{scope}[shift={(-0.6,0)}]
        \draw[] (5,-3.4) to (5.4,-3) to (5,-2.6) to (5,-3.4);
        \draw[] (5.4,-3) to [out=60, in=120](5.8,-3) to [out=-120, in=-60](5.4,-3);
        \draw[] (5.4,-3) to (5.3, -2.6);
        \draw[] (5.4,-3) to (5.5, -2.6);
        \draw[] (4.9, -3.7) to (5, -3.4);
        \draw[] (4.7, -3.5) to (5, -3.4);
        \draw[] (4.8,- 2.4) to (5, -2.6);
        \draw[] (6,-3) to (5.8,-3);

        \node[] at (4.55,-3.5) {\scriptsize $6$};
        \node[] at (4.8,-3.8) {\scriptsize $1$};
        \node[] at (5.3,-2.4) {\scriptsize $3$};
        \node[] at (5.5,-2.4) {\scriptsize $4$};
        \node[] at (6.2,-3) {\scriptsize $2$};
        \node[] at (4.7,-2.3) {\scriptsize $5$};
        
        \draw[RedOrange,thick,dashed] (5.2,-2.5) to (5.2,-3.5);
        \draw[Purple,thick,dashed] (4.8,-3) to [out=-30, in=180](5.4,-3.3) to [out=0, in=-100](5.7,-2.6);
        \end{scope}

        \begin{scope}[shift={(-0.5,0)}]
        \draw[] (7.8,-2.6) to (7.4,-3) to (7.8, -3.4) to (7.8,-2.6);
        \draw[] (7.4,-3) to [out=-120, in=-60] (7,-3) to [out=60, in=120] (7.4,-3);
        \draw[] (6.8,-3.1) to (7,-3);
        \draw[] (6.8,-2.9) to (7,-3);
        \draw[] (7.4,-3) to (7.4,-2.6);
        \draw[] (8,-3.6) to (7.8,-3.4);
        \draw[] (7.9, -2.3) to (7.8,-2.6) to (8.1,-2.5);

        \node[] at (6.7,-2.8) {\scriptsize $6$};
        \node[] at (6.7,-3.2) {\scriptsize $1$};
        \node[] at (8.1,-3.7) {\scriptsize $2$};
        \node[] at (8.2,-2.5) {\scriptsize $3$};
        \node[] at (7.9,-2.1) {\scriptsize $4$};
        \node[] at (7.4,-2.4) {\scriptsize $5$};

        \draw[RedOrange,thick,dashed] (7.7,-2.5) to (7.7,-3.5);
        \draw[Purple,thick,dashed] (7.1,-2.7) to [out=-80, in=-180](7.4,-3.3) to [out=0, in=-150](8.1,-2.9);
        \end{scope}

        \begin{scope}[shift={(-0.2,0)}]
        \draw[] (9.2,-3.4) to (8.8,-3) to (9.2, -2.6) to (9.2,-3.4);
        \draw[] (9.2,-3.4) to (10,-3.4) to (10,-2.6) to (9.2,-2.6);
        \draw[] (8.6,-3.1) to (8.8,-3);
        \draw[] (8.6,-2.9) to (8.8,-3);
        \draw[] (9,-2.4) to (9.2,-2.6);
        \draw[] (10,-3.4) to (10.2,-3.6);
        \draw[] (10.1,-2.3) to (10,-2.6) to (10.3,-2.5);

        \node[] at (8.5,-2.8) {\scriptsize $6$};
        \node[] at (8.5,-3.2) {\scriptsize $1$};
        \node[] at (10.3,-3.7) {\scriptsize $2$};
        \node[] at (10.4,-2.5) {\scriptsize $3$};
        \node[] at (10.1,-2.1) {\scriptsize $4$};
        \node[] at (8.9,-2.3) {\scriptsize $5$};

        \draw[RedOrange,thick,dashed] (8.7,-3.2) to (9.8,-2.4);
        \draw[Purple,thick,dashed] (8.8,-2.8) to [out=-50, in=-180](9.4,-3.65) to [out=30, in=-150](10.3,-2.8);
        \end{scope}

        \begin{scope}[shift={(0.2,0)}]
        \draw[] (11,-3.4) to (11.8, -3.4) to (11.8, -2.6);
        \draw[] (11.8, -2.6) to (11, -2.6) to (11, -3.4);
        \draw[] (10.7, -3.5) to (11, -3.4);
        \draw[] (10.9, -3.7) to (11, -3.4);
        \draw[] (10.8,- 2.4) to (11, -2.6);
        \draw[] (11.8,-3.4) to (12.2,-3) to (11.8,-2.6);
        \draw[] (12.2,-3) to (12.4,-3);
        \draw[] (11.9,-2.3) to (11.8,-2.6) to (12.1,-2.5);

        \node[] at (10.55,-3.5) {\scriptsize $6$};
        \node[] at (10.8,-3.8) {\scriptsize $1$};
        \node[] at (12.5,-3) {\scriptsize $2$};
        \node[] at (12.2,-2.4) {\scriptsize $3$};
        \node[] at (12,-2.1) {\scriptsize $4$};
        \node[] at (10.7,-2.3) {\scriptsize $5$};

        \draw[RedOrange,thick,dashed] (11.3,-2.3) to (12.3,-3.4);
        \draw[Purple,thick,dashed] (10.7,-2.8) to [out=-30, in=-150](11.8,-3.6) to [out=40, in=-120](12.2,-2.6);
        \end{scope}
        
        \end{tikzpicture}

        \begin{tikzpicture}[overlay, remember picture,decoration={markings,mark=at position .99 with {\arrow[scale=1.4,>=stealth]{>}}}]
    \draw [solid,color=Black] ([yshift=0cm,xshift=7.7cm]{pic cs:minimal_cuts}) -- ([yshift=6cm,xshift=7.7cm]{pic cs:minimal_cuts});
    \draw [solid,color=Black] ([yshift=0cm,xshift=9.9cm]{pic cs:minimal_cuts}) -- ([yshift=6cm,xshift=9.9cm]{pic cs:minimal_cuts});
    \draw [solid,color=Black] ([yshift=0cm,xshift=12.3cm]{pic cs:minimal_cuts}) -- ([yshift=6cm,xshift=12.3cm]{pic cs:minimal_cuts});
    \draw [dashed,color=Black] ([yshift=6.1cm,xshift=5.7cm]{pic cs:minimal_cuts}) -- ([yshift=6.1cm,xshift=14.7cm]{pic cs:minimal_cuts});
    \draw [dashed,color=Black] ([yshift=-0.1cm,xshift=5.7cm]{pic cs:minimal_cuts}) -- ([yshift=-0.1cm,xshift=14.7cm]{pic cs:minimal_cuts});
    \end{tikzpicture}

        \caption{A one-mass pentabox integral, which supports three cuts in the $s_{345}$ channel and four cuts in the $s_{234}$ channel. When combined in all possible ways, these cuts give rise to the twelve partially-overlapping cuts shown on the right.}
    \label{fig:pentabox_cuts_1}
\end{figure}
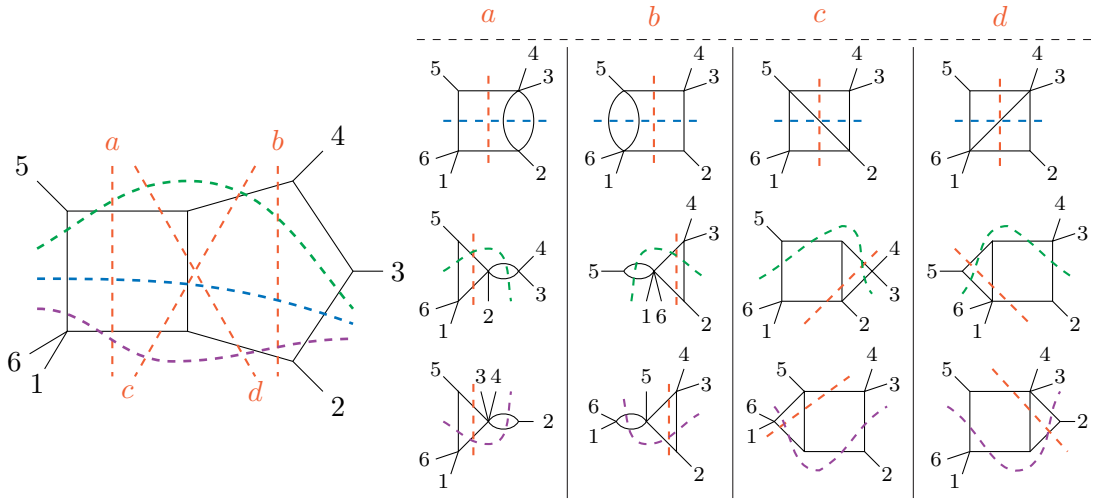

As an example, consider the one-mass pentabox diagram shown in Figure~\ref{fig:pentabox_cuts_1}. This graph supports three minimal cuts in the $s_{345}$ channel (shown in green, blue, and purple) and four minimal cuts in the $s_{234}$ channel (shown in orange). Combining these minimal cuts in all possible ways, we can construct twelve partially-overlapping cuts, as depicted on the right side of the figure.\footnote{As highlighted in section~\ref{sec:minimal_cuts}, some of these partially-overlapping cuts are not minimal, in the sense that one or more of their on-shell conditions can be relaxed while still partitioning the external momenta into the necessary subsets. These cuts can be discarded, as they can be contracted to other partially-overlapping cuts that involve fewer propagators.} As none of these partially-overlapping cuts match those from~\eqref{eq:2loopViolations}, we predict that $\text{Disc}_{s_{345}} ( \text{Disc}_{s_{234}}(\mathcal{I})) = \text{Disc}_{s_{234}} ( \text{Disc}_{s_{345}}(\mathcal{I})) = 0$
for any Feynman integral with this topology. In fact, the same prediction holds for any Feynman diagram that this one-mass pentabox diagram can be contracted to, since there is no way to generate one of the cuts in~\eqref{eq:2loopViolations} by contracting additional lines. We have confirmed that this prediction is indeed borne out in the complete family of integrals associated with this one-mass pentabox topology~\cite{Abreu:2020jxa}.

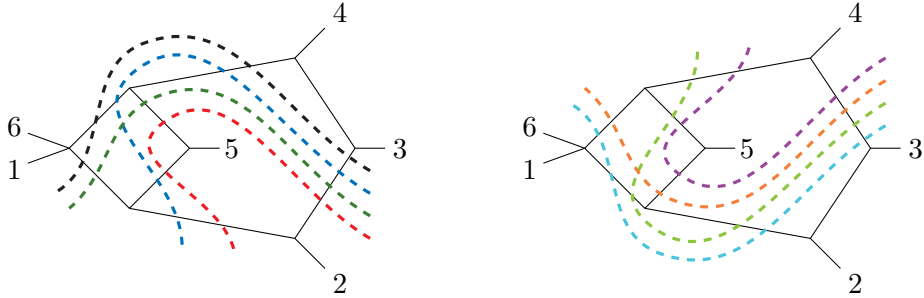
\begin{figure}
    \centering
        \begin{tikzpicture}
        \coordinate (a) at (-.8,0);
        \coordinate (b) at (0,.8);
        \coordinate (c) at (.8,0);
        \coordinate (d) at (0,-.8);
        \coordinate (e) at (2.2, 1.2);
        \coordinate (f) at (3, 0);
        \coordinate (g) at (2.2, -1.2);

        \coordinate (a1) at (-1.35, 0.2);
        \coordinate (a2) at (-1.35, -0.2);
        \coordinate (c1) at (1.2, 0);
        \coordinate (e1) at (2.6, 1.6);
        \coordinate (f1) at (3.4, 0);
        \coordinate (g1) at (2.6, -1.6);

        \node[] at (-1.53,.28) {$6$};
        \node[] at (-1.53,-.28) {$1$};
        \node[] at (1.36,0) {$5$};
        \node[] at (2.8,1.8) {$4$};
        \node[] at (3.6,0) {$3$};
        \node[] at (2.8,-1.8) {$2$};
        
        \draw[] (a) to (b) 
                    to (c) 
                    to (d)
                    to (a);
        \draw[] (b) to (e)
                    to (f)
                    to (g)
                    to (d);
        
        \draw[] (a1) to (a);
        \draw[] (a2) to (a);
        \draw[] (c1) to (c);
        \draw[] (e1) to (e);
        \draw[] (f1) to (f);
        \draw[] (g1) to (g);

        \draw[Black,thick,dashed,line width=1.2pt] (3.2,-0.3) to [out=150,in=20] (0.2,1.4) to [out=200,in=30] (-1.,-0.6);
        \draw[RoyalBlue,thick,dashed,line width=1.2pt] (3.2,-0.6) to [out=150,in=40] (0.04,1.) to [out=220,in=90] (0.7,-1.3);
        \draw[OliveGreen,thick,dashed,line width=1.2pt] (3.2,-0.9) to [out=150,in=20] (0.4,0.7) to [out=200,in=40] (-0.8,-0.8);
        \draw[Red,thick,dashed,line width=1.2pt] (3.2,-1.2) to [out=150,in=40] (0.4,0.3) to [out=220,in=100] (1.4,-1.4);
        \end{tikzpicture}
        \qquad 
        \quad 
\begin{tikzpicture}
        \coordinate (a) at (-.8,0);
        \coordinate (b) at (0,.8);
        \coordinate (c) at (.8,0);
        \coordinate (d) at (0,-.8);
        \coordinate (e) at (2.2, 1.2);
        \coordinate (f) at (3, 0);
        \coordinate (g) at (2.2, -1.2);

        \coordinate (a1) at (-1.35, 0.2);
        \coordinate (a2) at (-1.35, -0.2);
        \coordinate (c1) at (1.2, 0);
        \coordinate (e1) at (2.6, 1.6);
        \coordinate (f1) at (3.4, 0);
        \coordinate (g1) at (2.6, -1.6);

        \node[] at (-1.53,.28) {$6$};
        \node[] at (-1.53,-.28) {$1$};
        \node[] at (1.36,0) {$5$};
        \node[] at (2.8,1.8) {$4$};
        \node[] at (3.6,0) {$3$};
        \node[] at (2.8,-1.8) {$2$};
        
        \draw[] (a) to (b) 
                    to (c) 
                    to (d)
                    to (a);
        \draw[] (b) to (e)
                    to (f)
                    to (g)
                    to (d);
        
        \draw[] (a1) to (a);
        \draw[] (a2) to (a);
        \draw[] (c1) to (c);
        \draw[] (e1) to (e);
        \draw[] (f1) to (f);
        \draw[] (g1) to (g);

        \draw[SkyBlue,thick,dashed,line width=1.2pt] (3.2,0.3) to [out=-150,in=-20] (0.2,-1.4) to [out=-200,in=-30] (-1.,0.6);
        \draw[LimeGreen,thick,dashed,line width=1.2pt] (3.2,0.6) to [out=-150,in=-40] (0.04,-1.) to [out=-220,in=-90] (0.7,1.3);
        \draw[Orange,thick,dashed,line width=1.2pt] (3.2,0.9) to [out=-150,in=-20] (0.4,-0.7) to [out=-200,in=-40] (-0.8,0.8);
        \draw[Purple,thick,dashed,line width=1.2pt] (3.2,1.2) to [out=-150,in=-40] (0.4,-0.3) to [out=-220,in=-100] (1.4,1.4);
        \end{tikzpicture}
        
        \caption{The minimal cuts that appear for one of the one-mass hexabox topologies considered in~\cite{Abreu:2021smk} in the $s_{25}$ channel (left), and in the $s_{45}$ channel (right).}
    \label{fig:hexabox_cuts}
\end{figure}

A similar example is depicted in Figure~\ref{fig:hexabox_cuts}, for channels $s_{24}$ and $s_{25}$ of a one-mass nonplanar hexabox topology. As shown there, this diagram supports four minimal cuts in both of these channels. One can thus construct sixteen partially-overlapping cuts by combining these cuts in all possible ways, as we did in the last example. However, in this case it is quicker to notice that all possible partially-overlapping $s_{24}$, $s_{25}$ cuts will partition the diagram into five regions, each of which contains at least one external momentum. This is already enough to see that none of the cuts from~\eqref{eq:2loopViolations} will appear, so we predict that $\text{Disc}_{s_{24}} ( \text{Disc}_{s_{25}}(\mathcal{I})) = \text{Disc}_{s_{25}} ( \text{Disc}_{s_{24}}(\mathcal{I})) = 0$ for any Feynman integral associated with this topology or its contractions. Again, we have confirmed this prediction for all integrals in the corresponding one-mass hexabox family, as computed in~\cite{Abreu:2021smk}.

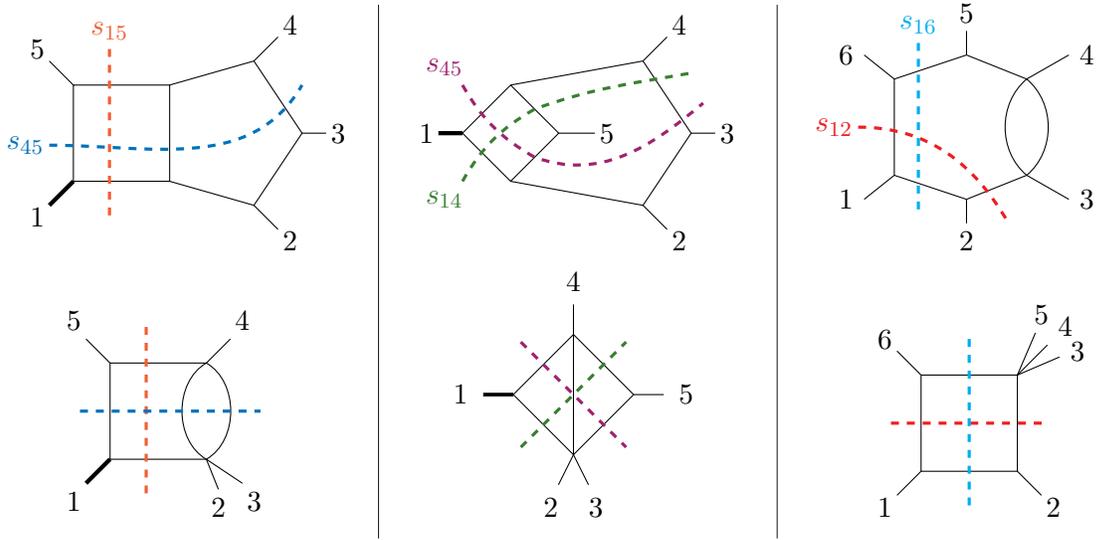
\begin{figure}[t]
\centering
    \tikzmark{three_examples}
    \begin{subfigure}[t]{.3\textwidth}
        \begin{tikzpicture}[scale=.8]
        \coordinate (a) at (-.8,.8);
        \coordinate (b) at (.8,.8);
        \coordinate (c) at (.8,-.8);
        \coordinate (d) at (-.8,-.8);
        \coordinate (e) at (2.2, 1.2);
        \coordinate (f) at (3, 0);
        \coordinate (g) at (2.2, -1.2);

        \coordinate (d1) at (-1.2, -1.2);
        \coordinate (e1) at (2.6, 1.6);
        \coordinate (f1) at (3.4, 0);
        \coordinate (g1) at (2.6, -1.6);
        
        \draw[] (a) to (b) to (c) to (d)to (a);
        \draw[] (b) to (e) to (f) to (g) to (c);
        
        \draw[] (-1.2,1.2) to (a);
        \draw[line width=1.6pt] (-1.2, -1.2) to (d);
        \draw[] (e1) to (e);
        \draw[] (f1) to (f);
        \draw[] (g1) to (g);

        \node[] at (-1.4,-1.4) {$1$};
        \node[] at (2.8,-1.8) {$2$};
        \node[] at (3.6,0) {$3$};
        \node[] at (2.8,1.8) {$4$};
        \node[] at (-1.4, 1.4) {$5$};

        \node[RedOrange,thick] at (-0.2,1.7) {$s_{15}$};
        \node[RoyalBlue,thick] at (-1.6,-0.2) {$s_{45}$};
        \draw[RedOrange,thick,dashed,line width=1.2pt] (-0.2,1.4) to (-0.2,-1.4);
        \draw[RoyalBlue,thick,dashed, line width=1.2pt] (-1.2,-0.2) to [out=0, in=-120] (3,0.8);
        \end{tikzpicture}
    \end{subfigure}
      \hfill
    \begin{subfigure}[t]{.3\textwidth}
        \centering
        \begin{tikzpicture}[scale=0.8]

        \coordinate (a) at (-.8,0);
        \coordinate (b) at (0,.8);
        \coordinate (c) at (.8,0);
        \coordinate (d) at (0,-.8);
        \coordinate (e) at (2.2, 1.2);
        \coordinate (f) at (3, 0);
        \coordinate (g) at (2.2, -1.2);

        \coordinate (a1) at (-1.2, 0);
        \coordinate (c1) at (1.4, 0);
        \coordinate (e1) at (2.6, 1.6);
        \coordinate (f1) at (3.4, 0);
        \coordinate (g1) at (2.6, -1.6);

        \node[] at (-1.4,0) {$1$};
        \node[] at (1.6,0) {$5$};
        \node[] at (2.8,1.8) {$4$};
        \node[] at (3.6,0) {$3$};
        \node[] at (2.8,-1.8) {$2$};
        
        \draw[] (a) to (b) 
                    to (c) 
                    to (d)
                    to (a);
        \draw[] (b) to (e)
                    to (f)
                    to (g)
                    to (d);
        
        \draw[line width=1.6pt] (a1) to (a);
        \draw[] (c1) to (c);
        \draw[] (e1) to (e);
        \draw[] (f1) to (f);
        \draw[] (g1) to (g);

        \draw[RedViolet,thick,dashed,line width=1.2pt] (-0.8,0.8) to [out=-60,in=-210] (0.4,-0.4) to [out=-20,in=-140] (3.2,0.5);
        \node[RedViolet] at (-1.1,1.1) {$s_{45}$};
        \draw[OliveGreen,thick,dashed,line width=1.2pt] (-0.8,-0.8) to [out=60,in=210] (0.4,0.4) to [out=20,in=-170] (3,1);
        \node[OliveGreen] at (-1.1,-1.1) {$s_{14}$};
        \end{tikzpicture}
    \end{subfigure}
      \hfill
    \begin{subfigure}[t]{.3\textwidth}
    \begin{tikzpicture}[scale=0.8]
        \coordinate (a) at (-.4,.8);
        \coordinate (b) at (.8,1.2);
        \coordinate (c) at (.8,-1.2);
        \coordinate (d) at (-.4,-.8);
        \coordinate (e) at (1.8, .8);
        \coordinate (f) at (1.8, -.8);

        \coordinate (a1) at (-0.9, 1.2);
        \coordinate (d1) at (-0.9, -1.2);
        \coordinate (e1) at (2.5, 1.2);
        \coordinate (f1) at (2.5, -1.2);
        \coordinate (b1) at (.8,1.6);
        \coordinate (c1) at (.8, -1.6);
        
        \draw[] (b) to (a) 
                    to (d) 
                    to (c);
        \draw[] (b) to (e)
                    to [out=-140, in = 140] (f);
        \draw[] (e) to [out=-40, in = 40] (f)    
                    to (c);
        
        \draw[] (a1) to (a);
        \draw[] (d1) to (d);
        \draw[] (e1) to (e);
        \draw[] (f1) to (f);
        \draw[] (b1) to (b);
        \draw[] (c1) to (c);

        \node[] at (-1.2,1.2) {$6$};
        \node[] at (0.8,1.9) {$5$};
        \node[] at (2.8,1.2) {$4$};
        \node[] at (2.8,-1.2) {$3$};
        \node[] at (0.8,-1.9) {$2$};
        \node[] at (-1.2,-1.2) {$1$};

        \draw[Red,thick,dashed,line width=1.2pt] (-1,0.) to [out=0,in=120] (1.5,-1.6);
        \draw[cyan,thick,dashed,line width=1.2pt] (0,1.4) to (0,-1.4);

        \node[Red] at (-1.4,0) {$s_{12}$};
        \node[cyan] at (0,1.7) {$s_{16}$};

    \end{tikzpicture} 
        
    \end{subfigure}
    \hfill
    \begin{subfigure}[b]{.3\textwidth}
    \hspace{0.8cm}
        \begin{tikzpicture}[scale=.8]
        \coordinate (a) at (-.8,.8);
        \coordinate (b) at (.8,.8);
        \coordinate (c) at (.8,-.8);
        \coordinate (d) at (-.8,-.8);

        \draw[] (a) to (b) to [out=-30,in=30](c) to (d)to (a);
        \draw[] (b) to [out=-150,in=150](c);
        
        \draw[] (-1.2,1.2) to (a);
        \draw[] (1.2,1.2) to (b);
        \draw[] (1,-1.3) to (c);
        \draw[] (1.4, -1.2) to (c);
        \draw[line width=1.6pt] (-1.2, -1.2) to (d);

        \node[] at (-1.4,-1.5) {$1$};
        \node[] at (-1.4, 1.5) {$5$};
        \node[] at (1.4,1.5) {$4$};
        \node[] at (1,-1.6) {$2$};
        \node[] at (1.6,-1.5) {$3$};

        \draw[RedOrange,thick,dashed,line width=1.2pt] (-0.2,1.4) to (-0.2,-1.4);
        \draw[RoyalBlue,thick,dashed, line width=1.2pt] (-1.3,0) to (1.7,0);
        
        \end{tikzpicture}
    \end{subfigure}
    \hfill
    \begin{subfigure}[b]{0.3\linewidth}
    \hspace{.5cm}
        \begin{tikzpicture}[scale=1]
        \coordinate (a) at (0,.8);
        \coordinate (b) at (.8,0);
        \coordinate (c) at (0,-.8);
        \coordinate (d) at (-.8,0);

        \draw[] (a) to (b) to (c) to (d) to (a);
        \draw[] (a) to (c);
        \draw[line width=1.4pt] (-1.2,0) to (d);
        \draw[] (0,1.2) to (a);
        \draw[] (1.2,0) to (b);
        \draw[] (-0.2,-1.2) to (c);
        \draw[] (0.2,-1.2) to (c);
        

        \node[] at (-1.5,0) {$1$};
        \node[] at (0,1.5) {$4$};
        \node[] at (1.5,0) {$5$};
        \node[] at (-0.3,-1.5) {$2$};
        \node[] at (0.3,-1.5) {$3$};

        \draw[OliveGreen,thick,dashed, line width=1.2pt] (-0.7,-0.7) to (0.7,0.7);
        \draw[RedViolet,thick,dashed,line width=1.2pt] (-0.7,0.7) to (0.7, -0.7);
        \end{tikzpicture}
    \end{subfigure}
    \hfill
    \begin{subfigure}[b]{0.3\linewidth}
     \centering
    \begin{tikzpicture}[scale=0.8]
        \coordinate (a) at (0.5,.8);
        \coordinate (c) at (2.1,-.8);
        \coordinate (d) at (0.5,-.8);
        \coordinate (b) at (2.1,0.8);

        \coordinate (a1) at (0.1, 1.2);
        \coordinate (d1) at (0.1, -1.2);
        \coordinate (b1) at (2.8, 1.1);
        \coordinate (b2) at (2.6, 1.3);
        \coordinate (b3) at (2.4, 1.5);
        \coordinate (c1) at (2.5, -1.2);
        
        \draw[] (a1) to (a);
        \draw[] (d1) to (d);
        \draw[] (b1) to (b);
        \draw[] (b2) to (b);
        \draw[] (b3) to (b);
        \draw[] (c1) to (c);
        \draw[] (a) to (b) to (c) to (d) to (a);

        \node[] at (2.9,1.6) {$4$};
        \node[] at (2.5,1.8) {$5$};
        \node[] at (3.1,1.2) {$3$};
        \node[] at (2.7,-1.4) {$2$};
        \node[] at (-0.1,1.4) {$6$};
        \node[] at (-0.1,-1.4) {$1$};

        \draw[Red,thick,dashed,line width=1.2pt] (0,0) to (2.6,0);
        \draw[cyan,thick,dashed,line width=1.2pt] (1.3,1.4) to (1.3,-1.4);
    \end{tikzpicture}  
    \end{subfigure}
    
    \begin{tikzpicture}[overlay, remember picture,decoration={markings,mark=at position .99 with {\arrow[scale=1.4,>=stealth]{>}}}]
    \draw [solid,color=Black] ([yshift=3.4cm,xshift=5.2cm]{pic cs:three_examples}) -- ([yshift=-3.7cm,xshift=5.2cm]{pic cs:three_examples});
    \draw [solid,color=Black] ([yshift=3.4cm,xshift=10.5cm]{pic cs:three_examples}) -- ([yshift=-3.7cm,xshift=10.5cm]{pic cs:three_examples});
    \end{tikzpicture}
    
    \caption{Three examples of Steinmann-violating pairs of channels. We depict both the partially-overlapping cut of the two-loop integral topology (above), and the corresponding cut graph that appears in~\eqref{eq:2loopViolations} (below).}
    \label{fig:two_loop_steinmann_violation_examples}
\end{figure}

\begin{table}[b]
\centering
\begin{tabular}{|c||c|c|c|c|c|c|c|}
\hline
\makecell{Family \\ Topology \\[.3cm] \ } &     \begin{tikzpicture}[scale=0.3]
        \coordinate (a) at (-.8,.8);
        \coordinate (b) at (.8,.8);
        \coordinate (c) at (.8,-.8);
        \coordinate (d) at (-.8,-.8);
        \coordinate (e) at (2.2, 1.2);
        \coordinate (f) at (3, 0);
        \coordinate (g) at (2.2, -1.2);

        \coordinate (a1) at (-1.2, 1.2);
        \coordinate (d1) at (-1.2, -1.2);
        \coordinate (e1) at (2.6, 1.6);
        \coordinate (f1) at (3.4, 0);
        \coordinate (g1) at (2.6, -1.6);
        
        \draw[] (a) to (b) 
                    to (c) 
                    to (d)
                    to (a);
        \draw[] (b) to (e)
                    to (f)
                    to (g)
                    to (c);
        
        \draw[] (a1) to (a);
        \draw[] (d1) to (d);
        \draw[] (e1) to (e);
        \draw[] (f1) to (f);
        \draw[] (g1) to (g);

        \coordinate (ff) at (0,-1.9);
        \node at (ff)[circle,fill=white,inner sep=1pt]{};

    \end{tikzpicture} &
    \begin{tikzpicture}[scale=0.3]
        \coordinate (a) at (-.8,0);
        \coordinate (b) at (0,.8);
        \coordinate (c) at (.8,0);
        \coordinate (d) at (0,-.8);
        \coordinate (e) at (2.2, 1.2);
        \coordinate (f) at (3, 0);
        \coordinate (g) at (2.2, -1.2);

        \coordinate (a1) at (-1.2, 0);
        \coordinate (c1) at (1.4, 0);
        \coordinate (e1) at (2.6, 1.6);
        \coordinate (f1) at (3.4, 0);
        \coordinate (g1) at (2.6, -1.6);
        
        \draw[] (a) to (b) 
                    to (c) 
                    to (d)
                    to (a);
        \draw[] (b) to (e)
                    to (f)
                    to (g)
                    to (d);
        
        \draw[] (a1) to (a);
        \draw[] (c1) to (c);
        \draw[] (e1) to (e);
        \draw[] (f1) to (f);
        \draw[] (g1) to (g);

        \coordinate (ff) at (0,-1.9);
        \node at (ff)[circle,fill=white,inner sep=1pt]{};

    \end{tikzpicture}  &
    \begin{tikzpicture}[scale=0.3]
        \coordinate (a) at (-.8,.8);
        \coordinate (b) at (.8,.8);
        \coordinate (c) at (.8,-.8);
        \coordinate (d) at (-.8,-.8);
        \coordinate (e) at (2.4, .8);
        \coordinate (f) at (2.4, -.8);
        \coordinate (g) at (.8,0);

        \coordinate (a1) at (-1.2, 1.2);
        \coordinate (d1) at (-1.2, -1.2);
        \coordinate (e1) at (2.8, 1.2);
        \coordinate (f1) at (2.8, -1.2);
        \coordinate (g1) at (1.3,0);
        
        \draw[] (a) to (b) 
                    to (c) 
                    to (d)
                    to (a);
        \draw[] (b) to (e)
                    to (f)
                    to (c);
        
        \draw[] (a1) to (a);
        \draw[] (d1) to (d);
        \draw[] (e1) to (e);
        \draw[] (f1) to (f);
        \draw[] (g1) to (g);

        \coordinate (ff) at (0,-1.9);
        \node at (ff)[circle,fill=white,inner sep=1pt]{};

    \end{tikzpicture} &
    \begin{tikzpicture}[scale=0.3]
        \coordinate (a) at (-.8,.8);
        \coordinate (b) at (.8,.8);
        \coordinate (c) at (.8,-.8);
        \coordinate (d) at (-.8,-.8);
        \coordinate (e) at (2.2, 1.2);
        \coordinate (f) at (3, 0);
        \coordinate (g) at (2.2, -1.2);

        \coordinate (a1) at (-1.2, 1.2);
        \coordinate (d1) at (-1.2, -1.2);
        \coordinate (e1) at (2.6, 1.6);
        \coordinate (f1) at (3.4, 0);
        \coordinate (g1) at (2.6, -1.6);
        
        \draw[] (a) to (b) 
                    to (c) 
                    to (d)
                    to (a);
        \draw[] (b) to (e)
                    to (f)
                    to (g)
                    to (c);
        
        \draw[] (a1) to (a);
        \draw[] (d1) to (d);
        \draw[] (e1) to (e);
        \draw[] (f1) to (f);
        \draw[] (g1) to (g);

        \coordinate (ff) at (0,-1.9);
        \node at (ff)[circle,fill=white,inner sep=1pt]{};

    \end{tikzpicture} &
    \begin{tikzpicture}[scale=0.3]
        \coordinate (a) at (-.8,.8);
        \coordinate (b) at (.8,.8);
        \coordinate (c) at (.8,-.8);
        \coordinate (d) at (-.8,-.8);
        \coordinate (e) at (2.4, .8);
        \coordinate (f) at (2.4, -.8);

        \coordinate (a1) at (-1.2, 1.2);
        \coordinate (d1) at (-1.2, -1.2);
        \coordinate (e1) at (2.8, 1.2);
        \coordinate (f1) at (2.8, -1.2);
        \coordinate (b1) at (.8,1.2);
        \coordinate (c1) at (.8, -1.2);
        
        \draw[] (a) to (b) 
                    to (c) 
                    to (d)
                    to (a);
        \draw[] (b) to (e)
                    to (f)
                    to (c);
        
        \draw[] (a1) to (a);
        \draw[] (d1) to (d);
        \draw[] (e1) to (e);
        \draw[] (f1) to (f);
        \draw[] (b1) to (b);
        \draw[] (c1) to (c);

        \coordinate (ff) at (0,-1.9);
        \node at (ff)[circle,fill=white,inner sep=1pt]{};

    \end{tikzpicture} & 
    \begin{tikzpicture}[scale=0.3]
        \coordinate (a) at (-.8,1.2);        
        \coordinate (b) at (-1.4,0);
        \coordinate (c) at (-.8,-1.2);
        \coordinate (e) at (.8,.8);
        \coordinate (d) at (.8,-.8);
        \coordinate (g) at (2, 0);

        \coordinate (a1) at (-1.4, 1.4);
        \coordinate (b1) at (-2, 0);
        \coordinate (c1) at (-1.4, -1.4);
        \coordinate (e1) at (1.2, 1.2);
        \coordinate (d1) at (1.2, -1.2);
        \coordinate (g1) at (2.4, 0);
        
        \draw[] (a) to (b) 
                    to (c) 
                    to (d)
                    to (e)
                    to (a);
        \draw[] (e) to (g)
                    to (d);
        
        \draw[] (a1) to (a);
        \draw[] (d1) to (d);
        \draw[] (e1) to (e);
        \draw[] (g1) to (g);
        \draw[] (b1) to (b);
        \draw[] (c1) to (c);

        \coordinate (ff) at (0,-1.9);
        \node at (ff)[circle,fill=white,inner sep=1pt]{};

    \end{tikzpicture}&
    \begin{tikzpicture}[scale=0.3]
        \coordinate (a) at (-.6,.8);
        \coordinate (b) at (.8,1.3);
        \coordinate (c) at (.8,-1.3);
        \coordinate (d) at (-.6,-.8);
        \coordinate (e) at (2, .8);
        \coordinate (f) at (2, -.8);

        \coordinate (a1) at (-1.2, 1.2);
        \coordinate (d1) at (-1.2, -1.2);
        \coordinate (e1) at (2.8, 1.2);
        \coordinate (f1) at (2.8, -1.2);
        \coordinate (b1) at (.8,1.8);
        \coordinate (c1) at (.8, -1.8);
        
        \draw[] (b) to (a) 
                    to (d) 
                    to (c);
        \draw[] (b) to (e)
                    to [out=-140, in = 140] (f);
        \draw[] (e) to [out=-40, in = 40] (f)    
                    to (c);
        
        \draw[] (a1) to (a);
        \draw[] (d1) to (d);
        \draw[] (e1) to (e);
        \draw[] (f1) to (f);
        \draw[] (b1) to (b);
        \draw[] (c1) to (c);

        \coordinate (ff) at (0,-1.9);
        \node at (ff)[circle,fill=white,inner sep=1pt]{};

        \coordinate (fff) at (0,1.9);
        \node at (fff)[circle,fill=white,inner sep=1pt]{};

    \end{tikzpicture}
    \\[-.8cm] \hline
    \makecell{Number of \\ External Masses} & 1 & 1 & 1 & 2 & 2 & 0 & 0 \\\hline
    \makecell{Number of \\ Master Integrals} & 235 & 307 & 321 & 621 & 66 & 44 & 32 \\\hline
    \makecell{Number of \\ Unique Topologies} & 163 & 191 & 167 & 547 & 49 & 23 & 8 \\\hline
\end{tabular}

\caption{Summary of the families of integrals (drawn from~\cite{Abreu:2020jxa,Abreu:2021smk,Abreu:2024yit,Henn:2024ngj}) against which we have tested our two-loop Steinmann violation conjecture. When we indicate the number of external masses, we mean that all families of integrals with this many massive external momenta (placed at any inequivalent set of vertices) were considered.}
\label{tab:two-loop-data}
\end{table}

We have carried out similar checks in all of the 1,626 master integrals listed above; these families of integrals are summarized in Table~\ref{tab:two-loop-data}. We find that our two-loop conjecture is obeyed in every one of these integrals, with respect to every partially-overlapping channel. Note that this set of integrals includes many examples that involve numerators and propagators raised to higher powers, so this is not a special property of scalar Feynman integrals. Moreover, the predictions of this conjecture become stronger when considering subsector integrals, since the partially-overlapping cuts that appear in~\eqref{eq:2loopViolations} often disappear when further propagators are contracted. Even so, we find many examples in which Steinmann is violated; three such examples are depicted in Figure~\ref{fig:two_loop_steinmann_violation_examples}.

In fact, for the set of integrals described by Table~\ref{tab:two-loop-data}, we find a stronger result than implied by our two-loop conjecture. Namely, the Steinmann relations are violated \emph{if and only if} one of the partially-overlapping cuts in~\eqref{eq:2loopViolations} is present. More specifically, although the presence of one of the cuts from~\eqref{eq:2loopViolations} does not imply Steinmann violation at every order in the $\epsilon$ expansion (and correspondingly, at every weight), we find that all possible violations do in fact occur in these integrals by weight five. (In fact, for some of the families --- such as the six-point pentatriangle and hexabubble, and two of the nonplanar one-mass hexabox configurations ---  this saturation is already achieved by weight four.) Thus, while it is possible to construct Feynman integrals in which one of the minimal cuts in~\eqref{eq:2loopViolations} appears, but in which Steinmann is not violated to any order in $\epsilon$, it seems that such examples do not naturally occur.\footnote{As an example, in the two-mass hard integral in Figure~\ref{fig:2mhard1}, multiplying by the numerator $\ell_1 \cdot p_2$ removes all Steinmann-violating terms. Intuitively, this is expected because this numerator vanishes precisely on the relevant Landau solution, $\ell_1^{\mu} = 0$. More explicitly, one can rewrite
$\ell_1 \cdot p_2 = \tfrac{1}{2}\big[(\ell_1 + p_2)^2 - \ell_1^2\big]$, so that the integral decomposes into a sum of two triangle integrals. Each triangle contains only one of the singularities, $s_{12}$ or $s_{234}$, but not both. Since no single diagram contains both singularities simultaneously, there is no possibility for the Steinmann relations to be violated.}

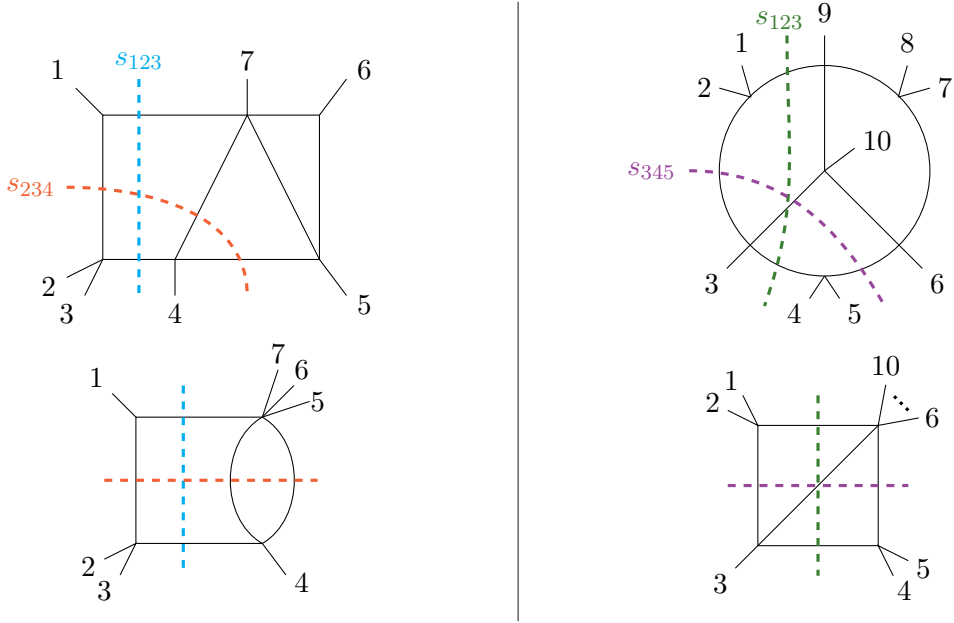
\begin{figure}
    \centering
    \tikzmark{3loop_examples}
\begin{subfigure}[t]{.45\textwidth}
\centering
        \begin{tikzpicture}[scale=1.2]
        \coordinate (a) at (-.8,.8);
        \coordinate (b) at (0.8,.8);
        \coordinate (c) at (1.6,.8);
        \coordinate (d) at (1.6,-.8);
        \coordinate (e) at (0,-.8);
        \coordinate (f) at (-.8,-.8);

        \draw[] (a) to (b) to (c) to (d) to (e) to (f) to (a);
        \draw[] (e) to (b) to (d);

        \draw[] (-1.1,1.1) to (-0.8,.8);
        \draw[] (0.8,1.2) to (0.8,0.8);
        \draw[] (1.9,1.2) to (1.6,0.8);
        \draw[] (1.9,-1.2) to (1.6,-0.8);
        \draw[] (0,-1.2) to (0,-0.8);
        \draw[] (-1.2,-1) to (-0.8,-0.8);
        \draw[] (-1,-1.2) to (-0.8,-0.8);

        \node[] at (-1.3,1.3) {$1$};
        \node[] at (0.8,1.4) {$7$};
        \node[] at (2.1,1.3) {$6$};
        \node[] at (2.1,-1.3) {$5$};
        \node[] at (0,-1.4) {$4$};
        \node[] at (-1.2,-1.4) {$3$};
        \node[] at (-1.4,-1.1) {$2$};

        \draw[RedOrange,thick,dashed,line width=1.2pt] (-1.2,0) to [out=0,in=90](0.8,-1.2);
        \draw[cyan,thick,dashed,line width=1.2pt] (-0.4,1.2) to (-0.4,-1.2);

        \node[RedOrange] at (-1.6,0) {$s_{234}$};
        \node[cyan] at (-0.4,1.4) {$s_{123}$};

        \end{tikzpicture}
\end{subfigure}
\hfill
\begin{subfigure}[t]{.45\textwidth}
\centering
\hspace{-1cm}
    \begin{tikzpicture}[scale=1]
        \draw (0,0) circle (1.4);
        \draw (0,1.4) to (0,1.8);
        \draw (0,-1.4) to (-0.2,-1.7);
        \draw (0,-1.4) to (0.2,-1.7);
        \draw (-0.98,0.98) to (-1.4,1.1);
        \draw (-0.98,0.98) to (-1.1,1.4);
        \draw (0.98,0.98) to (1.4,1.1);
        \draw (0.98,0.98) to (1.1,1.4);
        \draw (-0.98,-0.98) to (-1.3,-1.3);
        \draw (-0.98,-0.98) to (0,0);
        \draw (0.98,-0.98) to (0,0);
        \draw (0.98,-0.98) to (1.3,-1.3);
        \draw (0,0) to (0,1.4);
        \draw (0,0) to (.4,.3);

        \node[] at (-1.1,1.7) {$1$};
        \node[] at (-1.6,1.1) {$2$};
        \node[] at (-1.5,-1.5) {$3$};
        \node[] at (-0.4,-1.9) {$4$};
        \node[] at (0.4,-1.9) {$5$};
        \node[] at (1.5,-1.5) {$6$};
        \node[] at (1.1,1.7) {$8$};
        \node[] at (1.6,1.1) {$7$};
        \node[] at (0,2.1) {$9$};
        \node[] at (.7,.4) {$10$};

        \node[Purple] at (-2.3,0) {$s_{345}$};
        \node[OliveGreen] at (-0.6,2) {$s_{123}$};

        \draw[OliveGreen,thick,dashed,line width=1.2pt] (-0.5,1.8) to [out=-90,in=70](-0.8,-1.8);
        \draw[Purple,thick,dashed,line width=1.2pt] (-1.8,0) to [out=0,in=120](0.8,-1.8);
        
    \end{tikzpicture}
\end{subfigure}
\hfill
\begin{subfigure}[b]{.45\textwidth}
\centering
\hspace{0.5cm}
    \begin{tikzpicture}[scale=1.05]
        \coordinate (a) at (-.8,.8);
        \coordinate (b) at (.8,.8);
        \coordinate (c) at (.8,-.8);
        \coordinate (d) at (-.8,-.8);

        \draw[] (a) to (b) to [out=-30,in=30](c) to (d)to (a);
        \draw[] (b) to [out=-150,in=150](c);
        
        \draw[] (-1.1,1.1) to (a);
        \draw[] (1.2,1.2) to (b);
        \draw[] (1.4,1) to (b);
        \draw[] (1,1.4) to (b);
        \draw[] (1.1, -1.2) to (c);
        \draw[] (-1.2, -1) to (d) to (-1,-1.2);

        \node[] at (-1.2,-1.4) {$3$};
        \node[] at (-1.4,-1.1) {$2$};
        \node[] at (-1.3, 1.3) {$1$};
        \node[] at (1.3,1.4) {$6$};
        \node[] at (1.5,1) {$5$};
        \node[] at (1,1.6) {$7$};
        \node[] at (1.3,-1.3) {$4$};

        \draw[cyan,thick,dashed,line width=1.2pt] (-0.2,1.2) to (-0.2,-1.2);
        \draw[RedOrange,thick,dashed, line width=1.2pt] (-1.2,0) to (1.5,0);
    \end{tikzpicture}
\end{subfigure}
\hfill
\begin{subfigure}[b]{.45\textwidth}
\centering
    \begin{tikzpicture}[scale=1]
        \coordinate (a) at (-.8,.8);
        \coordinate (b) at (.8,.8);
        \coordinate (c) at (.8,-.8);
        \coordinate (d) at (-.8,-.8);
        
        \draw[] (a) to (b) to (c) to (d) to (a);
        \draw[] (d) to (b);

        \draw[] (-1,1.2) to (a) to (-1.2,1);
        \draw[] (-1.1,-1.1) to (d);
        \draw[] (1.2,-1) to (c) to (1,-1.2);
        \draw[] (1.34,0.9) to (b) to (0.9,1.34);

        \node[] at (-1.15,1.4) {$1$};
        \node[] at (-1.4,1.1) {$2$};
        \node[] at (-1.3,-1.3) {$3$};
        \node[] at (1.15,-1.4) {$4$};
        \node[] at (1.4,-1.1) {$5$};
        \node[] at (1.52,0.9) {$6$};
        \node[] at (0.9,1.6) {$10$};
        \draw [dotted, line width=1.2pt] (1.2, 1) -- (1,1.2);

        \draw[OliveGreen,thick,dashed,line width=1.2pt] (0,1.2) to (0,-1.2);
        \draw[Purple,thick,dashed,line width=1.2pt] (-1.2,0) to (1.2,0);
          
    \end{tikzpicture}
\end{subfigure}

 \begin{tikzpicture}[overlay, remember picture,decoration={markings,mark=at position .99 with {\arrow[scale=1.4,>=stealth]{>}}}]
    \draw [solid,color=Black] ([yshift=4.4cm,xshift=8cm]{pic cs:3loop_examples}) -- ([yshift=-4.8cm,xshift=8cm]{pic cs:three_examples});
    \end{tikzpicture}

        \caption{Two three-loop diagrams that contain partially-overlapping channels which only have a single minimal  cut. This minimal cut corresponds to a two-loop diagram, but does not match any of the cuts in~\eqref{eq:2loopViolations}.}
    \label{fig:three_loop_cuts}
\end{figure}

Our two-loop conjecture also holds implications for higher-loop diagrams. For instance, consider the pair of three-loop diagrams shown in Figure~\ref{fig:three_loop_cuts}.\footnote{Note that the second of these diagrams is a nonplanar version of the wheel diagram considered in~\cite{Bourjaily:2019hmc}, which cannot be evaluated in terms of multiple polylogarithms. We emphasize that the Steinmann relations (and other classes of constraints that apply to the sequential discontinuities of Feynman integrals) can be shown to hold no matter what types of special functions appear.} These diagrams contain partially-overlapping channels that have a unique minimal cut. Since these minimal cuts correspond to two-loop diagrams, but do not match any of the cuts in~\eqref{eq:2loopViolations}, our conjecture implies that all of the integrals associated with these topologies should respect the Steinmann relations for these channels. Clearly, it will be possible to find similar predictions at all loop orders for Feynman integrals that have amenable topologies. 

Currently, we know of no examples of Steinmann violation that occur in the absence of a partially-overlapping cut in the class~\eqref{eq:violating_minimal_cuts}. Thus, we remain optimistic that a much stronger version of this two-loop conjecture actually holds --- namely, that Steinmann violation only occurs when a minimal cut of the form~\eqref{eq:violating_minimal_cuts} is present. In the absence of the same abundance of three- and higher-loop data to study, however, we leave this appealing possibility to future exploration.

\newpage
\section{Conclusions}
\label{sec:conclusions}
The Steinmann relations have played a central role in the modern amplitude bootstrap program. Even so, these relations have --- in practice --- only been employed to constrain double discontinuities with respect to three- and higher-particle channels, since the (na\"ive versions of the) Steinmann relations are known to be violated by discontinuities in massless two-particle channels. In this paper, we have sought to improve this situation by developing a more precise understanding of when (and why) the Steinmann relations are violated. We have done this by analyzing the singularities that lead to this violation at one loop, and identifying analogous singularities at higher loop orders. We then went on to conjecture an even stronger statement: that no further Steinmann-violating singularities appear at two loops. Leveraging this conjecture, we were able to outline a simple graphical test for when Steinmann should be violated at two loops, and at higher loops when only two-loop partially-overlapping cuts appear.

We expect this new method for identifying instances of Steinmann violation will prove useful for bootstrap approaches, as well as in the context of differential equation methods when formulating boundary conditions. We highlight, however, that our conjecture only applies to integrals that involve massless virtual particles. As mentioned in Section~\ref{sec:landau_review}, the Steinmann relations are also known to be violated in integrals with massive propagators. It would therefore be interesting to explore whether there exists a similar story involving minimal cuts for when Steinmann is violated in massive integrals.

Even in the massless case, our predictions beyond one loop remain conjectural. We thus invite further exploration of whether Steinmann-violating singularities appear at higher loops, that go beyond the cut configurations in~\eqref{eq:violating_minimal_cuts}. Alternately, it would be interesting if the absence of such singularities could be rigorously established; such a result would imply we could reliably extend our two-loop graphical test for massless Steinmann violation to all loop orders, and predict that the Steinmann relations will not be violated whenever there are no cuts of the form~\eqref{eq:violating_minimal_cuts} present (to any loop order). 

Finally, let us highlight some further mysteries whose investigation may prove insightful. Why do all the Steinmann-violating singularities we have identified lie so close to infrared singularities? For instance, recall the one-loop $s_{234} = 0$ singular configuration we found in~\eqref{eq:momentumspace2mhardsol2}. This solution sits on top of another solution to the Landau equations, which also sets 
\begin{gather}
    \ell^{\mu}=0\, ,\ \alpha_{2} = \alpha_3 = \alpha_4 =0 \, .\label{eq:2mhardIR}
 \end{gather}
This new solution imposes no constraints on the external kinematics, and hence corresponds to an infrared divergence. However, importantly, solution~\eqref{eq:2mhardIR} only requires the propagators associated with $\alpha_1$, $\alpha_2$, and $\alpha_4$ to be on shell. Thus, the $s_{234}=0$ solution to the Landau equations only appears when the fourth propagator is placed on shell. Similar infrared-divergent momentum configurations can be found that intersect the other Steinmann-violating singularities found in Sections~\ref{sec:steinmann_viiolating_solutions} and~\ref{sec:all_loop_solutions}; however, we have not yet discerned whether or not there exists a deep connection between these kinematic and infrared singularities. Relatedly, it is worth noting that non-factorizing contributions to one-loop gauge theory amplitudes arise in conjunction with precisely the same class of two-mass-hard box configurations that give rise to Steinmann violation~\cite{Bern:1995ix}. It would be interesting to understand the precise connection between these two phenomena, and whether they can be traced back to the same underlying physics. We leave these interesting questions for future investigation.

\section*{Acknowledgments}

We thank Samuel Abreu, Zvi Bern, Claude Duhr, Einan Gardi, and Cristian Vergu for insightful discussions.
H.S.H. gratefully acknowledges funding provided by the J. Robert Oppenheimer Endowed Fund of the Institute for Advanced Study, while AJM and LL acknowledge support from the Royal Society grant
URF{\textbackslash}R1{\textbackslash}221233. AJM additionally acknowledges support from the European Research Council (ERC) under the European Union’s Horizon Europe research and innovation program grant agreement 101163627 (ERC Starting Grant ``AmpBoot''). This material is based upon work supported by the U.S. Department of Energy, Office of Science, Office of High Energy Physics under Award Number DE-SC0009988.

\newpage

\bibliographystyle{apsrev4-1}
\bibliography{refs}

@article{araki:1961,
	author = {Araki,Huzihiro},
	doi = {10.1063/1.1703695},
	journal = {Journal of Mathematical Physics},
	number = {2},
	pages = {163-177},
	title = {Generalized Retarded Functions and Analytic Function in Momentum Space in Quantum Field Theory},
	url = {https://doi.org/10.1063/1.1703695},
	volume = {2},
	year = {1961}}

@article{Dixon:2022rse,
    author = "Dixon, Lance J. and Gurdogan, Omer and McLeod, Andrew J. and Wilhelm, Matthias",
    title = "{Bootstrapping a stress-tensor form factor through eight loops}",
    eprint = "2204.11901",
    archivePrefix = "arXiv",
    primaryClass = "hep-th",
    reportNumber = "SLAC-PUB-17653, CERN-TH-2022-039",
    doi = "10.1007/JHEP07(2022)153",
    journal = "JHEP",
    volume = "07",
    pages = "153",
    year = "2022"
}

@article{Ellis:2007qk,
    author = "Ellis, R. Keith and Zanderighi, Giulia",
    title = "{Scalar one-loop integrals for QCD}",
    eprint = "0712.1851",
    archivePrefix = "arXiv",
    primaryClass = "hep-ph",
    reportNumber = "FERMILAB-PUB-07-633-T, OUTP-07-16P",
    doi = "10.1088/1126-6708/2008/02/002",
    journal = "JHEP",
    volume = "02",
    pages = "002",
    year = "2008"
}

@article{Dennen_2017,
   title={Landau singularities from the amplituhedron},
   volume={2017},
   ISSN={1029-8479},
   url={http://dx.doi.org/10.1007/JHEP06(2017)152},
   DOI={10.1007/jhep06(2017)152},
   number={6},
   journal={Journal of High Energy Physics},
   publisher={Springer Science and Business Media LLC},
   author={Dennen, T. and Prlina, I. and Spradlin, M. and Stanojevic, S. and Volovich, A.},
   year={2017},
   month=jun }

@article{Hannesdottir:2024hke,
    author = "Hannesdottir, Holmfridur and McLeod, Andrew and Schwartz, Matthew D. and Vergu, Cristian",
    title = "{The Landau Bootstrap}",
    eprint = "2410.02424",
    archivePrefix = "arXiv",
    primaryClass = "hep-ph",
    month = "10",
    year = "2024"
}

@book{Hannesdottir:2022bmo,
    author = "Hannesdottir, Holmfridur Sigridar and Mizera, Sebastian",
    title = "{What is the i\ensuremath{\varepsilon} for the S-matrix?}",
    eprint = "2204.02988",
    archivePrefix = "arXiv",
    primaryClass = "hep-th",
    doi = "10.1007/978-3-031-18258-7",
    isbn = "978-3-031-18257-0, 978-3-031-18258-7",
    publisher = "Springer",
    series = "SpringerBriefs in Physics",
    month = "1",
    year = "2023"
}

@article{Hannesdottir:2024cnn,
    author = "Hannesdottir, Holmfridur S. and Lippstreu, Luke and McLeod, Andrew J. and Polackova, Maria",
    title = "{Minimal Cuts and Genealogical Constraints on Feynman Integrals}",
    eprint = "2406.05943",
    archivePrefix = "arXiv",
    primaryClass = "hep-th",
    month = "6",
    year = "2024"
}

@article{Caron-Huot:2023ikn,
    author = "Caron-Huot, Simon and Giroux, Mathieu and Hannesdottir, Holmfridur S. and Mizera, Sebastian",
    title = "{Crossing beyond scattering amplitudes}",
    eprint = "2310.12199",
    archivePrefix = "arXiv",
    primaryClass = "hep-th",
    doi = "10.1007/JHEP04(2024)060",
    journal = "JHEP",
    volume = "04",
    pages = "060",
    year = "2024"
}

@article{Mizera:2021icv,
    author = "Mizera, Sebastian and Telen, Simon",
    title = "{Landau discriminants}",
    eprint = "2109.08036",
    archivePrefix = "arXiv",
    primaryClass = "math-ph",
    doi = "10.1007/JHEP08(2022)200",
    journal = "JHEP",
    volume = "08",
    pages = "200",
    year = "2022"
}

@article{nakanishi1959,
    author = {Nakanishi, Noboru},
    title = "{Ordinary and Anomalous Thresholds in Perturbation Theory}",
    journal = {Progress of Theoretical Physics},
    volume = {22},
    number = {1},
    pages = {128-144},
    year = {1959},
    month = {07},
    issn = {0033-068X},
    doi = {10.1143/PTP.22.128},
    url = {https://doi.org/10.1143/PTP.22.128}
}

@article{Dixon:2022xqh,
    author = {Dixon, Lance J. and G\"urdo\u{g}an, \"Omer and Liu, Yu-Ting and McLeod, Andrew J. and Wilhelm, Matthias},
    title = "{Antipodal Self-Duality for a Four-Particle Form Factor}",
    eprint = "2212.02410",
    archivePrefix = "arXiv",
    primaryClass = "hep-th",
    reportNumber = "CERN-TH-2022-190, SLAC-PUB-17711",
    doi = "10.1103/PhysRevLett.130.111601",
    journal = "Phys. Rev. Lett.",
    volume = "130",
    number = "11",
    pages = "111601",
    year = "2023"
}

@article{Fevola:2023fzn,
    author = "Fevola, Claudia and Mizera, Sebastian and Telen, Simon",
    title = "{Principal Landau Determinants}",
    eprint = "2311.16219",
    archivePrefix = "arXiv",
    primaryClass = "math-ph",
    month = "11",
    year = "2023"
}

@article{Abreu:2023rco,
    author = "Abreu, Samuel and Chicherin, Dmitry and Ita, Harald and Page, Ben and Sotnikov, Vasily and Tschernow, Wladimir and Zoia, Simone",
    title = "{All Two-Loop Feynman Integrals for Five-Point One-Mass Scattering}",
    eprint = "2306.15431",
    archivePrefix = "arXiv",
    primaryClass = "hep-ph",
    doi = "10.1103/PhysRevLett.132.141601",
    journal = "Phys. Rev. Lett.",
    volume = "132",
    number = "14",
    pages = "141601",
    year = "2024"
}

@article{Abreu:2024yit,
    author = "Abreu, Samuel and Chicherin, Dmitry and Sotnikov, Vasily and Zoia, Simone",
    title = "{Two-loop five-point two-mass planar integrals and double Lagrangian insertions in a Wilson loop}",
    eprint = "2408.05201",
    archivePrefix = "arXiv",
    primaryClass = "hep-th",
    reportNumber = "CERN-TH-2024-136, LAPTH-043/24, ZU-TH 40/24",
    doi = "10.1007/JHEP10(2024)167",
    journal = "JHEP",
    volume = "10",
    pages = "167",
    year = "2024"
}

@article{Vergu:2025mag,
    author = "Vergu, C.",
    title = "{Landau Analysis in Momentum Space with Massless Particles: an Amuse Bouche}",
    eprint = "2512.12763",
    archivePrefix = "arXiv",
    primaryClass = "hep-th",
    month = "12",
    year = "2025"
}

@article{Bern:1995ix,
    author = "Bern, Zvi and Chalmers, Gordon",
    title = "{Factorization in one loop gauge theory}",
    eprint = "hep-ph/9503236",
    archivePrefix = "arXiv",
    reportNumber = "UCLA-95-TEP-6",
    doi = "10.1016/0550-3213(95)00226-I",
    journal = "Nucl. Phys. B",
    volume = "447",
    pages = "465--518",
    year = "1995"
}

@article{Fevola:2023kaw,
    author = "Fevola, Claudia and Mizera, Sebastian and Telen, Simon",
    title = "{Landau Singularities Revisited: Computational Algebraic Geometry for Feynman Integrals}",
    eprint = "2311.14669",
    archivePrefix = "arXiv",
    primaryClass = "hep-th",
    doi = "10.1103/PhysRevLett.132.101601",
    journal = "Phys. Rev. Lett.",
    volume = "132",
    number = "10",
    pages = "101601",
    year = "2024"
}

@article{Lippstreu:2023oio,
    author = "Lippstreu, Luke and Spradlin, Marcus and Yelleshpur Srikant, Akshay and Volovich, Anastasia",
    title = "{Landau singularities of the 7-point ziggurat. Part II}",
    eprint = "2305.17069",
    archivePrefix = "arXiv",
    primaryClass = "hep-th",
    doi = "10.1007/JHEP01(2024)069",
    journal = "JHEP",
    volume = "01",
    pages = "069",
    year = "2024"
}

@article{He:2024fij,
    author = "He, Song and Jiang, Xuhang and Liu, Jiahao and Yang, Qinglin",
    title = "{Landau-based Schubert analysis}",
    eprint = "2410.11423",
    archivePrefix = "arXiv",
    primaryClass = "hep-th",
    month = "10",
    year = "2024"
}

@article{Berghoff:2022mqu,
    author = "Berghoff, Marko and Panzer, Erik",
    title = "{Hierarchies in relative Picard-Lefschetz theory}",
    eprint = "2212.06661",
    archivePrefix = "arXiv",
    primaryClass = "math-ph",
    month = "12",
    year = "2022"
}

@article{pham,
     author = {Pham, Fr\'ed\'eric},
     title = {Singularit\'es des processus de diffusion multiple},
     journal = {Annales de l'I.H.P. Physique th\'eorique},
     pages = {89--204},
     publisher = {Gauthier-Villars},
     volume = {6},
     number = {2},
     year = {1967},
     zbl = {0154.46102},
     mrnumber = {214341},
     language = {fr},
     url = {http://www.numdam.org/item/AIHPA_1967__6_2_89_0/}
}

@book{pham2011singularities,
  title={Singularities of integrals: Homology, hyperfunctions and microlocal analysis},
  author={Pham, F.},
  isbn={9780857296030},
  series={Universitext},
  url={https://books.google.com/books?id=8Nvg4MpCPrUC},
  year={2011},
  publisher={Springer London}
}

@article{Dixon:2023kop,
    author = "Dixon, Lance J. and Liu, Yu-Ting",
    title = "{An eight loop amplitude via antipodal duality}",
    eprint = "2308.08199",
    archivePrefix = "arXiv",
    primaryClass = "hep-th",
    reportNumber = "SLAC-PUB-17693",
    doi = "10.1007/JHEP09(2023)098",
    journal = "JHEP",
    volume = "09",
    pages = "098",
    year = "2023"
}

@article{Basso:2024hlx,
    author = "Basso, Benjamin and Dixon, Lance J. and Tumanov, Alexander G.",
    title = "{The three-point form factor of Tr \ensuremath{\phi}$^{3}$ to six loops}",
    eprint = "2410.22402",
    archivePrefix = "arXiv",
    primaryClass = "hep-th",
    doi = "10.1007/JHEP02(2025)034",
    journal = "JHEP",
    volume = "02",
    pages = "034",
    year = "2025"
}

@article{Kotikov:1990kg,
    author = "Kotikov, A. V.",
    title = "{Differential equations method: New technique for massive Feynman diagrams calculation}",
    reportNumber = "ITF-90-31E",
    doi = "10.1016/0370-2693(91)90413-K",
    journal = "Phys. Lett. B",
    volume = "254",
    pages = "158--164",
    year = "1991"
}

@article{Kotikov:1991pm,
    author = "Kotikov, A. V.",
    title = "{Differential equation method: The Calculation of N point Feynman diagrams}",
    doi = "10.1016/0370-2693(91)90536-Y",
    journal = "Phys. Lett. B",
    volume = "267",
    pages = "123--127",
    year = "1991",
    note = "[Erratum: Phys.Lett.B 295, 409--409 (1992)]"
}

@article{Lippstreu:2022bib,
    author = "Lippstreu, Luke and Spradlin, Marcus and Volovich, Anastasia",
    title = "{Landau singularities of the 7-point ziggurat. Part I}",
    eprint = "2211.16425",
    archivePrefix = "arXiv",
    primaryClass = "hep-th",
    doi = "10.1007/JHEP07(2024)024",
    journal = "JHEP",
    volume = "07",
    pages = "024",
    year = "2024"
}

@article{Caron-Huot:2020bkp,
    author = {Caron-Huot, Simon and Dixon, Lance J. and Drummond, James M. and Dulat, Falko and Foster, Jack and G\"urdo\u{g}an, \"Omer and von Hippel, Matt and McLeod, Andrew J. and Papathanasiou, Georgios},
    title = "{The Steinmann Cluster Bootstrap for $N$ = 4 Super Yang-Mills Amplitudes}",
    eprint = "2005.06735",
    archivePrefix = "arXiv",
    primaryClass = "hep-th",
    reportNumber = "DESY-20-087",
    doi = "10.22323/1.376.0003",
    journal = "PoS",
    volume = "CORFU2019",
    pages = "003",
    year = "2020"
}

@article{Landshoff1966,
    Author = {Landshoff, P. V. and Olive, D. I. and Polkinghorne, J. C.},
    Da = {1966/05/01},
    Doi = {10.1007/BF02752870},
    Id = {Landshoff1966},
    Isbn = {1826-9869},
    Journal = {Il Nuovo Cimento A (1971-1996)},
    Number = {2},
    Pages = {444--453},
    Title = {The hierarchical principle in perturbation theory},
    Ty = {JOUR},
    Url = {https://doi.org/10.1007/BF02752870},
    Volume = {43},
    Year = {1966}}

@article{boyling1968homological,
  title={{A homological approach to parametric Feynman integrals}},
  author={Boyling, JB},
  journal={Il Nuovo Cimento A (1965-1970)},
  volume={53},
  number={2},
  pages={351--375},
  year={1968},
  publisher={Springer}
}

@article{Cahill:1973qp,
	author = {Cahill, Kevin E. and Stapp, Henry P.},
	doi = {10.1016/0003-4916(75)90006-8},
	journal = {Annals Phys.},
	pages = {438},
	reportnumber = {LBL-2428},
	slaccitation = {%%CITATION = APNYA,90,438;%%},
	title = {{Optical Theorems and Steinmann Relations}},
	volume = {90},
	year = {1975}}

@article{Remiddi:1997ny,
	archiveprefix = {arXiv},
	author = {Remiddi, Ettore},
	eprint = {hep-th/9711188},
	journal = {Nuovo Cim. A},
	pages = {1435--1452},
	reportnumber = {DFUB-97-15},
	doi={10.1007/BF03185566},
	title = {{Differential equations for Feynman graph amplitudes}},
	volume = {110},
	year = {1997}}

@article{Gehrmann:1999as,
	archiveprefix = {arXiv},
	author = {Gehrmann, T. and Remiddi, E.},
	doi = {10.1016/S0550-3213(00)00223-6},
	eprint = {hep-ph/9912329},
	journal = {Nucl. Phys. B},
	pages = {485--518},
	reportnumber = {TTP-99-49},
	title = {{Differential equations for two loop four point functions}},
	volume = {580},
	year = {2000}}

@article{Abreu:2020jxa,
    author = "Abreu, Samuel and Ita, Harald and Moriello, Francesco and Page, Ben and Tschernow, Wladimir and Zeng, Mao",
    title = "{Two-Loop Integrals for Planar Five-Point One-Mass Processes}",
    eprint = "2005.04195",
    archivePrefix = "arXiv",
    primaryClass = "hep-ph",
    doi = "10.1007/JHEP11(2020)117",
    journal = "JHEP",
    volume = "11",
    pages = "117",
    year = "2020"
}

@article{Abreu:2021smk,
    author = "Abreu, Samuel and Ita, Harald and Page, Ben and Tschernow, Wladimir",
    title = "{Two-loop hexa-box integrals for non-planar five-point one-mass processes}",
    eprint = "2107.14180",
    archivePrefix = "arXiv",
    primaryClass = "hep-ph",
    doi = "10.1007/JHEP03(2022)182",
    journal = "JHEP",
    volume = "03",
    pages = "182",
    year = "2022"
}

@article{Caron-Huot:2016owq,
	archiveprefix = {arXiv},
	author = {Caron-Huot, Simon and Dixon, Lance J. and McLeod, Andrew and von Hippel, Matt},
	doi = {10.1103/PhysRevLett.117.241601},
	eprint = {1609.00669},
	journal = {Phys. Rev. Lett.},
	number = {24},
	pages = {241601},
	primaryclass = {hep-th},
	reportnumber = {SLAC-PUB-16811},
	title = {{Bootstrapping a Five-Loop Amplitude Using Steinmann Relations}},
	volume = {117},
	year = {2016}}

@article{Dixon:2016nkn,
	archiveprefix = {arXiv},
	author = {Dixon, Lance J. and Drummond, James and Harrington, Thomas and McLeod, Andrew J. and Papathanasiou, Georgios and Spradlin, Marcus},
	doi = {10.1007/JHEP02(2017)137},
	eprint = {1612.08976},
	journal = {JHEP},
	pages = {137},
	primaryclass = {hep-th},
	slaccitation = {%%CITATION = ARXIV:1612.08976;%%},
	title = {{Heptagons from the Steinmann Cluster Bootstrap}},
	volume = {02},
	year = {2017}}

@article{Cutkosky:1960sp,
	author = {Cutkosky, R. E.},
	doi = {10.1063/1.1703676},
	journal = {J. Math. Phys.},
	pages = {429--433},
	title = {{Singularities and discontinuities of Feynman amplitudes}},
	volume = {1},
	year = {1960}}

@article{Prlina:2018ukf,
	archiveprefix = {arXiv},
	author = {Prlina, Igor and Spradlin, Marcus and Stanojevic, Stefan},
	doi = {10.1103/PhysRevLett.121.081601},
	eprint = {1805.11617},
	journal = {Phys. Rev. Lett.},
	number = {8},
	pages = {081601},
	primaryclass = {hep-th},
	title = {{All-loop singularities of scattering amplitudes in massless planar theories}},
	volume = {121},
	year = {2018}}

@article{Benincasa:2020aoj,
	archiveprefix = {arXiv},
	author = {Benincasa, Paolo and McLeod, Andrew J. and Vergu, Cristian},
	doi = {10.1103/PhysRevD.102.125004},
	eprint = {2009.03047},
	journal = {Phys. Rev. D},
	pages = {125004},
	primaryclass = {hep-th},
	title = {{Steinmann Relations and the Wavefunction of the Universe}},
	volume = {102},
	year = {2020}}

@article{Weinzierl:2022eaz,
	archiveprefix = {arXiv},
	author = {Weinzierl, Stefan},
	eprint = {2201.03593},
	month = {1},
	primaryclass = {hep-th},
	title = {{Feynman Integrals}},
	year = {2022}}

@article{Bourjaily:2019hmc,
	archiveprefix = {arXiv},
	author = {Bourjaily, Jacob L. and McLeod, Andrew J. and Vergu, Cristian and Volk, Matthias and Von Hippel, Matt and Wilhelm, Matthias},
	doi = {10.1007/JHEP01(2020)078},
	eprint = {1910.01534},
	journal = {JHEP},
	pages = {078},
	primaryclass = {hep-th},
	title = {{Embedding Feynman Integral (Calabi-Yau) Geometries in Weighted Projective Space}},
	volume = {01},
	year = {2020}}

@article{Bourjaily:2020wvq,
    author = "Bourjaily, Jacob L. and Hannesdottir, Holmfridur and McLeod, Andrew J. and Schwartz, Matthew D. and Vergu, Cristian",
    title = "{Sequential Discontinuities of Feynman Integrals and the Monodromy Group}",
    eprint = "2007.13747",
    archivePrefix = "arXiv",
    primaryClass = "hep-th",
    doi = "10.1007/JHEP01(2021)205",
    journal = "JHEP",
    volume = "01",
    pages = "205",
    year = "2021"
}

@article{Drummond:2018caf,
	archiveprefix = {arXiv},
	author = {Drummond, James and Foster, Jack and G{\"u}rdo{\u{g}}an, {\"O}mer and Papathanasiou, Georgios},
	doi = {10.1007/JHEP03(2019)087},
	eprint = {1812.04640},
	journal = {JHEP},
	pages = {087},
	primaryclass = {hep-th},
	reportnumber = {DESY-18-214},
	slaccitation = {%%CITATION = ARXIV:1812.04640;%%},
	title = {{Cluster adjacency and the four-loop NMHV heptagon}},
	volume = {03},
	year = {2019}}

@article{Elvang:2013cua,
	archiveprefix = {arXiv},
	author = {Elvang, Henriette and Huang, Yu-tin},
	eprint = {1308.1697},
	primaryclass = {hep-th},
	slaccitation = {%%CITATION = ARXIV:1308.1697;%%},
	title = {{Scattering Amplitudes}},
	year = {2013}}

@article{Steinmann,
	author = {Steinmann, O},
	journal = {Helv. Physica Acta},
	pages = {257},
	title = {{\"Uber den Zusammenhang zwischen den Wightmanfunktionen und der retardierten Kommutatoren}},
	volume = {33},
	year = {1960}}

@article{Steinmann2,
	author = {Steinmann, O},
	journal = {Helv. Physica Acta},
	pages = {347},
	title = {{Wightman-Funktionen und retardierten Kommutatoren. II}},
	volume = {33},
	year = {1960}}

@article{Drummond:2017ssj,
	archiveprefix = {arXiv},
	author = {Drummond, James and Foster, Jack and G{\"u}rdo{\u{g}}an, {\"O}mer},
	doi = {10.1103/PhysRevLett.120.161601},
	eprint = {1710.10953},
	journal = {Phys. Rev. Lett.},
	number = {16},
	pages = {161601},
	primaryclass = {hep-th},
	slaccitation = {%%CITATION = ARXIV:1710.10953;%%},
	title = {{Cluster Adjacency Properties of Scattering Amplitudes in $\mathcal{N}=4$ Supersymmetric Yang-Mills Theory}},
	volume = {120},
	year = {2018}}

@article{Caron-Huot:2019bsq,
	archiveprefix = {arXiv},
	author = {Caron-Huot, Simon and Dixon, Lance J. and Dulat, Falko and Von Hippel, Matt and McLeod, Andrew J. and Papathanasiou, Georgios},
	doi = {10.1007/JHEP09(2019)061},
	eprint = {1906.07116},
	journal = {JHEP},
	pages = {061},
	primaryclass = {hep-th},
	reportnumber = {DESY 19-062, DESY-19-062, HU-EP-19/05, SLAC--PUB--17414},
	slaccitation = {%%CITATION = ARXIV:1906.07116;%%},
	title = {{The Cosmic Galois Group and Extended Steinmann Relations for Planar $\mathcal{N} = 4$ SYM Amplitudes}},
	volume = {09},
	year = {2019}}

@article{Caron-Huot:2018dsv,
	archiveprefix = {arXiv},
	author = {Caron-Huot, Simon and Dixon, Lance J. and von Hippel, Matt and McLeod, Andrew J. and Papathanasiou, Georgios},
	doi = {10.1007/JHEP07(2018)170},
	eprint = {1806.01361},
	journal = {JHEP},
	pages = {170},
	primaryclass = {hep-th},
	reportnumber = {DESY 18-009, SLAC-PUB-17228, DESY-18-041},
	title = {{The Double Pentaladder Integral to All Orders}},
	volume = {07},
	year = {2018}}

@article{Hannesdottir:2022xki,
    author = "Hannesdottir, Holmfridur S. and McLeod, Andrew J. and Schwartz, Matthew D. and Vergu, Cristian",
    title = "{Constraints on sequential discontinuities from the geometry of on-shell spaces}",
    eprint = "2211.07633",
    archivePrefix = "arXiv",
    primaryClass = "hep-th",
    reportNumber = "CERN-TH-2022-189",
    doi = "10.1007/JHEP07(2023)236",
    journal = "JHEP",
    volume = "07",
    pages = "236",
    year = "2023"}

@article{Henn:2024ngj,
    author = "Henn, Johannes M. and Matija\v{s}i\'c, Antonela and Miczajka, Julian and Peraro, Tiziano and Xu, Yingxuan and Zhang, Yang",
    title = "{A computation of two-loop six-point Feynman integrals in dimensional regularization}",
    eprint = "2403.19742",
    archivePrefix = "arXiv",
    primaryClass = "hep-ph",
    reportNumber = "MPP-2024-53, USTC-ICTS/PCFT-24-11",
    month = "3",
    year = "2024"
}

@article{Stapp:1971hh,
    author = "Stapp, H.P.",
    title = "{Inclusive Cross-Sections are Discontinuities}",
    doi = "10.1103/PhysRevD.3.3177",
    journal = "Phys. Rev. D",
    volume = "3",
    pages = "3177--3184",
    year = "1971"
}

@article{Hannesdottir:2021kpd,
    author = "Hannesdottir, Holmfridur S. and McLeod, Andrew J. and Schwartz, Matthew D. and Vergu, Cristian",
    title = "{Implications of the Landau equations for iterated integrals}",
    eprint = "2109.09744",
    archivePrefix = "arXiv",
    primaryClass = "hep-th",
    doi = "10.1103/PhysRevD.105.L061701",
    journal = "Phys. Rev. D",
    volume = "105",
    number = "6",
    pages = "L061701",
    year = "2022"}

@phdthesis{Bjorken:1959fd,
	author = "Bjorken, James Daniel",
	title = "{Experimental tests of Quantum electrodynamics and spectral representations of Green's functions in perturbation theory}",
	reportNumber = "RX-1037",
	school = "Stanford U.",
	year = "1959"
}

@article{Landau:1959fi,
	author = "Landau, L.D.",
	title = "{On analytic properties of vertex parts in quantum field theory}",
	doi = "10.1016/B978-0-08-010586-4.50103-6",
	journal = "Nucl. Phys.",
	volume = "13",
	number = "1",
	pages = "181--192",
	year = "1960"
}
\end{document}